\documentclass[acmsmall,natbib=true]{acmart}
\setcitestyle{numbers,sort&compress}

\usepackage{multirow}
\usepackage{graphicx}
\usepackage{array}
\usepackage{pbox}
\usepackage{amsmath}
\usepackage{amsthm}
\newtheorem{example}{Example}[section]
\newtheorem{problem}{Problem}
\theoremstyle{remark}
\newtheorem{remark}{\textbf{Remark}}
\usepackage{algorithm}
\usepackage{algpseudocode}
\usepackage{booktabs}
\usepackage{tabularx}
\usepackage{ragged2e} 
\usepackage{subcaption}

\usepackage{tcolorbox}

\usepackage{listings}
\usepackage{xcolor}

\usepackage{enumitem}
\setlist[itemize]{leftmargin=*, nosep, topsep=0pt, partopsep=0pt}
\newcolumntype{C}[1]{>{\centering\arraybackslash}m{#1}}  % For table 1

\definecolor{codebg}{RGB}{248,248,248}
\definecolor{codecomment}{RGB}{0,128,0}
\definecolor{codekeyword}{RGB}{0,0,180}
\definecolor{codestring}{RGB}{163,21,21}

\usepackage{changepage}

\usepackage{enumitem}
\setlist[itemize]{leftmargin=*, topsep=1pt, itemsep=1pt, parsep=0pt}

\AtBeginDocument{%
  }
\setcopyright{acmlicensed}
\copyrightyear{2018}
\acmYear{2018}
\acmDOI{XXXXXXX.XXXXXXX}

\acmJournal{JACM}
\acmVolume{37}
\acmNumber{4}
\acmArticle{111}
\acmMonth{8}

\begin{document}

\title{Learning to Optimize by Differentiable Programming}

\author{Liping Tao}
\email{liping.tao@163.com}
\orcid{0000-0002-9830-4656}
\affiliation{%
  \institution{Nanyang Technological University}
  \city{Singapore}
  \country{Singapore}
}

\author{Xindi Tong}
\email{to0002di@e.ntu.edu.sg}
\orcid{0009-0007-1706-4472}
\affiliation{%
  \institution{Nanyang Technological University}
  \city{Singapore}
  \country{Singapore}
}

\author{Chee Wei Tan}
\authornote{$^{*}$ (Corresponding \ Author)}
\email{cheewei.tan@ntu.edu.sg}
\orcid{0000-0002-6624-9752}
\affiliation{%
  \institution{Nanyang Technological University}
  \city{Singapore}
  \country{Singapore}
}

\renewcommand{\shortauthors}{Tao, Tong and Tan}

\begin{abstract}
Solving massive-scale optimization problems requires scalable first-order methods with low per-iteration cost. This tutorial highlights a shift in optimization: using differentiable programming not only to execute algorithms but to learn how to design them. Modern frameworks such as PyTorch, TensorFlow, and JAX enable this paradigm through efficient automatic differentiation. Embedding first-order methods within these systems allows end-to-end training that improves convergence and solution quality. Guided by Fenchel-Rockafellar duality, the tutorial demonstrates how duality-informed iterative schemes such as the alternating direction method of multipliers, and the primal–dual hybrid gradient can be learned and adapted through representative case studies.
\end{abstract}




\received{20 February 2007}
\received[revised]{12 March 2009}
\received[accepted]{5 June 2009}

\maketitle

\section{Introduction}
\label{sec:intro}
Optimization problems are fundamental in operations research, economics, engineering, and computer science, covering a wide range of formulations and applications \cite{ahmaditeshnizi2023optimus, Avigad2023}. Although advances in optimization algorithms have greatly improved reliability and efficiency, scalability remains a key challenge as computational costs grow and convergence becomes harder to guarantee with increasing problem size. Large-scale and complex optimization tasks face two main difficulties, effective problem formulation and high computational cost \cite{careem2024deep}. Solving such problems often demands significant human expertise and computational resources, since traditional solvers struggle to maintain efficiency as complexity increases \cite{kotary2021end}. 

In recent years, \textit{learning to optimize} has emerged as a promising paradigm at the intersection of machine learning and optimization \cite{tang2024learn}. By leveraging data-driven models, particularly deep learning, it learns optimization strategies from data, enabling efficient generalization across problem instances and often accelerating large-scale or nonconvex optimization compared with traditional solvers \cite{chen2022learning}. The rapid advancement of machine learning has further stimulated the integration of artificial intelligence and optimization technologies~\cite{van2024ai4opt}, leading to new learning-based approaches for solving challenging optimization problems \cite{bengio2021machine}.

Within this landscape, first-order methods ~\cite{beck2017first}, such as gradient descent and its variants, remain fundamental to modern machine learning and artificial intelligence. Proximal algorithms~\cite{parikh2014proximal}, as extensions of first-order methods, are particularly effective for handling constraints and non-smooth regularization. Approaches like the proximal gradient method~\cite{yun2021adaptive}, the Alternating Direction Method of Multipliers (ADMM)~\cite{boyd2011distributed}, and their accelerated variants efficiently solve problems combining smooth and non-smooth components through iterative subproblem decomposition~\cite{wang2019global}. However, proximal algorithms face challenges including computational overhead, hyperparameter sensitivity, and limited scalability in high-dimensional or non-convex settings~\cite{lai2023prox}. The learning-to-optimize framework offers a promising remedy by learning proximal updates or adaptive rules from data~\cite{lai2023prox}, thereby accelerating convergence, enhancing robustness, and extending proximal methods to larger and more complex problems. Nonetheless, a central question persists: how close are the solutions produced by machine learning models to the true optimum? Evaluating and certifying solution quality remains a fundamental issue.

\begin{figure}[!t]
    \centering
    \includegraphics[width=0.96\linewidth, height=3.3cm]{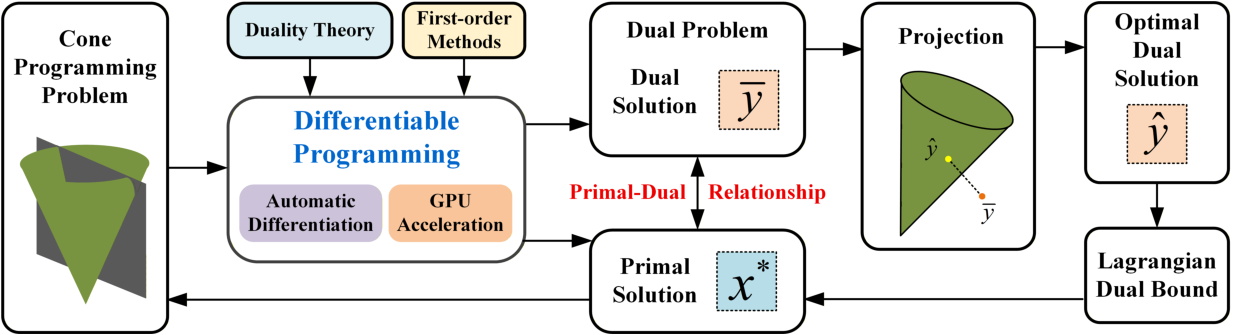}
    \caption{Learning to Optimize via Differentiable Programming by Combining Duality and First-order Methods.}
    \label{fig:diffprog}
\end{figure}

Duality theory provides a principled framework for evaluating, certifying, and improving learned optimization solutions through complementary primal and dual formulations~\cite{tanneau2024dual, tran2021differentially}. Classical Lagrangian duality introduces dual variables to reformulate constrained optimization problems, enabling principled assessment of primal optimality under strong duality while revealing structural relationships between primal and dual variables~\cite{bertsekas2014constrained, komodakis2015playing, ekholm2023duality}. Complementing this perspective, Fenchel-Rockafellar duality~\cite{boyd2004convex, fenchel1949conjugate, rockafellar1997convex, borwein2006convex} provides variational representations for unconstrained and composite objectives, forming the foundation of proximal operators, regularization techniques, and many differentiable optimization layers. Together, these duality principles support modern dual learning, guiding solution refinement and efficient optimization for conic, nonlinear, and large-scale problems~\cite{mandi2024decision}. These properties make duality theory a natural foundation for differentiable programming, where optimization algorithms are embedded into end-to-end trainable computational graphs.

Although first-order methods and duality theory have substantially advanced learning-to-optimize frameworks, embedding optimization within end-to-end trainable systems remains computationally challenging, requiring gradients to propagate through optimization solvers, iterative algorithms, and structured constraints that are not natively differentiable. Recent advances integrate \textit{differentiable programming} with duality theory and first-order methods, establishing a scalable and efficient paradigm for learning to optimize~\cite{blondel2024elements, sapienza2024differentiable}. Differentiable programming addresses this challenge by combining computer programs with automatic differentiation~\cite{paszke2017automatic, baydin2018automatic, blondel2024elements}, treating programs with control flow, recursion, and complex structures as composable differentiable modules. By embedding optimization layers and iterative algorithms directly into differentiable computational graphs, it enables seamless gradient propagation and efficient end-to-end optimization, overcoming the limitations of traditional black-box optimization and numerical differentiation~\cite{zhao2025fully, sapienza2024differentiable}. Representative applications include differentiable earth-system calibration~\cite{gelbrecht2023differentiable}, large-scale physical simulation~\cite{hu2019difftaichi}, and wireless network optimization via DIFFRACT~\cite{tan2026diffract}, demonstrating the broad applicability of this paradigm. Despite the rapid growth of differentiable programming, existing surveys primarily emphasize automatic differentiation, software frameworks, or differentiable optimization layers in isolation. As illustrated in Fig.~\ref{fig:diffprog}, this tutorial presents a unified perspective that integrates differentiable programming, duality theory, and first-order methods into a scalable framework for learning to optimize.

To illustrate how differentiable programming, first-order methods, and duality theory can be integrated within learning-to-optimize frameworks, we use the classical Nonnegative Least Squares (NNLS) problem~\cite{diakonikolas2022fast, esser2013method} as a running example throughout the paper. Given $\mathbf{A}\!\in\!\mathbb{R}^{m\times n}$ and $\mathbf{b}\!\in\!\mathbb{R}^m$, NNLS is a Quadratic Cone Program (QCP) \cite{healey2026differentiating} that seeks a nonnegative vector $\mathbf{x}$ minimizing the residual $\|\mathbf{A}\mathbf{x}-\mathbf{b}\|_2^2$:
\begin{align}\tag{NNLS}
\label{prob:Intro-NNLS-primal}
\min_{\mathbf{x}\ge0}\;\tfrac{1}{2}\|\mathbf{A}\mathbf{x}-\mathbf{b}\|_2^2.
\end{align}
Our main contributions are summarized as follows:
\begin{adjustwidth}{1.5em}{0em}
\begin{itemize}
    \item \textit{Unified theoretical framework.} We present a systematic treatment of differentiable programming for learning to optimize, integrating cone programming, Lagrangian and Fenchel-Rockafellar duality, first-order methods, and representative software frameworks and optimization packages within a unified perspective.

    \item \textit{PyTorch-based differentiable implementations.} We demonstrate how optimization problems, duality theory, and fundamental first-order methods can be embedded into end-to-end differentiable computational graphs through PyTorch-based implementations of representative cone programs, accompanied by publicly available source code.\footnote{Source code is available at \url{https://github.com/convexsoft/diffprog}.}

    \item \textit{Reproducible case studies.} We present representative case studies spanning Linear Programming (LP), Second-Order Cone Programming (SOCP), Exponential Cone Programming (ECP), and Quadratic Programming (QP), illustrating how differentiable programming, duality theory, and first-order methods can be unified to solve these optimization problems through reproducible PyTorch implementations.
\end{itemize}
\end{adjustwidth}

\vspace{1mm}
\noindent \textbf{Organization.} The paper is organized as follows. Section \ref{sec:intro} introduces the background and outlines the main contributions. Section \ref{sec:dp} reviews the fundamentals of differentiable programming, including software frameworks and packages. Section \ref{sec:dp-for-op} explores the integration of differentiable programming with cone programming, duality theory, first-order methods, implicit differentiation, and corresponding implementation. Section \ref{sec:case-studies} presents PyTorch-based case studies unifying differentiable programming, duality theory, and first-order methods for end-to-end optimization, and Section \ref{sec:conclu} concludes the paper.

\section{Differentiable Programming}
\label{sec:dp}
Differentiable programming is a paradigm that treats differentiability as a first-class abstraction, enabling complex programs to be composed of differentiable functions and optimized end-to-end via automatic differentiation \cite{blondel2024elements, wang2018backpropagation}. By modeling computations as differentiable computation graphs, this approach unifies programming and optimization to allow for efficient gradient and higher-order derivative computation. Recently, Blondel et al. \cite{blondel2024elements} formalizes this unification by systematizing the design of programs including control flow, solvers, and data structures that remain fully trainable. Their framework integrates implicit layers, proximal mappings, and probabilistic estimators with classical optimization tools such as Fenchel-Rockafellar duality and Bregman divergences, establishing a rigorous foundation for embedding complex algorithmic procedures into neural systems. Consequently, modern machine learning leverages these differentiable programs through first-order methods for scalable optimization and second-order variants that incorporate curvature information, such as Hessians, to enhance convergence in precision-critical problems \cite{pan2024bpqp}. In the following sections, we review the theoretical foundations, software frameworks, and optimization software packages that support these implementations.

\subsection{Theoretical Foundations of Differentiable Programming}
This section presents some theoretical foundations of differentiable programming, covering gradient calculus, computation graphs, and automatic differentiation, and demonstrates these principles through a detailed analysis of the problem~\eqref{prob:Intro-NNLS-primal}.

\subsubsection{Gradient Calculus}
Consider a differentiable function $f : \mathbb{R}^P \to \mathbb{R}$. The gradient of $f$ at a point $w \in \mathbb{R}^P$ is defined as the column vector of its partial derivatives:
\begin{align}
\nabla f(w) :=
\begin{pmatrix}
\frac{\partial f}{\partial w_1}(w) \\
\vdots \\
\frac{\partial f}{\partial w_P}(w)
\end{pmatrix}
=
\begin{pmatrix}
\partial f(w)[e_1] \\
\vdots \\
\partial f(w)[e_P]
\end{pmatrix}
\in \mathbb{R}^P, 
\end{align}
where $\{e_1, \dots, e_P\}$ denote the standard basis of $\mathbb{R}^P$. By the linearity of the differential, the directional derivative of $f$ at $w$ along a direction $v = \sum_{i=1}^{P} v_i e_i$ is given by $\partial f(w)[v] = \sum_{i=1}^{P} v_i \partial f(w)[e_i] = \langle v, \nabla f(w) \rangle \in \mathbb{R}$, where $\langle \cdot, \cdot \rangle$ denotes the standard inner product \cite{boyd2004convex}, i.e., $\langle v, \nabla f(w) \rangle = v^{\top}\nabla f(w)$.

%%%---add: hessian matrix

\vspace{1mm}
\textbf{Jacobian Matrix.} Consider a differentiable function $f : \mathbb{R}^P \to \mathbb{R}^M$ with component functions $f_i : \mathbb{R}^P \to \mathbb{R}$ defined by $f(w) := (f_1(w), \dots, f_M(w))$. The Jacobian matrix of $f$ at a point $w \in \mathbb{R}^P$ is defined as the matrix of all first-order partial derivatives:
\begin{align}
\partial f(w) :=
\begin{pmatrix}
\frac{\partial f_1}{\partial w_1}(w) & \cdots & \frac{\partial f_1}{\partial w_P}(w) \\
\vdots & \ddots & \vdots \\
\frac{\partial f_M}{\partial w_1}(w) & \cdots & \frac{\partial f_M}{\partial w_P}(w)
\end{pmatrix}
\in \mathbb{R}^{M \times P}.
\end{align}

Equivalently, the Jacobian can be written as a stack of transposed component gradients, $\partial f(w) = \begin{pmatrix}\nabla f_1(w)^{\top}, \dots, \nabla f_M(w)^{\top}\end{pmatrix}^{\top} \in \mathbb{R}^{M \times P}$. By linearity of the differential, the directional derivative of $f$ at $w$ along $v = \sum_{i=1}^{P} v_i e_i \in \mathbb{R}^P$ is $\partial f(w)[v] = \sum_{i=1}^{P} v_i\,\partial_i f(w) = \partial f(w)\,v \in \mathbb{R}^M$, where $v_i \in \mathbb{R}$ and:
\begin{align}
\partial_i f(w)
:=
\partial f(w)[e_i]
=
\left(
\frac{\partial f_1}{\partial w_i}(w),
\dots,
\frac{\partial f_M}{\partial w_i}(w)
\right)^\top
\in \mathbb{R}^M,
\end{align}
denotes the vector of partial derivatives of $f$ with respect to the $i$-th coordinate of $w$ \cite{blondel2024elements}.

\vspace{1mm}
\textbf{Chain Rule.} Consider two differentiable functions $f : \mathbb{R}^P \to \mathbb{R}^M$ and $g : \mathbb{R}^M \to \mathbb{R}^R$. If $f$ is differentiable at $w \in \mathbb{R}^P$ and $g$ is differentiable at $f(w) \in \mathbb{R}^M$, then the composition $g \circ f$ is differentiable at $w$, and its Jacobian is given by the chain rule:
\begin{align}
\partial (g \circ f)(w)
= \partial g(f(w)) \, \partial f(w). 
\end{align}

\subsubsection{Computation Graph}
Computation graphs~\cite{bauer1974computational, blondel2024elements}, also known as Kantorovich graphs \cite{kantorovich1957system, fourer2010convexity}, are Directed Acyclic Graphs (DAGs)~\cite{digitale2022tutorial} that represent dependencies among inputs, intermediate variables, and outputs. Nodes correspond to quantities such as inputs, parameters, or operations such as addition, multiplication, and nonlinear activations, while edges encode data dependencies or atomic operations connecting them \cite{schulman2015gradient}. Once constructed, forward propagation efficiently computes the output of program~\cite{sapienza2024differentiable}. Specifically, derivatives are obtained using the Bauer formula~\cite{oktay2020randomized}, which recursively expresses $\partial v_j / \partial v_i$ through parent-node gradients:
\begin{align}
\frac{\partial v_j}{\partial v_i} = 
\sum_{w \to v_j} \frac{\partial v_j}{\partial w} \cdot \frac{\partial w}{\partial v_i},
\end{align}
which provides an exact graph-based form of the chain rule and enables backpropagation, the foundation of automatic differentiation. Extending naturally to higher-order derivatives, computation graphs form the backbone of modern frameworks such as PyTorch, TensorFlow, and JAX, and serve as a core abstraction in differentiable programming for large-scale optimization.

\subsubsection{Automatic Differentiation}
Automatic Differentiation (AD) is a key computational technique for efficiently and accurately computing derivatives of functions represented as computer programs~\cite{baydin2018automatic}. Unlike numerical differentiation, which introduces approximation errors, or symbolic differentiation, which suffers from expression explosion, automatic differentiation systematically applies the chain rule over a computational graph by extending variables with derivative information and redefining operators accordingly~\cite{margossian2019review}. Serving as the foundation of differentiable programming, automatic differentiation unifies calculus and programming, enabling scalable gradient-based optimization in frameworks such as PyTorch~\cite{paszke2017automatic, blondel2024elements, halim2024introduction}. Its reach extends beyond simple operation chains to programs with control flow, data structures, and computational effects, broadening the scope of differentiable modeling~\cite{abadi2019simple}.

Among the various modes of automatic differentiation, reverse-mode automatic differentiation~\cite{hogan2014fast}, which underlies backpropagation, plays a central role. It constructs a computational graph during the forward pass and propagates gradients backward, making it especially effective for large-scale optimization problems with many parameters~\cite{moses2021reverse}. Recent research has significantly strengthened the theoretical underpinnings of reverse-mode automatic differentiation. For example, Abadi et al.~\cite{abadi2019simple} introduced a minimal differentiable programming language; Innes et al.~\cite{innes2019differentiable} demonstrated automatic differentiation as a first-class language feature supporting custom adjoints and mixed-mode differentiation; and Wang et al.~\cite{wang2019demystifying, wang2018backpropagation} recast automatic differentiation through the lens of continuations, enabling dynamic, modular, and composable differentiation.

While backpropagation is often described as simply the chain rule, it is not merely the repeated application of the analytic chain rule to a symbolic expression \cite{griewank2008evaluating}. Rather, it can be derived by viewing a program as a system of equality constraints on intermediate variables: the stationarity conditions of the associated Lagrangian yield a triangular linear system in the dual variables, whose back-substitution is exactly the backpropagation recursion \cite{lecun1988theoretical}. In Lagrangian view, the forward pass enforces the constraints that define intermediate values, while the backward pass computes the Lagrange multipliers that propagate sensitivities backward through the computational graph. This perspective explains why reverse-mode automatic differentiation evaluates gradients at a cost comparable to a single forward execution and why it naturally extends to programs with branching, loops, and implicit computations. Next, we revisit the problem~\eqref{prob:Intro-NNLS-primal} and analyze backpropagation from both perspectives: the computation-graph chain-rule view and the Lagrangian view.

\begin{example}[Backpropagation via Chain Rule and Lagrangian]
\label{eg:Backpropagation-via-ChainRule-and-Lagrangian}
Consider the NNLS problem:
\begin{align}
\label{prob:AD-NNLS-primal}
\min_{\mathbf{x}\ge0}\;\tfrac{1}{2}\|\mathbf{A}\mathbf{x}-\mathbf{b}\|_2^2.
\end{align}

Introduce the auxiliary variables $\mathbf{z}=\mathbf{A}\mathbf{x}$, $\mathbf{y}=\mathbf{z}-\mathbf{b}$, and $f=\tfrac{1}{2}\|\mathbf{y}\|_2^2$. Temporarily ignoring the nonnegativity constraint $\mathbf{x}\ge0$, problem~\eqref{prob:AD-NNLS-primal} can be rewritten as:
\begin{equation}\label{prob:AD-NNLS-primal-rewrite}
\begin{aligned}
\min_{\mathbf{x,y,z},f} \quad  &f \\
\text{s.t.}\quad 
&\mathbf{z}=\mathbf{A}\mathbf{x}, \quad
\mathbf{y}=\mathbf{z}-\mathbf{b}, \quad
f=\tfrac{1}{2}\|\mathbf{y}\|_2^2. 
\end{aligned}
\end{equation}

\textbf{(1) Backpropagation via the Chain Rule:}
As shown in Figure~\ref{fig:AD-NNLS-PyTorch-Reverse-AD-Chain-Rule}, applying the chain rule along the computational graph yields:
\begin{align}
\label{eq:AD-chain-rule-x-gradient}
\nabla_\mathbf{x} f 
&\;=\;
\frac{\partial f}{\partial \mathbf{y}}
\frac{\partial \mathbf{y}}{\partial \mathbf{z}}
\frac{\partial \mathbf{z}}{\partial \mathbf{x}}
\;=\; \mathbf{A}^{\top}(\mathbf{A}\mathbf{x}-\mathbf{b}),
\end{align}
where $\mathbf{I}$ denotes the identity operator.  The conventional backpropagation path is along $f \!\to\! \mathbf{y} \!\to\! \mathbf{z} \!\to\! \mathbf{x}$.

\begin{figure}[!t]
    \centering
    \includegraphics[width=0.69\linewidth, height=1.4cm]{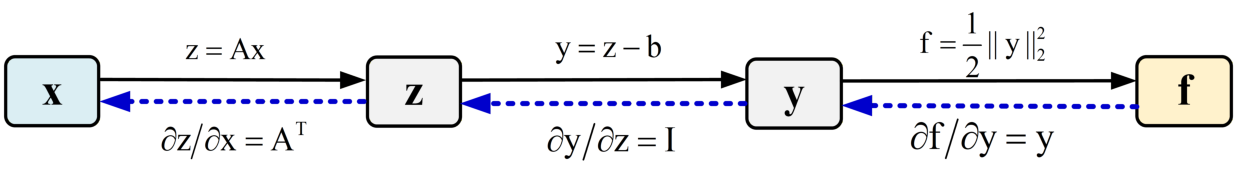}
    \caption{Backpropagation in PyTorch via the Chain Rule: The forward pass computes $z=Ax$, $y=z-b$, and $f=\tfrac{1}{2}\|y\|_2^2$, while the backward pass propagates gradients from $f$ to $x$ to obtain $\nabla_x f = A^\top(Ax-b)$.}
    \label{fig:AD-NNLS-PyTorch-Reverse-AD-Chain-Rule}
\end{figure}

\textbf{(2) Backpropagation as a Lagrangian System:}
Introduce dual variables $\boldsymbol{\lambda}$, $\boldsymbol{\mu}$, and $\rho$ for the constraints in \eqref{prob:AD-NNLS-primal-rewrite}. The Lagrangian is:
\begin{align}
\label{eq:BG-NNLS-Lagrangian}
\mathcal{L}(\mathbf{x}, \mathbf{z}, \mathbf{y}, f; \boldsymbol{\lambda}, \boldsymbol{\mu}, \rho)
\;=\; f 
+ \boldsymbol{\lambda}^{\top}(\mathbf{A}\mathbf{x}-\mathbf{z})
+ \boldsymbol{\mu}^{\top}(\mathbf{z}-\mathbf{b}-\mathbf{y})
+ \rho\bigl(\tfrac{1}{2}\|\mathbf{y}\|_2^2 -f\bigr).
\end{align}
Taking derivatives with respect to the dual variables recovers the forward pass:
\begin{align}
\nabla_{\boldsymbol{\lambda}} \mathcal{L} = \mathbf{A}\mathbf{x}-\mathbf{z}=0, \qquad
\nabla_{\boldsymbol{\mu}} \mathcal{L} = \mathbf{z}-\mathbf{b}-\mathbf{y}=0, \qquad
\nabla_{\rho} \mathcal{L} = \tfrac{1}{2}\|\mathbf{y}\|_2^2 -f =0,
\end{align}
corresponding exactly to the constraints in \eqref{prob:AD-NNLS-primal-rewrite}. Next, differentiating with respect to the intermediate variables $\{\mathbf{z},\mathbf{y},f\}$ gives the adjoint (backward) equations:
\begin{align}
\label{eq:adjoint-equations}
\nabla_\mathbf{z} \mathcal{L} = -\boldsymbol{\lambda} + \boldsymbol{\mu} = 0, \qquad
\nabla_\mathbf{y} \mathcal{L} = -\boldsymbol{\mu} + \rho\,\mathbf{y} = 0, \qquad
\nabla_f \mathcal{L} = 1 - \rho = 0.
\end{align}
Solving \eqref{eq:adjoint-equations} yields $\boldsymbol{\lambda}=\boldsymbol{\mu}=\rho\,\mathbf{y}$ and $\rho=1$. Finally, the gradient with respect to~$\mathbf{x}$ becomes:
\begin{align}
\label{eq:AD-Larangian-x-gradient}
\nabla_\mathbf{x} \mathcal{L} 
\;=\; \mathbf{A}^{\top}\boldsymbol{\lambda}
\;=\; \mathbf{A}^{\top}\boldsymbol{\mu}
\;=\; \mathbf{A}^{\top}(\rho\,\mathbf{y})
\;=\; \mathbf{A}^{\top}\mathbf{y}
\;=\; \mathbf{A}^{\top}(\mathbf{A}\mathbf{x}-\mathbf{b}),
\end{align}
which coincides exactly with the chain-rule computation in~\eqref{eq:AD-chain-rule-x-gradient}.
\end{example}

Example \ref{eg:Backpropagation-via-ChainRule-and-Lagrangian} shows that backpropagation can be seen either as repeated chain-rule application or as solving a Lagrangian-based adjoint system. Both yield the same gradients, with the adjoint view clarifying how backpropagation extends naturally to complex programs, implicit layers, and constrained computations.

\subsection{Differentiable Programming Software Frameworks}
This section introduces key software frameworks that enable differentiable programming, including PyTorch \cite{ketkar2021introduction}, JAX \cite{bradbury2018jax}, TensorFlow \cite{abadi2016tensorflow}, and the machine learning compilation framework Apache TVM \cite{chen2018tvm}. These frameworks, summarized in Table \ref{tab:dp-frameworks}, provide automatic differentiation and efficient numerical computation, forming the practical foundation of differentiable programming.

\subsubsection{PyTorch}
Developed by Facebook’s AI Research lab, PyTorch \cite{paszke2019pytorch} is an open-source deep learning framework that combines a flexible programming interface, efficient tensor computation, and seamless GPU acceleration \cite{ketkar2021introduction}. Its dynamic computation graph enables intuitive model construction, debugging, and experimentation, while the autograd engine provides powerful reverse-mode automatic differentiation for large-scale numerical computation. Core modules such as torch.nn for neural network components, torch.optim for optimization routines, and torch.utils.data for data handling support a broad spectrum of machine learning and scientific workflows. Beyond its role as a deep learning library, PyTorch serves as a unifying platform for differentiable programming by integrating numerical computation, automatic differentiation, and optimization into a coherent system \cite{johnson2025pytorch}. With autograd allowing gradients to flow through arbitrary control flow, loops, and user-defined data structures \cite{paszke2017automatic, hu2021theory}, PyTorch enables end-to-end differentiation of complex and parameterized programs, including iterative solvers, optimization algorithms, and physical simulations. This capability transforms traditional programming into an optimization-aware paradigm, providing both the theoretical expressiveness and practical performance necessary to develop differentiable algorithms, learning-enhanced numerical methods, and general optimization frameworks \cite{wang2018backpropagation, lai2023prox}. As a result, PyTorch has become a cornerstone of modern machine learning, scientific computing, and differentiable programming research.

\subsubsection{JAX}
Developed by Google Research, JAX \cite{bradbury2018jax, frostig2019compiling} is a high-performance framework for numerical computing and machine learning. Built upon NumPy \cite{van2011numpy, harris2020array}, it can be viewed as a differentiable extension of the NumPy system that integrates numerical computation, automatic differentiation, and compilation-driven optimization into an efficient framework. By combining the flexibility of the Python scientific computing ecosystem with robust automatic differentiation and accelerated compilation, JAX delivers performance and composability for research as well as engineering applications \cite{sapunov2024deep, wang2025integrating}. Within the field of differentiable programming, JAX plays a central role. Through support for automatic differentiation, just-in-time compilation, and parallel execution primitives, the framework enables efficient differentiation of arbitrary Python functions and facilitates large-scale parallel computation \cite{blondel2024elements}. The automatic differentiation system in JAX is based on function transformations that apply to any function expressed through JAX primitives. These transformations, including grad, jacfwd, jacrev, vmap, and jit, can be combined to construct complex differentiable models and algorithms with minimal implementation effort \cite{schoenholz2020jax}. Beyond deep neural network models \cite{bouziani2024differentiable}, JAX supports end-to-end differentiation of optimization algorithms, physical simulations, probabilistic models, and numerical solvers. JAX has become a core tool in differentiable programming, widely used in automatic-differentiation–based optimization, scientific computing, and differentiable physics simulations \cite{lu2024mpax}.

\subsubsection{TensorFlow}
TensorFlow \cite{abadi2016tensorflow} is a large-scale machine learning system designed for heterogeneous environments that utilizes a unified dataflow graph to represent both computation and mutable state. This architecture allows graph nodes to be mapped seamlessly across diverse hardware, including CPUs, GPUs, and TPUs, by encoding parameters and optimization algorithms directly as graph operations. By generalizing the traditional parameter-server model, the framework enables users to implement custom layers and training strategies without modifying the underlying system. It adopts a deferred execution model where symbolic graphs are constructed and then executed by an optimized runtime, ensuring flexible deployment from mobile devices to massive distributed clusters \cite{shukla2018machine}. The framework further integrates automatic differentiation, dynamic control flow, and user-level checkpointing to support complex architectures like recurrent neural networks while maintaining production-grade fault tolerance \cite{pang2020deep}. Building on this foundation, TensorFlow Distributions \cite{dillon2017tensorflow} introduces probabilistic modeling tailored for end-to-end differentiable computation through two core abstractions: distributions for stable sampling and bijectors for composable transformations. A key innovation is its shape semantics, which partitions tensors into sample, batch, and event dimensions to facilitate efficient vectorization and broadcasting across heterogeneous hardware. By supporting reparameterization, this extension allows gradients to propagate through stochastic samples, providing the foundational infrastructure for deep probabilistic programming systems like Edward and generative models such as PixelCNN++.

\begin{table*}[t]
\caption{Summary of Representative Software, Frameworks, and Packages of Differentiable Programming.}
\label{tab:dp-frameworks}
\centering
\renewcommand{\arraystretch}{1.2}
\scriptsize

\newcolumntype{L}[1]{>{\RaggedRight\arraybackslash}p{#1}}
\newcolumntype{Y}{>{\RaggedRight\arraybackslash}X}

%\begin{tabularx}{\textwidth}{ |L{0.11\textwidth} | L{0.14\textwidth} | Y | Y |}
\begin{tabularx}{\textwidth}{ |L{0.11\textwidth} | C{0.12\textwidth} | Y | Y |}
\hline

\rule{0pt}{3.8ex}
\textbf{\footnotesize Category} & \textbf{\footnotesize Tool} & \textbf{\footnotesize Feature} & \textbf{\footnotesize Purpose} \\
\hline

% ============================================================
% SOFTWARE & FRAMEWORKS
% ============================================================

\multirow{16}{=}{\textbf{Differentiable Programming Software Frameworks}} 

& \textbf{PyTorch} \cite{paszke2019pytorch}
& \begin{itemize}
    \item Dynamic computation graph
    \item GPU-accelerated tensor operations
    \item Reverse-mode AD via autograd
  \end{itemize}
& \begin{itemize}
    \item End-to-end differentiation of programs
    \item Support for iterative solvers and optimization algorithms
  \end{itemize}
\\ \cline{2-4}

& \textbf{JAX} \cite{frostig2019compiling}
& \begin{itemize}
    \item Differentiable NumPy API
    \item JIT compilation via XLA
    \item Composable transforms: grad, vmap, jit
  \end{itemize}
& \begin{itemize}
    \item Differentiable numerical computing
    \item Scientific simulation
    \item Large-scale parallel differentiable programming
  \end{itemize}
\\ \cline{2-4}

& \textbf{TensorFlow} \cite{abadi2016tensorflow}
& \begin{itemize}
    \item Dataflow graph with AD
    \item Heterogeneous execution (CPU/GPU/TPU) and XLA compilation
    \item High-level API ecosystem and production deployment
  \end{itemize}
& \begin{itemize}
    \item Large-scale training and deployment of differentiable models
    \item Differentiable programming with dynamic control flow and checkpointing
  \end{itemize}
%\\ \cline{2-4}

\\
\hline

% ============================================================
% DIFFERENTIABLE OPTIMIZATION LAYERS
% ============================================================

\multirow{10}{=}{\textbf{Optimization Software Packages}} 

& \textbf{CVXPYLayers} \cite{agrawal2019differentiable, agrawal2020differentiating}
& \begin{itemize}
    \item Differentiable convex optimization layers
    \item KKT-based implicit differentiation
    \item CVXPY modeling interface
  \end{itemize}
& \begin{itemize}
    \item Optimization layers for neural networks
    \item End-to-end differentiable learning
  \end{itemize}
\\ \cline{2-4}

& \textbf{PyEPO} \cite{tang2024pyepo}
& \begin{itemize}
    \item Predict-then-optimize framework
    \item Differentiable LP/MIP solvers
    \item Surrogate gradients for discrete decisions
  \end{itemize}
& \begin{itemize}
    \item End-to-end learning for decision-focused optimization
  \end{itemize}
\\ \cline{2-4}

& \textbf{DDNs} \cite{gould2021deep, gould2016differentiating}
& \begin{itemize}
    \item Implicit optimization layers
    \item Gradients from KKT system and second-order structure
    \item Solver-free, scalable differentiable optimization
  \end{itemize}
& \begin{itemize}
    \item Integrating optimization steps into neural networks
    \item Differentiation via implicit function theory
  \end{itemize} 
\\
\hline

\end{tabularx}
\end{table*}

\subsubsection{Machine Learning Compilation Framework}
As deep learning models scale and differentiable programming frameworks become increasingly prevalent, machine learning workloads demand more advanced compilation and hardware acceleration infrastructures. Traditional frameworks often depend on vendor-specific operator libraries, such as cuDNN \cite{chetlur2014cudnn}, limiting performance portability across heterogeneous hardware back-ends. Although GPUs have emerged as the dominant computing platform and CUDA \cite{erik2008nvidia} provides a general-purpose parallel programming model for executing large-scale tensor operations and automatic differentiation on NVIDIA hardware, directly developing high-performance CUDA kernels remains complex and lacks cross-architectural portability. To address these challenges, machine learning compilation frameworks such as Apache TVM \cite{chen2018tvm} provide end-to-end automated optimization pipelines beyond traditional “black-box” libraries. TVM lowers high-level model representations into multi-level intermediate representations and integrates automated operator scheduling with cross-platform code generation to optimize execution from model specification to GPU deployment. By decoupling computation graphs from hardware-specific details, TVM automatically generates optimized CUDA kernels. On modern GPUs, it exploits CUDA’s hierarchical execution model and employs machine learning–based cost models to search large scheduling spaces, identifying efficient configurations for thread binding, shared memory utilization, and latency hiding. Within differentiable programming systems, such frameworks accelerate both forward inference and backward gradient computation, bridging high-level differentiable abstractions with high-performance hardware execution.

\subsection{Optimization Software Packages based on Differentiable Programming}
Differentiable programming reframes optimization in machine learning by embedding optimization problems as trainable, differentiable components within learning architectures~\cite{zhao2025fully}. Instead of functioning solely as terminal solvers, optimization layers integrate objectives, constraints, and algorithms into end-to-end gradient-based training through automatic differentiation~\cite{baydin2018automatic}. This enables scalable systems that avoid manual gradient derivations and extend differentiability to complex algorithmic structures such as control loops, discrete decisions, and physical simulations~\cite{sapienza2024differentiable, bouziani2024differentiable}. Treating optimization as a differentiable layer has led to wide-ranging applications, including predictor–verifier training~\cite{dvijotham2018training}, deep equilibrium models~\cite{bai2019deep}, differentiable tensor networks~\cite{liao2019differentiable}, differentiable vision pipelines~\cite{li2018differentiable}, differentiable MPC~\cite{amos2018differentiable}, bilevel hyperparameter learning~\cite{pedregosa2016hyperparameter}, robustness via convex relaxations~\cite{wong2018provable, zhang2019towards}, and differentiable sorting and ranking~\cite{blondel2020fast}. Beyond convex settings, Theseus~\cite{pineda2022theseus} extends this paradigm to nonlinear least squares, combining second-order solvers with sparse GPU-accelerated linear algebra and implicit differentiation for end-to-end learning in robotics and vision. These developments establish differentiable programming as a unifying paradigm linking optimization, control, and scientific computing. The following section introduces representative optimization software packages (cf. Table \ref{tab:dp-frameworks}) for differentiable programming frameworks.

\subsubsection{CVXPYLayers}
CVXPYLayers \cite{agrawal2019differentiable, agrawal2020differentiating} is a Python library that exposes the solution map of a parametrized convex program as a differentiable function, enabling gradients to propagate through optimization problems during backpropagation in frameworks such as PyTorch, JAX, and MLX \cite{feng2025profiling}. It builds upon CVXPY \cite{diamond2016cvxpy, agrawal2018rewriting}, a Domain-Specific Language (DSL) that verifies convexity via Disciplined Convex Programming (DCP) \cite{Grant2006dcp}, originating from the CVX system \cite{grant2008cvx, grantboyd2008graph, grant2020cvx} for MATLAB, and automatically canonicalizes problems into cone programs solved by backends such as SCS \cite{ODonoghue2016}, ECOS \cite{domahidi2013ecos}, OSQP \cite{stellato2020osqp}, and MOSEK \cite{andersen2000mosek}. CVXPYLayers introduces Disciplined Parametrized Programming (DPP), a DCP subset that guarantees an affine mapping from parameters to cone data, yielding the Affine-Solver-Affine (ASA) form \cite{agrawal2019differentiable}. Gradients are computed analytically by implicitly differentiating the KKT conditions via the diffcp package \cite{diffcp2019}, and the framework further supports log-log convex programs through Disciplined Geometric Programming (DGP) \cite{agrawal2020differentiating}. CVXPYLayers~1.0 extends the framework with dual variable outputs alongside primal solutions and introduces GPU acceleration via the CuClarabel \cite{chen2024CuClarabel} backend.

\subsubsection{PyEPO}
Introduced by the work in \cite{tang2024pyepo}, PyEPO is a unified PyTorch-based framework for end-to-end predict-then-optimize learning in linear and integer programming. Classical two-stage pipelines train a predictive model by minimizing a statistical loss before solving a downstream optimization problem, whereas PyEPO integrates the optimizer directly into the computational graph and exposes decision-aware gradients to upstream predictors. To support this capability, the framework consolidates several influential families of surrogate-gradient and perturbation-based techniques, including SPO+, differentiable black-box solvers, differentiable perturbed optimizers, and perturbed Fenchel–Young losses. These approaches approximate or smooth the inherently non-differentiable mapping from objective coefficients to optimal decisions, enabling optimization-aware training even in discrete or polyhedral decision spaces. PyEPO further provides high-level interfaces to commercial linear programming and mixed integer programming solvers such as Gurobi and COPT, supports batched forward and backward evaluations, and standardizes benchmark tasks including shortest-path, knapsack, and traveling-salesperson problems. By systematizing a broad class of differentiable decision-focused learning methods within a modern automatic differentiation framework, PyEPO serves as a comprehensive experimental platform that demonstrates the value of differentiable programming for large-scale decision-making and combinatorial optimization.

\subsubsection{DDNs}
Proposed in \cite{gould2021deep, gould2016differentiating}, Deep Declarative Networks (DDNs) are a unified framework that integrates optimization problems directly into deep learning architectures. In a DDN, a layer is defined not by an explicit forward computation, but implicitly as the solution to a parameterized optimization problem. Instead of unrolling an iterative solver, they derive closed-form differentiation rules using the implicit function theorem, enabling efficient backpropagation through both unconstrained and constrained optimization layers. Their work shows that the required Jacobian–vector products can be computed using the problem’s optimality conditions, often involving only linear systems that depend on the local curvature of the objective. This yields a highly scalable alternative to solver unrolling. Subsequent developments~\cite{gould2025lecture} draw connections between convex analysis, duality, and automatic differentiation, further reducing the computational cost of implicit differentiation and facilitating the integration of optimization layers into modern deep learning frameworks such as PyTorch. Thus, DDNs offer a principled way to embed optimization problems inside neural networks, combining modeling flexibility with computational efficiency.

\section{Differentiable Programming for Optimization}
\label{sec:dp-for-op}
Many optimization problems share recurring structural patterns, making the integration of differentiable programming and optimization increasingly powerful. For example, Park et al.~\cite{park2023self} proposed a self-supervised primal–dual framework that approximates optimal solutions by emulating augmented Lagrangian dynamics, illustrating how differentiable programming enables end-to-end optimization across objectives and constraints. Furthermore, recent works~\cite{agrawal2019differentiable,poganvcic2019differentiation,bertrand2022implicit,constante2026enforcing,healey2026differentiating,magoon2026differentiation} have transformed classical optimization solvers into scalable differentiable modules, extending optimization layers to quadratic cone programs while enabling solver-agnostic differentiation through black-box optimization solvers. These advances further facilitate the seamless integration of learning and optimization in end-to-end differentiable systems. We next examine this interplay through cone programming, duality theory, first-order methods, and implicit differentiation, using NNLS-based differentiable programming implementation examples as illustrative cases.

\subsection{Cone Programming}
Cone programming~\cite{boyd2004convex} is a specialized subclass of conic optimization that unifies many classical convex problem families, including Linear Programming (LP)~\cite{karloff2008linear, strayer2012linear}, Quadratic Programming (QP)~\cite{frank1956algorithm, kouzoupis2018recent}, Quadratic Cone Programs (QCPs) \cite{healey2026differentiating}, and Semidefinite Programming (SDP)~\cite{vandenberghe1996semidefinite, yurtsever2021scalable}. While every cone program is a conic optimization problem, the converse does not hold. In cone programming, both the objective and constraints are linear, and the standard form is:
\begin{equation}\label{prob:cone-programming}
\begin{aligned}
\min_{x} \quad & c^{\top}x \\
\text{s.t.} \quad & A x = b, \quad x \in \mathcal{K},
\end{aligned}
\end{equation}
where $\mathcal{K}$ is a convex cone, $x \in \mathbb{R}^n$ is the decision variable, $c \in \mathbb{R}^n$ is the cost vector, $A \in \mathbb{R}^{m \times n}$ is the constraint matrix, and $b \in \mathbb{R}^m$ is the right-hand-side vector. In particular, the convex cone~$\mathcal{K}$ is often expressed as a Cartesian product of orthants, second-order cones, and positive semidefinite cones~\cite{dur2021conic}. When $\mathcal{K}$ is the nonnegative orthant, $\mathbb{R}^n_{+} = \{x \in \mathbb{R}^n \mid x_i \ge 0\}$, the cone programming reduces to a standard linear program of the form $\min_x f(x)$ subject to $A x = b$ and $x \ge 0$ and can be directly solved by CVX \cite{grant2008cvx, grant2020cvx, grantboyd2008graph}. When $\mathcal{K}$ is the second-order cone, $\mathcal{L}_n = \{(x, z) \in \mathbb{R}^{n-1} \times \mathbb{R} \mid |x|_2 \le z\}$, the resulting problems are Second-Order Cone Programs (SOCPs) \cite{hang2020second, lobo1998applications}, widely used in engineering and physics. Another important cone is the exponential cone \cite{busseti2019solution, parikh2014proximal}, which gives rise to Exponential Cone Programs (ECPs). ECPs provide a conic representation for entropy, log-sum-exp, geometric-programming, and exponential-family constraints, thereby extending the modeling capability of standard LPs and SOCPs. A further generalization employs the semidefinite cone, $\mathcal{S}_n = \{X \in \mathbb{S}^n \mid X \succeq 0\}$, where $\mathbb{S}^n$ denotes symmetric $n \times n$ matrices. The resulting problems are known as SDPs \cite{vandenberghe1996semidefinite, yurtsever2021scalable}. With their structured convexity and strong duality, cone programs \eqref{prob:cone-programming} anchor modern convex optimization and especially primal–dual interior-point methods \cite{wright1997primal}, and they are central to differentiable programming where convex, differentiable optimization layers are required \cite{agrawal2019differentiable}.

\subsection{Differentiable Programming in Cone Programs}
In this section, we examine the applications of differentiable programming, with a particular focus on the PyTorch framework, in the context of cone programming following \cite{boyd2004convex}. Firstly, let us consider the following cone programming:
\begin{equation}\label{prob:ben-recht-with-inequality}
\begin{aligned}
\min_x & \quad f(x) \\ 
\text{s.t. } & \quad Ax = b, \quad x \ge 0,
\end{aligned}
\end{equation}
where \(f : \mathbb{R}^n \to \mathbb{R}\) is a convex and continuously differentiable objective function, \(x \in \mathbb{R}^n\) denotes the vector of decision variables, \(A \in \mathbb{R}^{p \times n}\) and \(b \in \mathbb{R}^p\) specify the linear equality constraints, and the \(x \ge 0\) is imposed elementwise to ensure nonnegativity of the solution.

\begin{remark}
\textit{PyTorch is built around automatic differentiation and naturally supports gradient-based optimization. In practice, constrained optimization problems are often reformulated through penalty methods, barrier functions, or reparameterization techniques to enable effective use of first-order solvers. This paradigm facilitates the incorporation of complex constraints while maintaining compatibility with differentiable programming frameworks.}
\end{remark}

\subsubsection{Penalty Reformulation}
The equality constraint \(Ax = b\) in \eqref{prob:ben-recht-with-inequality} can be enforced using a quadratic penalty \(\lambda\|Ax - b\|_2^2\), and the nonnegativity constraint \(x \ge 0\) is imposed by projection. This leads to the problem:
\begin{align}
\min_{x \ge 0}\; f(x) + \lambda\|Ax - b\|_2^2,
\end{align}
which encourages \(Ax \to b\) and \(x \to \mathbb{R}_+^n\). This penalty formulation integrates naturally with PyTorch, enabling optimizers such as Adam \cite{kingma2015adam, courbariaux2016binarized} or L-BFGS \cite{liu1989limited} to incorporate these constraints within an end-to-end differentiable framework.

\subsubsection{Variable Transformation Reformulation}
Assuming $\mathrm{rank}(A)=p$, any feasible point for $Ax=b$ can be parameterized as $x = Fz + u$, where $u$ is a particular solution satisfying $Au=b$ and $F \in \mathbb{R}^{n \times (n-p)}$ is an orthonormal basis for the null space of $A$. Substituting this into the objective eliminates the equality constraints \cite{recht2024art, wright2022optimization}:
\begin{align}
\min_{Fz + u \ge 0} \; f(Fz + u).
\end{align}
The basis $F$ can be computed once via Singular Value Decomposition (SVD)~\cite{klema1980singular}, after which optimization proceeds over the reduced variable $z$. In practice, the nonnegativity constraint can be handled using projected, penalty-based, or reparameterization techniques, enabling compatibility with gradient-based solvers in PyTorch.

\subsubsection{Duality Reformulation}
An alternative reformulation of~\eqref{prob:ben-recht-with-inequality} is to introduce dual variables and derive its dual problem. Let $\nu$ and $\lambda \succeq 0$ denote the dual variables associated with $Ax=b$ and $x \ge 0$, respectively. The dual function is:
\begin{align}
\label{eq:ben-recht-with-inequality-dual}
g(\nu, \lambda)
&\;=\; \inf_{x}\{f(x) + \nu^{\top}(Ax - b) - \lambda^{\top}x\} \notag\\
&\;=\; -b^{\top}\nu - \sup_{x}\{(\lambda - A^{\top}\nu)^{\top}x - f(x)\} \notag\\
&\;=\; -b^{\top}\nu - f^{*}(\lambda - A^{\top}\nu),
\end{align}
where \(f^{*}\) is the Fenchel-Rockafellar conjugate \cite{fenchel1949conjugate, rockafellar1997convex, borwein2006convex} of \(f\). The resulting dual problem is:
\begin{align}
\label{prob:ben-recht-with-inequality-dual}
\max_{\nu,\, \lambda \ge 0}\quad -b^{\top}\nu - f^{*}(\lambda - A^{\top}\nu).
\end{align}
When $f^{*}$ is tractable and differentiable, the dual problem can be optimized using gradient-based first-order methods, with the constraint $\lambda \succeq 0$ handled via projection or reparameterization. Under Slater’s condition~\cite{boyd2004convex}, strong duality holds, and primal solutions can be recovered from the optimal dual variables through the KKT conditions.

\subsection{Duality Theory}
Machine learning has transformed optimization by enabling data-driven approaches to solving increasingly complex optimization problems~\cite{bengio2021machine}. As learned models achieve strong primal performance across linear, discrete, nonlinear, and nonconvex settings~\cite{kotary2021end}, a fundamental question arises: how can the optimality and feasibility of these solutions be rigorously certified? Duality theory provides a principled framework for certifying solution quality through complementary primal and dual formulations. Representative examples include dual Lagrangian learning~\cite{tanneau2024dual}, which exploits conic structure and expressive models to construct dual-feasible certificates and tight bounds for conic quadratic programs. Building on this foundation, recent advances have integrated dual principles into differentiable architectures, giving rise to deep dual learning, which unifies optimization theory with data-driven models to produce feasible, near-optimal solutions in an end-to-end manner~\cite{fioretto2020lagrangian}. Representative applications span nonconvex power flow~\cite{kim2024unsupervised}, combinatorial scheduling~\cite{kotary2022fast}, dual-stabilized augmented Lagrangian methods~\cite{bai2024new}, and dual-centric neural architectures that predict dual variables and reconstruct primal solutions~\cite{kotary2024learning}. Beyond certifying optimality, duality also reveals structural properties and algorithmic insights that are often inaccessible from the primal formulation alone. Motivated by these developments, the remainder of this section reviews the duality principles most relevant to differentiable cone programming, focusing on Lagrangian and Fenchel-Rockafellar duality~\cite{fenchel1949conjugate, rockafellar1997convex, borwein2006convex}.

\subsubsection{Lagrangian Duality}
Given a closed cone $\mathcal{K} \subseteq \mathbb{R}^n$, its dual cone is:
\begin{align}
\mathcal{K}^* \;=\; \{\, v \in \mathbb{R}^n \mid \langle v, z \rangle \ge 0,\; \forall z \in \mathcal{K} \,\},
\end{align}
the set of all vectors having nonnegative inner product with every element of $\mathcal{K}$. The dual cone is always closed and convex, regardless of whether $\mathcal{K}$ itself is convex~\cite{boyd2004convex}. If $\mathcal{K}$ is closed and convex, it satisfies the biconjugate property $(\mathcal{K}^*)^* = \mathcal{K}$, and it coincides with $\mathcal{K}$ when the cone is self-dual. 

Consider now the cone programming~\eqref{prob:cone-programming}:
\begin{equation}\label{prob:cone-programming-primal}%\tag{Primal}
\begin{aligned}
\min_{x} \quad & c^{\top}x \\
\text{s.t.} \quad & A x = b,\quad x \in \mathcal{K}.
\end{aligned}
\end{equation}
Introducing dual variables \(\lambda \in \mathbb{R}^{m}\) for the equality constraint and \(\mu \in \mathcal{K}^*\) for the conic constraint, where \(\mathcal{K}^* = \{\, \mu \in \mathbb{R}^n \mid \langle \mu, x \rangle \ge 0,\; \forall x \in \mathcal{K} \,\}\), the Lagrangian is:
\begin{align}
\label{eq:cone-programming-Lagrangian}
\mathcal{L}(x, \lambda, \mu)
\;=\; c^{\top}x + \lambda^{\top}(b - A x) - \mu^{\top}x,
\quad \mu \in \mathcal{K}^*.
\end{align}
The corresponding dual problem of \eqref{prob:cone-programming-primal} is:
\begin{equation}\label{prob:cone-programming-dual}%\tag{Dual}
\begin{aligned}
\max_{\lambda,\, \mu} \quad & b^{\top}\lambda \\
\text{s.t.} \quad & c - A^{\top}\lambda = \mu,\quad \mu \in \mathcal{K}^*,
\end{aligned}
\end{equation}
where the primal constraint \(x \in \mathcal{K}\) gives rise to the dual variable \(\mu\), which lies in the dual cone \(\mathcal{K}^*\).

\begin{example}[Lagrangian Duality of NNLS]
Given \(\mathbf{A} \in \mathbb{R}^{m \times n}\) and \(\mathbf{b} \in \mathbb{R}^m\), the NNLS problem is given by:
\begin{equation}\label{prob:DPFO-NNLS-primal}%\tag{NNLS-Primal}
\begin{aligned}
\min_\mathbf{x \ge 0} &\quad \tfrac{1}{2}\|\mathbf{A}\mathbf{x} - \mathbf{b}\|_2^2.
\end{aligned}
\end{equation}

To simplify the notation, defining $\mathbf{y=Ax-b}$ and introducing dual variables $\boldsymbol{\lambda}$ for $\mathbf{y=Ax-b}$ and $\boldsymbol{\mu} \ge 0$ for $\mathbf{x} \ge 0$, we can rewrite the Lagrangian formulation of \eqref{prob:DPFO-NNLS-primal}:
\begin{align}
\label{eq:DPFO-NNLS-Lagrangian}
\mathcal{L}{(\mathbf{x}, \mathbf{y}, \boldsymbol{\lambda}, \boldsymbol{\mu})}
\;=\; \tfrac{1}{2}\|\mathbf{y}\|_2^2 + \boldsymbol{\lambda}^{\top}(\mathbf{y-Ax+b}) - \boldsymbol{\mu}^{\top}\mathbf{x}.
\end{align}

Then, the corresponding dual function is given by $g(\boldsymbol{\lambda, \mu}) = \inf_{\mathbf{x,y}} \mathcal{L}$, i.e., $g(\boldsymbol{\lambda, \mu}) = \inf_{\mathbf{x,y}} \{ \tfrac{1}{2}\|\mathbf{y}\|_2^2 + \boldsymbol{\lambda}^{\top}(\mathbf{y-Ax}) - \boldsymbol{\mu}^{\top}\mathbf{x} \} + \mathbf{{b}^{\top}}\boldsymbol{\lambda}$. Therefore, the dual problem of \eqref{prob:DPFO-NNLS-primal} is:
\begin{equation}\label{eq:DPFO-NNLS-Lagrangian-dual}\tag{NNLS-Lagrangian-Dual}
\begin{aligned}
\max_{\boldsymbol{\lambda}} &\quad -\tfrac{1}{2} \|\boldsymbol{\lambda}\|_2^2 - \mathbf{b}^{\top}\boldsymbol{\lambda}\\
\text{s.t.} &\quad  \mathbf{A}^{\top}\boldsymbol{\lambda} \geq 0.
\end{aligned}
\end{equation}
\end{example}

\subsubsection{Fenchel-Rockafellar Duality}
Given a function \(f\), its Fenchel-Rockafellar conjugate is defined as:
\begin{align}
f^{*}(y) \;= \sup_{x \in \operatorname{dom}(f)} ( y^{\top}x - f(x) ).
\end{align}
Here, $\operatorname{dom}(f)$ denotes the domain of $f$. The Fenchel--Young inequality follows directly:
\begin{align}
f^{*}(y) \ge y^{\top}x - f(x)
    \quad\Longrightarrow\quad
f(x) \ge y^{\top}x - f^{*}(y).
\end{align}
It then implies:
\begin{align}
f(x) \ge \sup_{y} \; ( y^{\top}x - f^{*}(y) ) = f^{**}(x).
\end{align}
If \(f\) is closed and convex, equality holds and we obtain the biconjugation identity \(f^{**} = f\).

Now consider the composite optimization problem:
\begin{align}
\label{prob:fenchel-duality-composite-problem}
\min_{x}\; f(Ax) + g(x).
\end{align}
Introducing an auxiliary variable $z$ allows us to rewrite it as:
\begin{equation}\label{prob:fenchel-duality-composite-problem-rewrite}
\begin{aligned}
\min_{x,z} & \quad f(z) + g(x) \\
\text{s.t.} &\quad Ax = z.
\end{aligned}
\end{equation}
The corresponding saddle-point formulation of \eqref{prob:fenchel-duality-composite-problem-rewrite} is:
\begin{align}
\min_{x,z}\; \max_{u}\;
\mathcal{L}(x,z,u)=f(z) + g(x) + u^{T}(Ax - z),
\end{align}
and applying the standard dual construction gives the Fenchel-Rockafellar dual problem:
\begin{align}
&\max_{u}\; \min_{x,z} \quad f(z) + g(x) + u^{\top}(Ax-z) \notag \\
&=\max_{u}\big\{-\sup_{x,z}\{u^{\top}z-f(z) +(-A^{T}u)^{\top}x-g(x)\}\big\} \notag \\
&=\max_{u}\big\{ -f^{*}(u) - g^{*}(-A^{T}u)\big\}.
\end{align}

In practice, composite optimization problem \eqref{prob:fenchel-duality-composite-problem} is particularly well suited to operator-splitting schemes such as ADMM \cite{boyd2011distributed} or PDHG \cite{zhu2008efficient, chambolle2011first} method, both of which exploit the separability and Fenchel structure of $f$ and $g$. These algorithms will be discussed in subsequent sections.

\begin{example}[Fenchel-Rockafellar Duality of NNLS]
Given \(\mathbf{A} \in \mathbb{R}^{m \times n}\) and \(\mathbf{b} \in \mathbb{R}^m\), the NNLS problem is:
\begin{equation}\label{prob:fenchel-NNLS-primal}
\begin{aligned}
\min_{\mathbf{x}\ge 0} &\quad \tfrac{1}{2}\|\mathbf{A}\mathbf{x} - \mathbf{b}\|_2^2.
\end{aligned}
\end{equation}

\noindent To streamline notation, define $g(\mathbf{A}\mathbf{x}) = \tfrac{1}{2}\|\mathbf{A}\mathbf{x} - \mathbf{b}\|_2^2$. Then problem~\eqref{prob:fenchel-NNLS-primal} can be expressed as:
\begin{align}
\label{prob:fenchel-NNLS-primal-rewrite}
\min_{\mathbf{x}\ge 0}\; g(\mathbf{A}\mathbf{x})
&= \inf_{\mathbf{x}\ge 0} \left\{
    -\mathbf{y}^{\top}(\mathbf{A}\mathbf{x}) 
    + g(\mathbf{A}\mathbf{x}) 
    + \mathbf{y}^{\top}(\mathbf{A}\mathbf{x})
\right\} \notag \\
&= -\sup_{\mathbf{x}\ge 0}\big\{ \mathbf{y}^{\top}(\mathbf{A}\mathbf{x}) - g(\mathbf{A}\mathbf{x})\big\}
    + \inf_{\mathbf{x}\ge 0} \big\{ \mathbf{y}^{\top}(\mathbf{A}\mathbf{x}) \big\} \notag \\
&= -g^{*}(\mathbf{y})
    + \inf_{\mathbf{x}\ge 0} \big\{ \mathbf{y}^{\top}(\mathbf{A}\mathbf{x}) \big\},
\end{align}
which yields the corresponding saddle-point representation:
\begin{align}
\label{prob:fenchel-NNLS-saddle-point}
\max_{\mathbf{y}} \;
\big\{
    - g^{*}(\mathbf{y}) 
    + \inf_{\mathbf{x}\ge 0} \{\mathbf{y}^{\top}\mathbf{A}\mathbf{x}\}
\big\}
\;=\;
\max_{\mathbf{y}}\; \min_{\mathbf{x}\ge 0} 
\left\{
    - g^{*}(\mathbf{y}) + \mathbf{y}^{\top}\mathbf{A}\mathbf{x}
\right\}.
\end{align}
Thus the Fenchel-Rockafellar dual problem is $\max_{\mathbf{y}} \; - g^{*}(\mathbf{y})$. Noting that $g(\mathbf{Ax})=\tfrac{1}{2}\|\mathbf{A}\mathbf{x} - \mathbf{b}\|_2^2$, we obtain $g^{*}(\mathbf{y}) = \tfrac{1}{2}\|\mathbf{y}\|_2^2 + \mathbf{b}^{\top}\mathbf{y}$. Therefore, the Fenchel-Rockafellar dual problem of \eqref{prob:fenchel-NNLS-primal} becomes:
\begin{equation}\label{prob:fenchel-NNLS-dual-final}\tag{NNLS-Fenchel-Rockafellar-Dual}
\begin{aligned}
\max_{\mathbf{y}} \; - g^{*}(\mathbf{y})
\;=\;
\max_{\mathbf{y}}
\left\{
    -\tfrac{1}{2}\|\mathbf{y}\|_2^2 - \mathbf{b}^{\top}\mathbf{y}
\right\},
\end{aligned}
\end{equation}
which agrees exactly with the dual obtained from Lagrangian duality \eqref{eq:DPFO-NNLS-Lagrangian-dual}. Finally, the associated saddle-point formulation of \eqref{prob:fenchel-NNLS-primal} becomes:
\begin{equation}\label{prob:fenchel-NNLS-saddle-point-final}%\tag{NNLS-Fenchel-Rockafellar-Saddle-Point}
\begin{aligned}
\max_{\mathbf{y}}\; \min_{\mathbf{x}\ge 0} \;
\left\{
    \mathbf{y}^{\top}\mathbf{A}\mathbf{x}
    - \tfrac{1}{2}\|\mathbf{y}\|_2^2
    - \mathbf{b}^{\top}\mathbf{y}
\right\}.
\end{aligned}
\end{equation}

\end{example}

\subsection{First-Order Methods in Differentiable Programming}
\label{subsec:first-order-method-dp}
\begin{figure*}[!t]
  \centering
  \begin{subfigure}{0.25\textwidth}
    \centering
    \includegraphics[width=\linewidth]{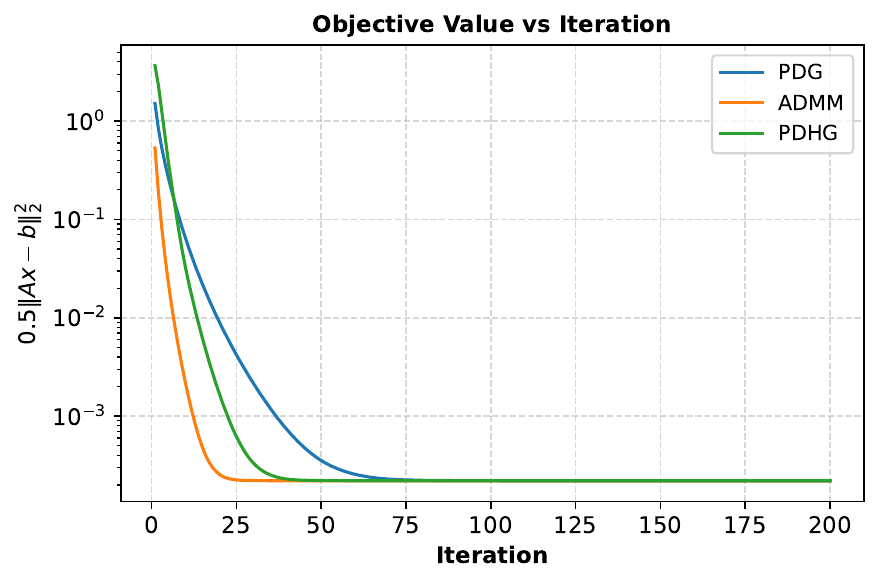}
    \caption{\footnotesize Objective value.}
    \label{fig:nnls_pdg_admm_pdhg_objective_value}
  \end{subfigure}\hfill
  %\hfill
  \begin{subfigure}{0.25\textwidth}
    \centering
    \includegraphics[width=\linewidth]{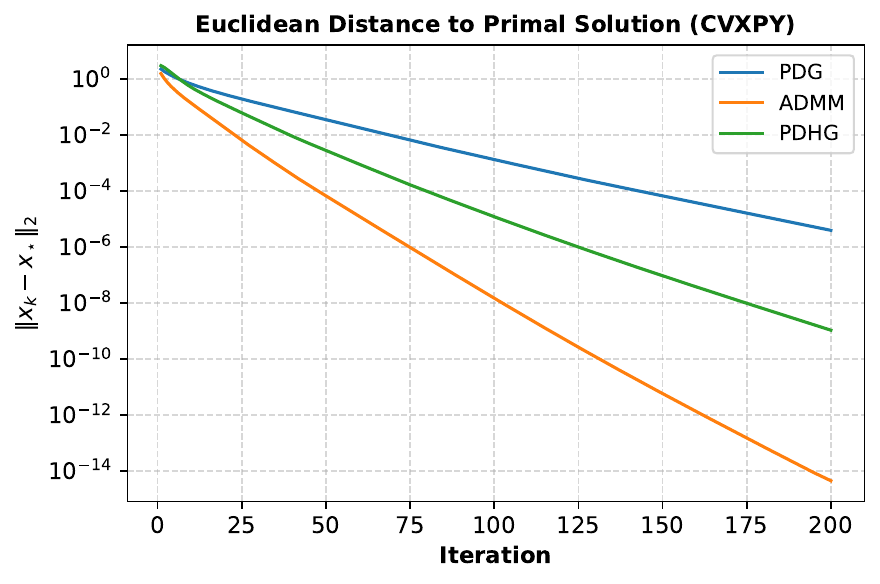}
    \caption{\footnotesize Primal solution.}
    \label{fig:nnls_pdg_admm_pdhg_distance_to_xstar}
  \end{subfigure}\hfill
  %\hfill
  \begin{subfigure}{0.25\textwidth}
    \centering
    \includegraphics[width=\linewidth]{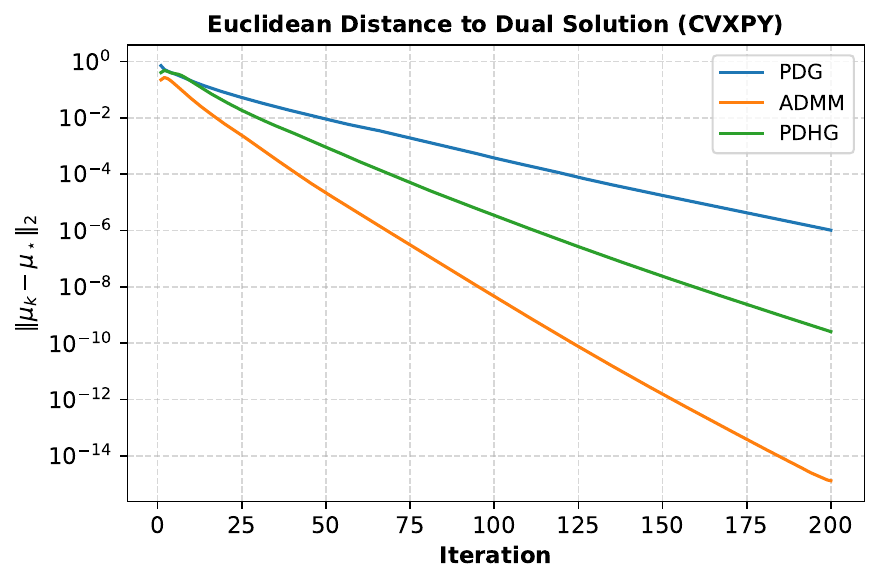}
    \caption{\footnotesize Dual solution.}
    \label{fig:nnls_pdg_admm_pdhg_distance_to_mustar}
  \end{subfigure}\hfill
  %\hfill
  \begin{subfigure}{0.25\textwidth}
    \centering
    \includegraphics[width=\linewidth]{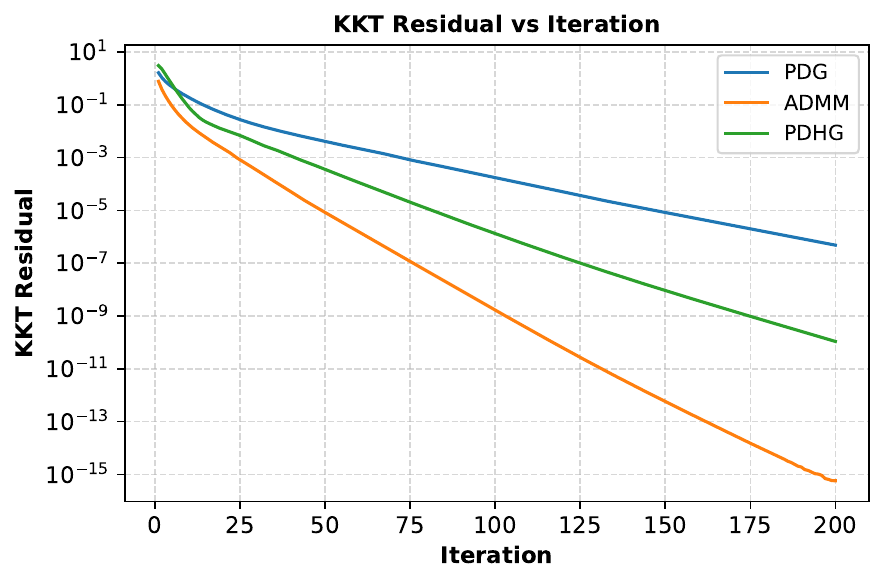}
    \caption{\footnotesize KKT residual.}
    \label{fig:nnls_pdg_admm_pdhg_kkt_residual}
  \end{subfigure}
  \caption{Convergence comparison of PDG, ADMM, and PDHG on the NNLS problem \eqref{prob:FOM-NNLS-primal}, measured by (a) objective value, (b) Euclidean distance to the primal solution, (c) Euclidean distance to the dual solution, and (d) KKT residual. Reference solutions are computed via CVXPY.}
  \label{fig:nnls_pdg_admm_pdhg_comparison}
\end{figure*}

First-order methods update variables using only gradient information \cite{teel2019first}, avoiding the second-derivative computations required by higher-order schemes such as Newton’s method~\cite{dembo1982inexact}. Their simplicity, low per-iteration cost, and scalability make them well suited to large-scale and high-dimensional problems~\cite{dvurechensky2021first}. In differentiable programming, they serve as core components by leveraging automatic differentiation for end-to-end trainable updates. Accelerated variants further improve convergence without added complexity~\cite{li2020accelerated}, with Nesterov’s method~\cite{nesterov2013gradient} being a key example. In this section, we introduce several first-order methods, including Primal–Dual Gradient method (PDG) \cite{boyd2004convex, du2019linear}, the Alternating Direction Method of Multipliers (ADMM)~\cite{boyd2011distributed}, and the Primal–Dual Hybrid Gradient (PDHG) algorithm~\cite{zhu2008efficient, chambolle2011first}. We then illustrate and compare these methods on the problem~\eqref{prob:FOM-NNLS-primal}, as shown in Figure~\ref{fig:nnls_pdg_admm_pdhg_comparison}, where \(\mathbf{A} \in \mathbb{R}^{m \times n}\) and \(\mathbf{b} \in \mathbb{R}^m\):
\begin{equation}\label{prob:FOM-NNLS-primal}
\begin{aligned}
\min_{\mathbf{x \ge 0}} \; \tfrac{1}{2}\|\mathbf{A}\mathbf{x} - \mathbf{b}\|_2^2.
\end{aligned}
\end{equation}

%%---Primal-Dual Gradient Method : PDG
%%---Projected Gradient Descent : PGD

\subsubsection{PDG}
PDG method~\cite{boyd2004convex, du2019linear} extends Gradient Descent (GD) \cite{ruder2016overview, andrychowicz2016learning} to constrained optimization through the Lagrangian saddle-point formulation. Given a differentiable objective $f(x)$, GD performs iterative updates:
\begin{align}
x_{t+1} = x_t - \gamma \nabla f(x_t),
\end{align}
where $\gamma>0$ denotes the step size. For convex functions with an $L$-Lipschitz continuous gradient, choosing $\gamma\in(0,1/L]$ yields a sublinear convergence rate of $\mathcal{O}(1/T)$, while linear convergence can be achieved when $f$ is additionally $\mu$-strongly convex.

Building upon GD, PDG solves constrained problems by jointly updating primal and dual variables. Specifically, gradient descent is applied to the primal variables, while gradient ascent is performed on the dual variables to enforce constraints and seek a saddle point of the Lagrangian \cite{kovalev2022accelerated}. Owing to their simplicity, scalability, and broad applicability, PDG methods are widely used in convex optimization, game theory, distributed learning, and differentiable programming \cite{nesterov2009primal}.

\begin{example}[PDG for NNLS]
\label{ex:fom-pdg-nnls}
Taking the NNLS problem~\eqref{prob:FOM-NNLS-primal} as an example, introducing a nonnegative dual variable \(\boldsymbol{\mu} \in \mathbb{R}^n\) gives the Lagrangian:
\begin{equation}
\label{eq:PDGM-NNLS-lagrangian}
\mathcal{L}(\mathbf{x}, \boldsymbol{\mu})
= \tfrac{1}{2}\|\mathbf{A}\mathbf{x} - \mathbf{b}\|_2^2
- \boldsymbol{\mu}^\top \mathbf{x}.
\end{equation}
The associated dual function is defined as:
\begin{align}
g(\boldsymbol{\mu}) = \inf_{\mathbf{x}} \; \mathcal{L}(\mathbf{x}, \boldsymbol{\mu}).
\end{align}

In the PDG method, the primal variable \(\mathbf{x}\) and dual variable \(\boldsymbol{\mu}\) are updated simultaneously by taking gradient steps on the Lagrangian~\eqref{eq:PDGM-NNLS-lagrangian}. The gradients are:
\begin{align}
\label{eq:PDGM-x-mu-gradient}
\nabla_{\mathbf{x}}\mathcal{L} = \mathbf{A}^\top(\mathbf{A}\mathbf{x}-\mathbf{b}) - \boldsymbol{\mu}, \qquad
\nabla_{\boldsymbol{\mu}}\mathcal{L} = -\mathbf{x}.
\end{align}
With step sizes \(\tau > 0\) and \(\sigma > 0\), the updates are:
\begin{align}
\label{eq:PDGM-x-update}
\mathbf{x}^{k+1}
= \Pi_{\mathbb{R}^n_{\ge 0}}
\!( \mathbf{x}^{k} - \tau \big( \mathbf{A}^\top (\mathbf{A}\mathbf{x}^{k} - \mathbf{b}) - \boldsymbol{\mu}^{k} \big) ), \qquad
\boldsymbol{\mu}^{k+1}
= \Pi_{\mathbb{R}^n_{\ge 0}} \!( \boldsymbol{\mu}^{k} - \sigma \mathbf{x}^{k+1} ).
\end{align}
Here, the primal update reduces the objective, while the dual update enforces the constraint \(\mathbf{x} \ge 0\). Their coupled iterations converge to a saddle point \((\mathbf{x}^\ast, \boldsymbol{\mu}^\ast)\) satisfying the KKT conditions. \(\Pi_{\mathbb{R}^n_{\ge 0}}\) denotes projection onto the nonnegative orthant, yielding the Projected Gradient Descent (PGD) method~\cite{choi2025convergence,gupta2018cnn}. More generally, when the primal variable must lie in a convex set \(\mathcal{C} \subseteq \mathbb{R}^n\), the update becomes $x^{k+1} = \Pi_{\{x \in \mathcal{C}\}} \!( x^{k} - \tau ( \mathbf{A}^\top (\mathbf{A}\mathbf{x}^{k} - \mathbf{b}) - \boldsymbol{\mu}^{k}) )$, where \(\Pi_{\mathcal{C}}(x) = \arg\min_{v \in \mathcal{C}} \|x - v\|_2^2\) is the Euclidean projection~\cite{boyd2004convex, liu2009efficient}. This ensures feasibility at every iteration while moving in a descent direction.
\end{example}

\begin{algorithm}[!tbp]
\caption{ADMM Algorithm}
\label{alg:admm}
\begin{algorithmic}[1]
\Require Initial values $\mathbf{x}^{0}$, $\mathbf{y}^{0}$, and $\boldsymbol{\lambda}^{0}$, tolerances $\epsilon_r$ and $\epsilon_s$.
\Repeat
    \State Update the variables $\mathbf{x}$, $\mathbf{y}$, and $\boldsymbol{\lambda}$ according to \eqref{eq:FOM-ADMM-NNLS-x-update}, \eqref{eq:FOM-ADMM-NNLS-y-update}, and \eqref{eq:FOM-ADMM-NNLS-lambda-update}, respectively;
\Until{the convergence criteria \eqref{eq:FOM-ADMM-NNLS-stop-criteria} is satisfied;}
\Ensure Optimal solutions $\mathbf{x}^\star$, $\mathbf{y}^\star$, and $\boldsymbol{\lambda}^\star$.
\end{algorithmic}
\end{algorithm}

\subsubsection{ADMM}
ADMM~\cite{boyd2011distributed} is an optimization algorithm that combines the decomposability of dual ascent with the method of multipliers. It is particularly well suited for large-scale and distributed convex problems with separable objectives~\cite{chang2016asynchronous}. It has become a foundational optimization tool with applications in statistical learning, signal processing, and network control.

\begin{example}[ADMM for NNLS]
\label{ex:fom-admm-nnls}
Taking the NNLS problem~\eqref{prob:FOM-NNLS-primal} to illustrate ADMM algorithm, we introduce an auxiliary variable \(\mathbf{y}\) and rewrite it as:
\begin{equation}\label{prob:FOM-ADMM-NNLS-primal}
\begin{aligned}
\min_{\mathbf{x,y}} &\quad \tfrac{1}{2}\|\mathbf{y}\|_2^2 \\
\text{s.t.}\;\; &\quad \mathbf{A}\mathbf{x} - \mathbf{b} = \mathbf{y},
\quad \mathbf{x} \ge 0,
\end{aligned}
\end{equation}
where \(\mathbf{A} \in \mathbb{R}^{m \times n}\), \(\mathbf{b} \in \mathbb{R}^m\), and \(\mathbf{y} \in \mathbb{R}^m\). Introducing a Lagrange multiplier \(\boldsymbol{\lambda} \in \mathbb{R}^m\) for $\mathbf{A}\mathbf{x} - \mathbf{b} = \mathbf{y}$, the augmented Lagrangian of the problem \eqref{prob:FOM-ADMM-NNLS-primal} becomes:
\begin{equation}
\label{eq:FOM-ADMM-NNLS-lagrangian}
\mathcal{L}(\mathbf{x}, \mathbf{y}, \boldsymbol{\lambda})
= \tfrac{1}{2}\|\mathbf{y}\|_2^2
+ \boldsymbol{\lambda}^\top(\mathbf{A}\mathbf{x} - \mathbf{b} - \mathbf{y})
+ \tfrac{\rho}{2}\|\mathbf{A}\mathbf{x} - \mathbf{b} - \mathbf{y}\|_2^2,
\end{equation}
where \(\rho > 0\) is a penalty parameter encouraging satisfaction of the coupling constraint.

Starting from initial values \((\mathbf{x}^0, \mathbf{y}^0, \boldsymbol{\lambda}^0)\), the updates of the primal variables \(\mathbf{x}\) and \(\mathbf{y}\), together with the dual variable \(\boldsymbol{\lambda}\), proceed as follows:
\begin{align}
\mathbf{x}^{k+1}
&= \arg\min_{\mathbf{x} \ge 0} \;
\mathcal{L}_\rho(\mathbf{x}, \mathbf{y}^k, \boldsymbol{\lambda}^k) 
=\arg\min_{\mathbf{x} \ge 0}
\{
(\boldsymbol{\lambda}^k)^\top \mathbf{A}\mathbf{x}
\;+\;
\frac{\rho}{2}\big\|
\mathbf{A}\mathbf{x} - \mathbf{b} - \mathbf{y}^k
\big\|_2^2
\}, \label{eq:FOM-ADMM-NNLS-x-update}\\
\mathbf{y}^{k+1}
&= \arg\min_{\mathbf{y}} \;
\mathcal{L}_\rho(\mathbf{x}^{k+1}, \mathbf{y}, \boldsymbol{\lambda}^k) 
= \arg\min_{\mathbf{y}}
\{
\frac{1}{2}\|\mathbf{y}\|_2^2
\;-\;
(\boldsymbol{\lambda}^k)^\top \mathbf{y}
\;+\;
\frac{\rho}{2}\big\|
\mathbf{A}\mathbf{x}^{k+1} - \mathbf{b} - \mathbf{y}
\big\|_2^2
\}, \label{eq:FOM-ADMM-NNLS-y-update}\\
\boldsymbol{\lambda}^{k+1}
&= \boldsymbol{\lambda}^k
+ \rho (\mathbf{A}\mathbf{x}^{k+1} - \mathbf{b} - \mathbf{y}^{k+1}) \label{eq:FOM-ADMM-NNLS-lambda-update}.
\end{align}
Given prescribed tolerances $\epsilon_r$ and $\epsilon_s$, convergence can be assessed via the primal and dual residuals:
\begin{align}
\label{eq:FOM-ADMM-NNLS-stop-criteria}
\mathbf{r}^{k+1} = \mathbf{A}\mathbf{x}^{k+1} - \mathbf{b} - \mathbf{y}^{k+1}, 
\quad \|\mathbf{r}^{k+1}\| \leq \epsilon_r; \qquad
\mathbf{s}^{k+1} = \rho \mathbf{A}^{\top}\!\big(\mathbf{y}^{k+1} - \mathbf{y}^{k}\big), \quad \|\mathbf{s}^{k+1}\| \leq \epsilon_s.
\end{align}
\end{example}

As shown in Algorithm \ref{alg:admm}, ADMM method stops when both residuals fall below prescribed tolerances, signaling satisfaction of the KKT conditions. The primal residual measures constraint violation, while the dual residual reflects deviation from stationarity and thus dual optimality. For convex problems, ADMM converges to the optimal solution \((\mathbf{x}^\star, \mathbf{y}^\star, \boldsymbol{\lambda}^\star)\). As a penalized alternating minimization method, ADMM balances primal and dual progress, supports parallel implementations, and combines the robustness of the method of multipliers with the flexibility of variable splitting, making it well suited to large-scale and distributed convex optimization.

\subsubsection{PDHG}
PDHG method \cite{zhu2008efficient, chambolle2011first} is a first-order primal–dual algorithm for large-scale convex and saddle-point problems. Each iteration jointly updates the primal and dual variables via simple proximal or gradient steps with linear operator evaluations, using inexpensive projections instead of costly linear solves~\cite{esser2010general}. PDHG excels in handling sparse and structured operators, is naturally parallelizable and preconditionable, and has found broad applications in imaging, total variation and $\ell_{1}$ regularization, optimal transport, and linear programming. With strong convexity or effective preconditioning, PDHG can be further accelerated for improved convergence speed and robustness. Specifically, the general problem addressed by PDHG is the generic saddle-point formulation:
\begin{align}
\label{prob:FOM-pdhg-saddle-point}
\min_{\mathbf{x}\in X}\max_{\mathbf{y}\in Y}
\big( \langle \mathbf{K}\mathbf{x}, \mathbf{y}\rangle 
    + G(\mathbf{x}) - F^{*}(\mathbf{y}) \big),
\end{align}
where $G : X \to [0,+\infty]$ and $F^{*} : Y \to [0,+\infty]$ are proper, convex, lower-semicontinuous functions. $F^{*}$ denotes the convex conjugate of a convex lower-semicontinuous function $F$. The corresponding primal and dual problems are:
\begin{align}
\label{prob:FOM-pdhg-primal-dual}
\min_{\mathbf{x}\in X} \; F(\mathbf{K}\mathbf{x}) + G(\mathbf{x}),
\qquad \qquad
\max_{\mathbf{y}\in Y} \; -\big(G^{*}(-\mathbf{K}^{*}\mathbf{y}) - F^{*}(\mathbf{y})\big).
\end{align}

Following the Chambolle--Pock framework~\cite{chambolle2011first}, the PDHG iterations (cf. Algorithm \ref{alg:pdhg}) are:
\begin{align}
\mathbf{y}^{\,n+1} &= 
\operatorname{prox}_{\sigma F^{*}}
\!\left(\mathbf{y}^{\,n} + \sigma \mathbf{K}\bar{\mathbf{x}}^{\,n}\right),
\qquad
\mathbf{x}^{\,n+1} = 
\operatorname{prox}_{\tau G}
\!\left(\mathbf{x}^{\,n} - \tau \mathbf{K}^{*}\mathbf{y}^{\,n+1}\right),
\label{eq:FOM-pdhg-update-y-x} \\
\bar{\mathbf{x}}^{\,n+1} &= 
\mathbf{x}^{\,n+1} 
+ \theta\big(\mathbf{x}^{\,n+1} - \mathbf{x}^{\,n}\big),
\label{eq:FOM-pdhg-update-xbar}
\end{align}
where $\sigma$ and $\tau$ are step sizes. $\mathbf{K}^{*}$ denotes the adjoint of $\mathbf{K}$, with the extrapolation parameter typically set to $\theta=1$, and the proximal operator \cite{parikh2014proximal} is:
\begin{align}
\operatorname{prox}_{\sigma F^{*}}(\mathbf{q})
= \arg\min_{\mathbf{y}}
(
\tfrac{1}{2}\|\mathbf{y}-\mathbf{q}\|_2^{2}
+ \sigma F^{*}(\mathbf{y})
).
\end{align}

Given prescribed tolerances $\epsilon_x$ and $\epsilon_y$, the PDHG iterations are terminated when the primal and dual residuals derived from the KKT conditions of~\eqref{prob:FOM-pdhg-saddle-point} satisfy:
\begin{align}
\mathbf{r}_x^{\,n}
=
\frac{1}{\tau} (\mathbf{x}^{\,n} - \operatorname{prox}_{\tau G} (\mathbf{x}^{\,n}-\tau \mathbf{K}^{*}\mathbf{y}^{\,n})), 
\qquad
\mathbf{r}_y^{\,n}
=
\frac{1}{\sigma} (\mathbf{y}^{\,n} - \operatorname{prox}_{\sigma F^{*}}
(\mathbf{y}^{\,n}+\sigma \mathbf{K}\mathbf{x}^{\,n})),
\end{align}
with stopping criteria:
\begin{align}
\label{eq:DP-FOM-pdhg-stop}
\|\mathbf{r}_x^{\,n}\|\le\epsilon_x,
\qquad
\|\mathbf{r}_y^{\,n}\|\le\epsilon_y.
\end{align}

In particular, an equivalent variant of PDHG replaces $\bar{\mathbf{x}}^{\,n+1}$ with $\bar{\mathbf{y}}^{\,n+1} = \mathbf{y}^{\,n+1} + \theta(\mathbf{y}^{\,n+1}-\mathbf{y}^{\,n})$, which is obtained by interchanging the primal and dual update steps.

\begin{algorithm}[!tbp]
\caption{PDHG Algorithm}
\label{alg:pdhg}
\begin{algorithmic}[1]
\Require 
Initial $\mathbf{x}^{0},\, \mathbf{y}^{0}$, and $\bar{\mathbf{x}}^{\,0}=\mathbf{x}^{0}$; step sizes $\tau,\sigma>0$ and extrapolation parameter $\theta\in[0,1]$.
\State Set iteration counter $k \gets 0$;
\Repeat
    \State Update dual variable $\mathbf{y}^{k+1}$ and primal variable $\mathbf{x}^{k+1}$ according to ~\eqref{eq:FOM-pdhg-update-y-x};
    \State Compute $\bar{\mathbf{x}}^{\,k+1}$ according to~\eqref{eq:FOM-pdhg-update-xbar};
    \State Check convergence criteria \eqref{eq:DP-FOM-pdhg-stop}; if satisfied, terminate;
    \State Set $k \gets k + 1$;
\Until{convergence;}
\Ensure Primal--dual solution $(\mathbf{x}^{*},\,\mathbf{y}^{*})$.
\end{algorithmic}
\end{algorithm}

\begin{example}[PDHG for NNLS]
\label{ex:fom-pdhg-nnls}
As an illustrative example, consider again the NNLS problem~\eqref{prob:FOM-NNLS-primal}:
\begin{equation}\label{prob:FOM-PDHG-NNLS-primal}
\min_{\mathbf{x}\ge 0}\; \tfrac{1}{2}\|\mathbf{A}\mathbf{x}-\mathbf{b}\|_{2}^{2}.
\end{equation}

Using the Fenchel-Rockafellar saddle-point representation previously derived in \eqref{prob:fenchel-NNLS-saddle-point-final}, we obtain the equivalent formulation:
\begin{equation}
\label{prob:FOM-PDHG-NNLS-saddle-point}
\min_{\mathbf{x}\ge0}\max_{\mathbf y}
\{
\langle \mathbf A\mathbf x,\mathbf y\rangle
-\frac12\|\mathbf y\|_2^2
-\mathbf b^\top\mathbf y
\}
=
\min_{\mathbf{x}}\max_{\mathbf y}
\left\{
\langle \mathbf A\mathbf x,\mathbf y\rangle
+G(\mathbf x)-F^*(\mathbf y)
\right\},
\end{equation}
where $G(\mathbf x)=\mathbb I_{\mathbb R^n_{\ge0}}(\mathbf x)$ and $F^*(\mathbf y)=\frac12\|\mathbf y\|_2^2+\mathbf b^\top\mathbf y$. The PDHG update of $\mathbf{x}$ associated with \eqref{prob:FOM-PDHG-NNLS-saddle-point} is:
\begin{align}
\mathbf{x}^{k+1}
&= \operatorname{prox}_{\tau G}\;(\mathbf{x}^{k}-\tau A^{\top}\mathbf{y}^{k}) \notag\\
&= \arg\min_{\mathbf{x}}
\left[
\tfrac12\|\mathbf{x}-\big(\mathbf{x}^{k}-\tau A^{\top}\mathbf{y}^{k}\big)\|_{2}^{2}
+\tau G(\mathbf{x})
\right] \notag\\
&= \arg\min_{\mathbf{x}\ge 0}
\left[
\tfrac12\|\mathbf{x}-\big(\mathbf{x}^{k}-\tau A^{\top}\mathbf{y}^{k}\big)\|_{2}^{2}
\right] 
\;=\; \Pi_{\mathbb{R}^n_{\ge 0}}\;(\mathbf{x}^{k}-\tau A^{\top}\mathbf{y}^{k}).
\end{align}
Similarly, the PDHG update of $\mathbf{y}$ is:
\begin{align}
\mathbf{y}^{k+1}
&=\operatorname{prox}_{\sigma F^{*}}
\;( \mathbf{y}^{k} + \sigma A\bar{\mathbf{x}}^{\,k+1} ) \notag\\
&= \arg\min_{\mathbf{y}}
\left[
\tfrac12\|\mathbf{y}-(\mathbf{y}^{k}+\sigma A\bar{\mathbf{x}}^{\,k+1})\|_{2}^{2}
+\sigma F^{*}(\mathbf{y})
\right] \notag\\
&= \arg\min_{\mathbf{y}}
\left[
\tfrac12(1+\sigma)\|\mathbf{y}\|_{2}^{2}
-(\mathbf{y}^{k}+\sigma A\bar{\mathbf{x}}^{\,k+1})^{\!\top}\mathbf{y}
+\sigma\,\mathbf{b}^{\top}\mathbf{y}
\right] \notag\\
&= \frac{\mathbf{y}^{k} + \sigma\big(A\bar{\mathbf{x}}^{\,k+1}-\mathbf{b}\big)}{1+\sigma} 
\;=\; \frac{\mathbf{y}^{k} + \sigma\big(A(2\mathbf{x}^{k+1}-\mathbf{x}^{k})-\mathbf{b}\big)}{1+\sigma},
\end{align}
where $\bar{\mathbf{x}}^{\,k+1}=2\mathbf{x}^{k+1}-\mathbf{x}^{k}$ is the over-relaxed primal iterate.
\end{example}

%\vspace{1mm}
\textbf{PDLP.}
Building upon the PDHG framework, several specialized variants have been developed to address the challenges of large-scale optimization. A prominent example is Google’s Primal–Dual Hybrid Gradient for Linear Programming (PDLP) \cite{applegate2021practical, applegate2026pdlp}, implemented in C++ and integrated into the open-source OR-Tools library \cite{google_ortools_github}. PDLP applies classical PDHG to a saddle-point formulation of linear programming and adopts a sophisticated two-level scheme consisting of outer restarts and inner PDHG-style iterations. While the previous version \cite{applegate2021practical} established its foundational advantages in scaling, the latest iteration \cite{applegate2026pdlp} further strengthens its robustness and efficiency on large-scale linear programming instances. PDLP incorporates several key enhancements, including diagonal preconditioning, adaptive step-size selection, dynamic primal–dual weighting, and an adaptive restarting mechanism that accelerates convergence. These refinements preserve the simplicity of projection-based first-order updates while improving numerical stability and high-accuracy performance. Moreover, its saddle-point structure admits a differentiable representation, making PDLP amenable to integration as a scalable optimization layer in end-to-end learning systems.

\subsection{Implicit Differentiation through Optimization Layers}
\label{subsec:implicit-diff}
The first-order methods discussed in Section~\ref{subsec:first-order-method-dp} provide an explicit way to differentiate through optimization algorithms: one may unroll a finite number of PDG, ADMM, or PDHG iterations and apply automatic differentiation to the resulting computation graph. This approach is natural in differentiable programming, but its memory and computational cost grow with the number of unrolled iterations. An alternative and often more solver-agnostic approach is implicit differentiation~\cite{kolter2020deep,blondel2024elements}, which differentiates the equations characterizing the solution rather than the internal iterations of a particular solver.

Many optimization-based components are naturally viewed as implicit functions. Unlike ordinary differentiable programs whose outputs are obtained by an explicit composition of elementary operations, an implicit function is defined through a relation that its output must satisfy, such as an optimality condition, a fixed-point equation, or a nonlinear system of equations~\cite{blondel2024elements}. In this setting, the output is written as $z^\star(\theta)$, where $\theta$ denotes problem data or learnable parameters, but $z^\star(\theta)$ is not available as an explicit formula. A generic implicit representation is:
\begin{equation}
    F(z^\star,\theta)=0,
\label{eq:implicit-equation}
\end{equation}
where $F$ encodes the equations characterizing the desired solution. 

The implicit function theorem~\cite{dontchev2009implicit,blondel2024elements} provides conditions under which this relation locally defines $z^\star$ as a differentiable function of $\theta$. In particular, if $F$ is continuously differentiable around $(z^\star,\theta)$ and the partial Jacobian $D_zF(z^\star,\theta)$ is nonsingular, then $z^\star(\theta)$ is locally differentiable. Since the defining relation holds along the solution path, i.e., $F(z^\star(\theta),\theta)=0$, differentiating both sides with respect to $\theta$ gives \cite{valenzuela2025centralized}:
\begin{equation}
    D_zF(z^\star,\theta)D_\theta z^\star + D_\theta F(z^\star,\theta) =0.
    \quad\Longrightarrow\quad
    D_\theta z^\star
    =
    -\left(D_zF(z^\star,\theta)\right)^{-1}
    D_\theta F(z^\star,\theta).
\label{eq:implicit-chain-rule}
\end{equation}

Therefore, the derivative of the solution map is obtained by solving a linear system involving the Jacobian of the defining equations, rather than by differentiating through every internal iteration of a solver. In optimization layers, $F(z^\star,\theta)=0$ is typically chosen as an optimality system; for a smooth parametrized convex program, $z=(x,\lambda,\nu)$ collects the primal and dual variables and $F$ denotes the KKT residual system. Under standard regularity conditions such as strong duality, differentiability, active-set nondegeneracy, and nonsingularity of the KKT Jacobian, the solution map is locally differentiable~\cite{barratt2018differentiability}, and the backward pass reduces to solving a KKT linear system.

A representative example is OptNet~\cite{amos2017optnet}, which embeds a quadratic program as a neural-network layer. The forward pass solves the QP, while the backward pass is obtained by differentiating the QP KKT conditions at the optimum. This idea extends beyond QPs to more general convex programs. In particular, CVXPYLayers~\cite{agrawal2019differentiable,agrawal2020differentiating} canonicalizes a disciplined parametrized convex program into cone-program data and computes derivatives of the solution map through implicit differentiation of the associated cone-program optimality system. Therefore, CVXPYLayers differs from unrolled first-order methods: the computation graph does not contain every interior-point or splitting iteration of the solver, but instead exposes the optimizer itself as a differentiable implicit layer.

Implicit differentiation also provides a useful bridge to bilevel learning. For hyperparameter optimization and meta-learning, one may either differentiate the exact lower-level optimality conditions or unroll a finite optimization trajectory. The latter view is adopted in bilevel programming frameworks where the inner dynamics are differentiated to compute hypergradients~\cite{franceschi2018bilevel}. Thus, unrolled differentiation and implicit differentiation represent two complementary mechanisms for differentiating through optimization: the former differentiates the algorithmic path, whereas the latter differentiates the fixed optimality system. Recent work further shows that, for large-scale structured optimization problems, implicit differentiation can exploit separability and coupling structure by decomposing the total sensitivity into local Jacobians and coupling corrections~\cite{valenzuela2025centralized}.

\subsection{Differentiable Programming Implementation for Cone Programs}
As shown in Figure~\ref{fig:diffprog}, differentiable programming provides a unified framework for solving cone programs with first-order methods. Primal variables are updated by gradient steps, while dual variables follow projected ascent in the dual space. Dual bounds and duality gaps quantify consistency between primal and dual iterates and guide refinement of the primal solution. Under strong duality, solving the dual problem not only certifies optimality, but also enables recovery of the primal solution via dual solution, thereby offering an alternative solution strategy. In this section, we revisit the problem \eqref{prob:Intro-NNLS-primal} to illustrate how differentiable programming, built on Lagrangian duality, provides a principled framework for analyzing cone programs and implementing them across multiple frameworks: PyTorch-based first-order methods and CVXPYLayers with automatic KKT differentiation for scalable end-to-end learning.

\subsubsection{Classical NNLS Problem Implementation in PyTorch}
This section implements efficient first-order algorithms in PyTorch and show that backpropagation enables learning to predict dual solutions, from which the corresponding primal solution can be recovered.

\begin{example}[Classical NNLS]
Given $\mathbf{A} \in \mathbb{R}^{m \times n}$ and $\mathbf{b} \in \mathbb{R}^m$, find \(\mathbf{x} \in \mathbb{R}_+^n\) by solving:
\begin{equation}\label{prob:DP-FOM-NNLS-primal}%\tag{NNLS-Primal}
\begin{aligned}
\min_\mathbf{x \ge 0} &\quad \tfrac{1}{2} \|\mathbf{Ax} - \mathbf{b}\|_2^2.
\end{aligned}
\end{equation}

Rewrite the corresponding dual problem \eqref{eq:DPFO-NNLS-Lagrangian-dual} to the minimum dual problem: 
\begin{equation}\label{prob:DP-FOM-NNLS-dual-rewrite}\tag{NNLS-Dual-Min}
\begin{aligned}
\min_{\boldsymbol{\lambda,\mu}} &\quad \tfrac{1}{2} \|\boldsymbol{\lambda}\|_2^2 + \mathbf{b}^{\top}\boldsymbol{\lambda} \\
\text{s.t.} &\quad \mathbf{A}^{\top}\boldsymbol{\lambda} - \boldsymbol{\mu} = 0, \quad \boldsymbol{\mu} \ge 0.
\end{aligned}
\end{equation}

The problem \eqref{prob:DP-FOM-NNLS-primal} is a convex cone program and satisfies both the Slater condition and the KKT conditions. Consequently, its optimal primal solution $\mathbf{x}^*$ can be derived from the optimal dual solution.
\end{example}

\begin{figure}[!t]
    \centering
    \includegraphics[width=0.70\linewidth, height=1.8cm]{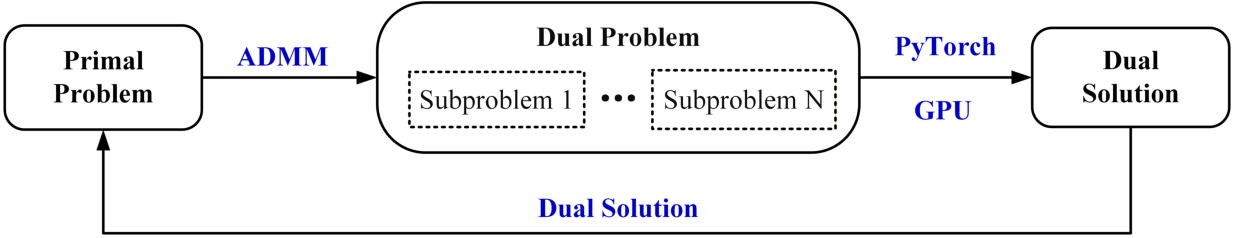}
    \caption{ADMM with PyTorch. The primal problem is reformulated into a decomposed dual problem, whose subproblems are solved in parallel on GPU, and the dual solution recovers the primal solution.}
    \label{fig:L2O-DP-PyTorch-FOM-with-GPU}
\end{figure}

\textbf{(1) ADMM with PyTorch:} Introducing the Lagrangian multiplier $\mathbf{z}$, i.e., dual variable, and a penalty parameter $\rho > 0$, the augmented Lagrangian of \eqref{prob:DP-FOM-NNLS-dual-rewrite} is given by: 
\begin{align}
\label{eq:DP-FOM-NNLS-Aug-Larangian}
\mathcal{L}_\rho(\boldsymbol{\lambda}, \boldsymbol{\mu}, \mathbf{z}) 
= \tfrac{1}{2} \|\boldsymbol{\lambda}\|_2^2 + \mathbf{b}^{\top}\boldsymbol{\lambda} 
+ \mathbf{z}^{\top}(\mathbf{A}^{\top}\boldsymbol{\lambda} - \boldsymbol{\mu})
+ \tfrac{\rho}{2}\|\mathbf{A}^{\top}\boldsymbol{\lambda} - \boldsymbol{\mu}\|_2^2.
\end{align}

By rearranging the Lagrangian function \eqref{eq:DP-FOM-NNLS-Aug-Larangian}, removing constant terms, and defining $\boldsymbol{\upsilon} = \tfrac{1}{\rho}\mathbf{z}$, the corresponding augmented Lagrangian in the scaled dual form can be obtained as:
\begin{align}
\label{eq:DP-FOM-NNLS-Aug-Larangian-Scaled}
\mathcal{L}_\rho(\boldsymbol{\lambda}, \boldsymbol{\mu}, \boldsymbol{\upsilon}) 
= \tfrac{1}{2} \|\boldsymbol{\lambda}\|_2^2 + \mathbf{b}^{\top}\boldsymbol{\lambda} 
+ \tfrac{\rho}{2}\|\mathbf{A}^{\top}\boldsymbol{\lambda} - \boldsymbol{\mu} + \boldsymbol{\upsilon}\|_2^2,
\end{align}

\noindent The ADMM iterative updates for solving the dual problem \eqref{prob:DP-FOM-NNLS-dual-rewrite} are given by: 
\begin{align}
\boldsymbol{\lambda}^{k+1} &= \arg\min_{\boldsymbol{\lambda}} 
\; \tfrac{1}{2} \|\boldsymbol{\lambda}\|_2^2 + \mathbf{b}^{\top}\boldsymbol{\lambda}
+\tfrac{\rho}{2}\|\mathbf{A}^{\top}\boldsymbol{\lambda} - \boldsymbol{\mu}^k + \boldsymbol{\upsilon}^k\|_2^2, \\
\boldsymbol{\mu}^{k+1} &= \arg\min_{\boldsymbol{\mu} \ge 0} 
\; \tfrac{\rho}{2}\|\mathbf{A}^{\top}\boldsymbol{\lambda}^{k+1} - \boldsymbol{\mu} + \boldsymbol{\upsilon}^k\|_2^2, \\
\boldsymbol{\upsilon}^{k+1} &= 
\boldsymbol{\upsilon}^k + (\mathbf{A}^{\top}\boldsymbol{\lambda}^{k+1} - \boldsymbol{\mu}^{k+1}).
\end{align}

The iterative procedure decomposes the dual problem \eqref{prob:DP-FOM-NNLS-dual-rewrite} into simple subproblems with respect to $\boldsymbol{\lambda}$, $\boldsymbol{\mu}$, and $\boldsymbol{\upsilon}$, each admitting a closed-form or efficiently computable update. This process is easily implemented in PyTorch to leverage GPU acceleration and parallel computation (cf. Figure~\ref{fig:L2O-DP-PyTorch-FOM-with-GPU}). As shown in the code below, with PyTorch’s efficient tensor operations and built-in projection support, the ADMM-based dual optimization pipeline achieves scalable, high-performance computation for large-scale NNLS problems within the differentiable programming framework.

\vspace{2mm}
\begin{lstlisting}[language=Python, basicstyle=\ttfamily\scriptsize]
device = torch.device("cuda" if torch.cuda.is_available() else "cpu")
def nnls_dual_admm(A, b, rho=1.0, iters=500):
    m, n = A.shape
    AT = A.T
    L = torch.linalg.cholesky(torch.eye(m, dtype=A.dtype, device=A.device) 
         + rho*(A@AT))
    lam, mu, v = [torch.zeros(sz, dtype=A.dtype, device=A.device) for sz in (m, n, n)]
    for _ in range(iters):
        lam = torch.cholesky_solve((-b + rho*A@(mu-v)).unsqueeze(1), L).squeeze(1)
        z   = AT@lam + v
        mu  = torch.clamp(z, min=0.0)
        v   = v + AT@lam - mu
    x = recover_x_from_mu_active_set(A, b, mu, tau, max_refine)
    return lam, x
\end{lstlisting}

\begin{figure}[!t]
    \centering
    \includegraphics[width=0.70\linewidth, height=2.1cm]{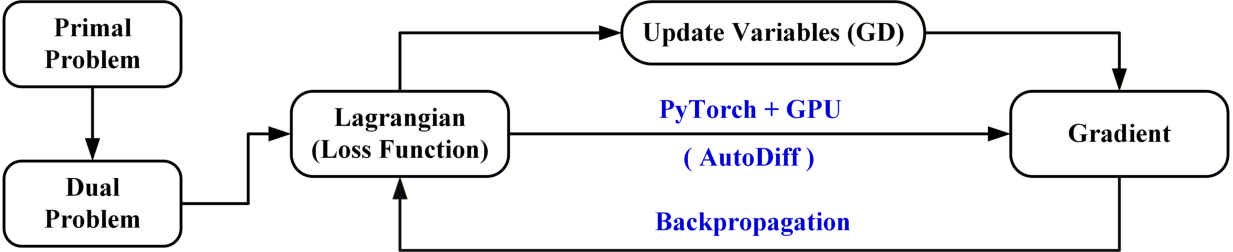}
    \caption{Learning with PyTorch. The dual problem is reformulated as minimizing its Lagrangian, treated as a loss function. Gradients are computed via reverse-mode automatic differentiation and variables are updated on GPU with projection steps, enabling end-to-end learning of the optimal solution.}
    \label{fig:L2O-DP-PyTorch-Learning-with-GPU}
\end{figure}

\vspace{2mm}
\textbf{(2) Learning with PyTorch:} 
An alternative approach to solving the dual problem \eqref{prob:DP-FOM-NNLS-dual-rewrite} is to treat its Lagrangian as a trainable loss and learn the optimal dual solution within PyTorch (cf. Figure~\ref{fig:L2O-DP-PyTorch-Learning-with-GPU}). Specifically, by introducing the auxiliary variable $\mathbf{z}$, the corresponding Lagrangian loss function is given by:
\vspace{1.5mm}
\begin{align}
\label{eq:DP-FOM-NNLS-Dual-Larangian}
\mathcal{L}(\boldsymbol{\lambda}, \boldsymbol{\mu}, \mathbf{z}) 
= \tfrac{1}{2} \|\boldsymbol{\lambda}\|_2^2 + \mathbf{b}^{\top}\boldsymbol{\lambda} 
+ \mathbf{z}^{\top}(\mathbf{A}^{\top}\boldsymbol{\lambda} - \boldsymbol{\mu}).
\end{align}
\vspace{1.5mm}

The nonnegativity constraint $\boldsymbol{\mu} \ge 0$ is enforced via projection. As illustrated in the code block below, once reformulated as a differentiable loss, the problem reduces to minimizing the Lagrangian. Through reverse-mode automatic differentiation combined with projection steps, PyTorch computes and backpropagates gradients to iteratively update the variables using built-in gradient-based optimizers until convergence, enabling end-to-end learning of the optimal solution. 

\begin{figure*}[!t]
  \centering
  \begin{subfigure}{0.45\textwidth}
    \centering
    \includegraphics[width=\linewidth]{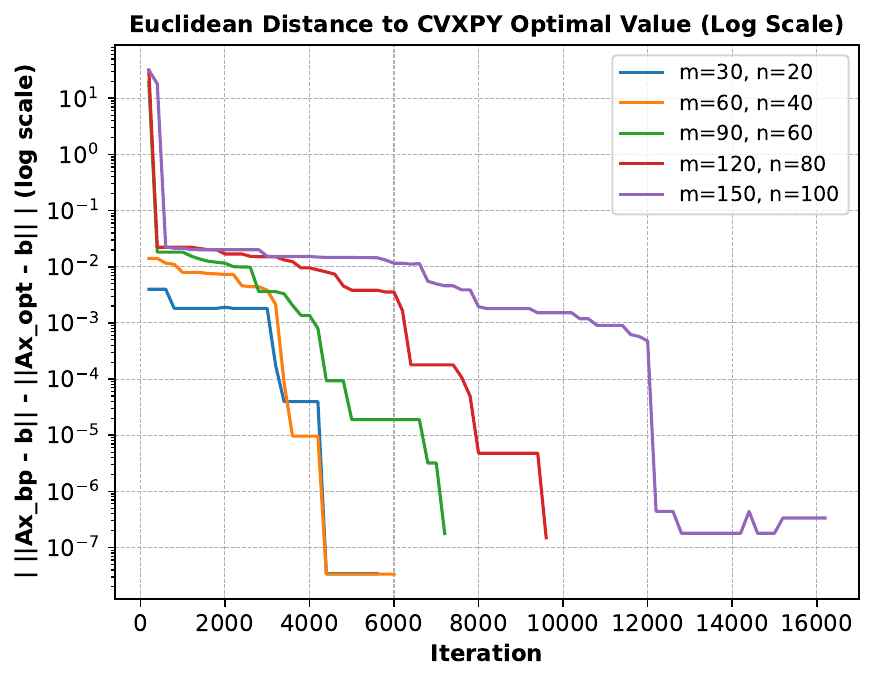}
    \caption{Euclidean distance to the optimal value.}
    \label{fig:nnls_learning_vs_cvxpy_obj_gap}
  \end{subfigure}
  %\hfill
  \begin{subfigure}{0.45\textwidth}
    \centering
    \includegraphics[width=\linewidth]{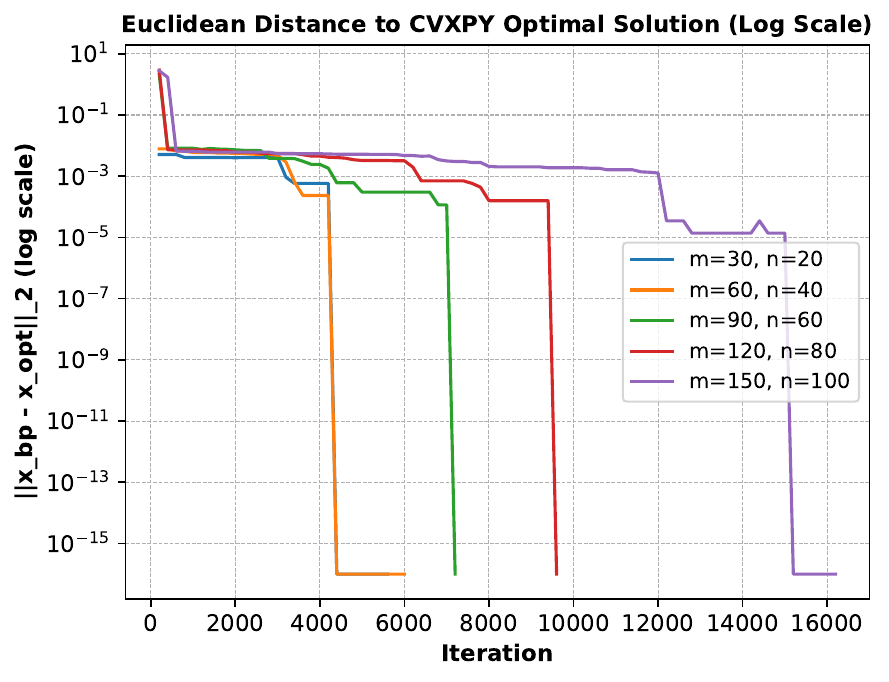}
    \caption{Euclidean distance to the optimal solution.}
    \label{fig:nnls_learning_vs_cvxpy_dist_to_opt}
  \end{subfigure}
  \caption{Convergence of the learning-based method on the NNLS problem \eqref{prob:DP-FOM-NNLS-primal}, evaluated by (a) the Euclidean distance to the CVXPY optimal value and (b) the Euclidean distance to the optimal solution computed by CVXPY. The method is implemented in PyTorch.}
  \label{fig:nnls_learning_vs_cvxpy}
\end{figure*}

As illustrated in Figure~\ref{fig:nnls_learning_vs_cvxpy}, we evaluate the convergence behavior of the learning-based solver across different problem scales. The solution obtained by CVXPY serves as the reference benchmark, allowing us to assess both the convergence speed and the solution accuracy of the proposed approach relative to the true optimum. This comparison provides insight into the optimization dynamics of the learning-based method as well as its ability to approximate the CVXPY solution with high fidelity.

\vspace{3mm}
\begin{lstlisting}[language=Python, basicstyle=\ttfamily\scriptsize]
device = torch.device("cuda" if torch.cuda.is_available() else "cpu")
def learn_nnls_dual_by_backprop(A, b, lr=1e-2, steps=3000, eps=1e-6):
    m, n = A.shape
    lam, mu, z = [torch.nn.Parameter(torch.zeros(sz, dtype=A.dtype, device=A.device)) for sz in (m, n, n)]
    L = lambda: 0.5*(lam@lam) + b@lam + z@(A.T@lam - mu)
    opt_min = torch.optim.Adam([lam, mu], lr=lr)
    opt_max = torch.optim.Adam([z], lr=lr)
    for _ in range(steps):
        opt_min.zero_grad();  L().backward();  opt_min.step()
        with torch.no_grad(): mu.clamp_(min=0.0)
        opt_max.zero_grad();  (-L()).backward();  opt_max.step()
        if torch.linalg.norm(A.T@lam - mu) <= eps and r_KKT <=eps: break
    x = recover_x_from_mu_active_set(A, b, mu, tau, max_refine)
    return lam.detach(), mu.detach(), z.detach(), x
\end{lstlisting}

\subsubsection{GPU-Based NNLS Problem Implementation in PyTorch}
We present the GPU-based distributed NNLS problem and demonstrate its parallel, scalable GPU-based implementation in PyTorch.

\begin{example}[GPU-Based Distributed NNLS]
Consider the NNLS problem in which the data $(\mathbf A_i,\mathbf b_i)$ are stored across $N$ computing nodes that jointly estimate a common nonnegative signal $\mathbf x$:
\begin{equation}
\label{prob:DP-FOM-Large-Scale-NNLS-Primal}
\min_{\mathbf x\ge 0}
\quad
\frac{1}{2}\sum_{i=1}^{N}
\|\mathbf A_i\mathbf x-\mathbf b_i\|_2^2 .
\end{equation}
Introducing auxiliary variables $\mathbf y_i=\mathbf A_i\mathbf x-\mathbf b_i$ with dual variables $\boldsymbol{\lambda}_i$, and the dual variable $\boldsymbol{\mu}\ge0$ for $\mathbf x\ge0$, the Lagrangian is:
\begin{equation}
\mathcal L
=
\sum_{i=1}^{N}
\left[
\tfrac12\|\mathbf y_i\|_2^2
+
\boldsymbol{\lambda}_i^\top
(\mathbf A_i\mathbf x-\mathbf b_i-\mathbf y_i)
\right]
-
\boldsymbol{\mu}^\top\mathbf x .
\end{equation}
Minimization over $\mathbf y_i$ and $\mathbf x$ yields $\mathbf y_i^\star=\boldsymbol{\lambda}_i$ and $\sum_i\mathbf A_i^\top\boldsymbol{\lambda}_i=\boldsymbol{\mu}\ge0$, respectively. The dual problem is:
\begin{equation}
\label{prob:DP-FOM-Large-Scale-NNLS-Dual}
\begin{aligned}
\min_{\boldsymbol{\lambda}=[\boldsymbol{\lambda}_1,\dots,\boldsymbol{\lambda}_N]}
\quad
& \frac{1}{2}\sum_{i=1}^{N}
\|\boldsymbol{\lambda}_i\|_2^2
+
\sum_{i=1}^{N}
\mathbf b_i^\top\boldsymbol{\lambda}_i \\
\text{s.t.}
\quad
& \sum_{i=1}^{N}
\mathbf A_i^\top\boldsymbol{\lambda}_i
\ge
\mathbf 0 .
\end{aligned}
\end{equation}
\end{example}

\textbf{(1) GPU-Based ADMM Implementation in PyTorch:}
Next, we solve~\eqref{prob:DP-FOM-Large-Scale-NNLS-Dual} via ADMM. Introducing local variables $z_i\in\mathbb R^n$ with $z_i=A_i^\top\lambda_i$, the dual problem can be written as:
\begin{equation}
\label{prob:DP-FOM-Large-Scale-NNLS-Dual-Consensus}
\begin{aligned}
\min_{\lambda=[\lambda_1,\dots,\lambda_N], \; z=[z_1, \dots, z_N]} \quad 
& \sum_{i=1}^{N}
\left(
\frac{1}{2}\|\lambda_i\|_2^2 
+ b_i^\top \lambda_i
\right) \\
\text{s.t.} \quad 
& A_i^\top \lambda_i = z_i,\quad i=1,\dots,N,\qquad \sum_{i=1}^{N} z_i \ge 0 .
\end{aligned}
\end{equation}
The corresponding scaled augmented Lagrangian is:
\begin{align}
\mathcal{L}_{\rho}
=
\sum_{i=1}^{N}
\left(
\frac{1}{2}\|\lambda_i\|_2^2 
+ b_i^{\top}\lambda_i
+ \frac{\rho}{2}
\|A_i^{\top}\lambda_i - z_i + u_i\|_2^2
\right)
+
\iota_{\ge 0}
\left(
\sum_{i=1}^{N}z_i
\right),
\end{align}
where $\rho>0$ is the penalty parameter, $u_i$ is the scaled dual variable, and $\iota_{\ge0}(\cdot)$ is the indicator function of the nonnegative orthant. The ADMM updates are given by:
\begin{align}
\lambda_i^{k+1}
&=
\arg\min_{\lambda_i}
\left\{
\frac{1}{2}\|\lambda_i\|_2^2
+ b_i^{\top}\lambda_i
+ \frac{\rho}{2}
\|A_i^{\top}\lambda_i - z_i^{k} + u_i^{k}\|_2^2
\right\}, \\
z_i^{k+1}
&=
v_i^k-\bar v^k+[\bar v^k]_+,
\qquad
v_i^k=A_i^\top\lambda_i^{k+1}+u_i^k,
\qquad
\bar v^k=\frac{1}{N}\sum_{i=1}^{N}v_i^k,\\
u_i^{k+1}
&=
u_i^k+A_i^\top\lambda_i^{k+1}-z_i^{k+1},
\quad i=1,\dots,N.
\end{align}

\begin{figure}[!t]
    \centering
    \includegraphics[width=0.8\linewidth, height=2.1cm]{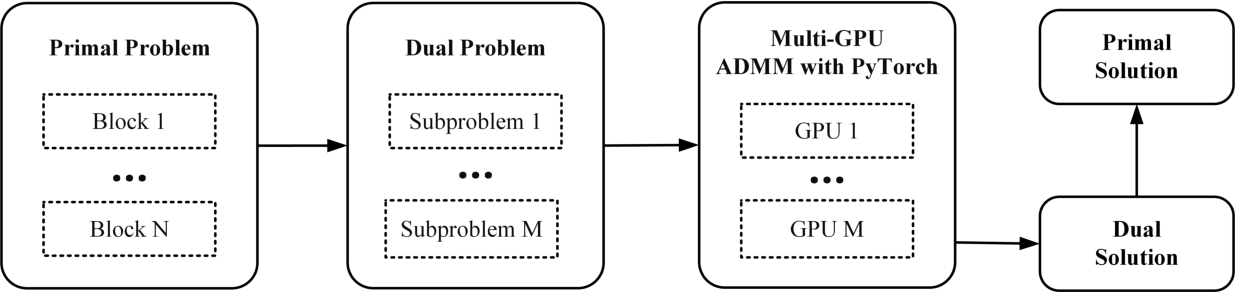}
    \caption{GPU-based ADMM implementation in PyTorch. The distributed NNLS dual problem is decomposed into block-wise subproblems, whose local updates are executed in parallel across GPUs. A global averaging/projection step enforces the coupling constraint $\sum_i \mathbf z_i \ge \mathbf 0$, and the primal solution is recovered after convergence via the KKT conditions.}
    \label{fig:L2O-DP-PyTorch-Large-Scale-with-Multi-GPU}
\end{figure}

%\vspace{3mm}
The iterative updates in the dual problem can be computationally demanding, especially when the data matrix is partitioned across many blocks. As illustrated in the code block below, PyTorch enables the block-wise $\boldsymbol{\lambda}_i$-updates to be executed in parallel on GPUs using batched linear algebra routines, while the coupling constraint is enforced through a lightweight global averaging/projection step. As illustrated in Figure~\ref{fig:L2O-DP-PyTorch-Large-Scale-with-Multi-GPU}, this local-parallel/global-reduction structure improves scalability and makes GPU-based distributed NNLS and related first-order methods practical. This observation aligns with recent evidence that dual iterations can scale efficiently on GPUs when fully vectorized~\cite{agrawal2022allocation}.

\vspace{2mm}
\begin{lstlisting}[language=Python, basicstyle=\ttfamily\scriptsize]
def dnnls_dual_admm_batch(A, B, rho=1.0, iters=1000):
    N, m, n = A.shape
    A_T = A.transpose(1, 2)
    I = torch.eye(m, device=A.device, dtype=A.dtype).unsqueeze(0)
    L = torch.linalg.cholesky(I + rho * torch.bmm(A, A_T))
    Lam = torch.zeros((N, m), device=A.device, dtype=A.dtype)
    Z_i = torch.zeros((N, n), device=A.device, dtype=A.dtype)
    U = torch.zeros_like(Z_i)
    for _ in range(iters):
        rhs = -B + rho * torch.bmm(A, (Z_i - U).unsqueeze(-1)).squeeze(-1)
        Lam = torch.cholesky_solve(rhs.unsqueeze(-1), L).squeeze(-1)
        A_Lam = torch.bmm(A_T, Lam.unsqueeze(-1)).squeeze(-1)
        V = A_Lam + U
        V_bar = V.mean(dim=0, keepdim=True)
        Z_i = V - V_bar + torch.clamp(V_bar, min=0.0)
        U = U + A_Lam - Z_i
    mu = torch.clamp(A_Lam.sum(dim=0), min=0.0)
    X = recover_x_from_mu_active_set(A, B, mu)
    return Lam, X, mu
\end{lstlisting}

\subsubsection{NNLS Differentiable Programming via CVXPYLayers}
We next introduce a differentiable programming implementation of \eqref{prob:Intro-NNLS-primal} using CVXPYLayers, where the matrix $\mathbf{A}$ is unknown. Signal recovery is adopted as the concrete instantiation.

\begin{figure}[!t]
    \centering
    \includegraphics[width=0.92\linewidth, height=2.5cm]{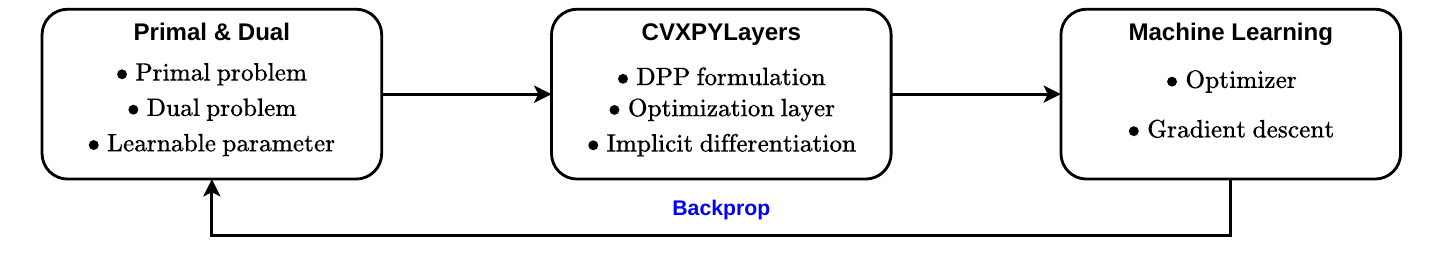}
    \caption{Differentiable programming implementation for the NNLS signal recovery problem via CVXPYLayers. The NNLS problem is formulated as a DPP and wrapped as a differentiable optimization layer. Implicit differentiation enables backpropagation through the optimization layer, allowing the reconstruction loss to update the learnable sensing matrix in an end-to-end differentiable learning framework.}
    \label{fig:nnls_signal_recovery_cvxpylayers}
\end{figure}

\begin{example}[NNLS Instantiation: Signal Recovery]
A true non-negative signal $\mathbf{x}_{\mathrm{true}} \in \mathbb{R}_+^n$ is observed through an unknown sensing matrix $\mathbf{P}_{\mathrm{true}}$, yielding measurement $\mathbf{b} = \mathbf{P}_{\mathrm{true}}\mathbf{x}_{\mathrm{true}}$. Given a learnable surrogate $\mathbf{P} \in \mathbb{R}^{m \times n}$ and observation $\mathbf{b} \in \mathbb{R}^m$, the signal is recovered by solving:
\begin{equation}
\label{prob:DP-CVXPYLayers-NNLS-primal}
\min_{\mathbf{x} \ge 0} \quad \tfrac{1}{2} \|\mathbf{Px} - \mathbf{b}\|_2^2,
\end{equation}
whose dual problem is:
\begin{equation}
\label{eq:DPFO-NNLS-dual-cvxpylayers}
\max_{\boldsymbol{\lambda}} \quad
-\tfrac{1}{2}\|\boldsymbol{\lambda}\|_2^2 + \mathbf{b}^{\!\top}\boldsymbol{\lambda}
\quad \text{s.t.} \quad \mathbf{P}^{\!\top}\boldsymbol{\lambda} \le \mathbf{0},
\end{equation}
which follows directly from the Lagrangian \eqref{eq:DPFO-NNLS-Lagrangian} without $\boldsymbol{\lambda}\leftarrow-\boldsymbol{\lambda}$ used in \eqref{eq:DPFO-NNLS-Lagrangian-dual}.

At optimum, the dual solution can be recovered from $\boldsymbol{\lambda}^\star = \mathbf b-\mathbf P\mathbf x^\star$. The sensing matrix $\mathbf{P}$ is learned by minimizing the reconstruction loss:
\begin{equation}
\label{eq:nnls-signal-recovery-loss}
  \mathcal{L}(\mathbf{P})
  = \tfrac{1}{2}\|\mathbf{x}^*(\mathbf{P}) - \mathbf{x}_{\mathrm{true}}\|_2^2,
\end{equation}
where minimizing $\mathcal{L}$ requires propagating gradients through the optimization layer, which is precisely the role of differentiable programming. Once $\mathbf{P}$ has converged, it serves as the data matrix $\mathbf{A}$ in the general NNLS formulation \eqref{prob:Intro-NNLS-primal}.
\end{example}

\begin{figure*}[!t]
  \centering
  \begin{subfigure}{0.27\textwidth}
    \centering
    \includegraphics[width=\linewidth]{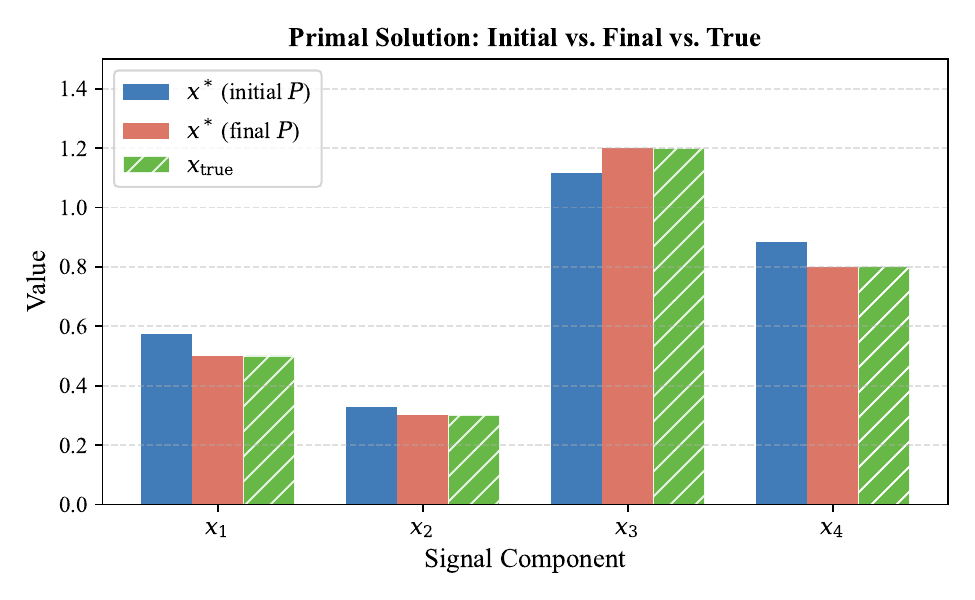}
    \caption{\footnotesize Primal solution.}
    \label{fig:nnls_signal_recovery_primal_solution}
  \end{subfigure}
  %\hfill
  \begin{subfigure}{0.25\textwidth}
    \centering
    \includegraphics[width=\linewidth]{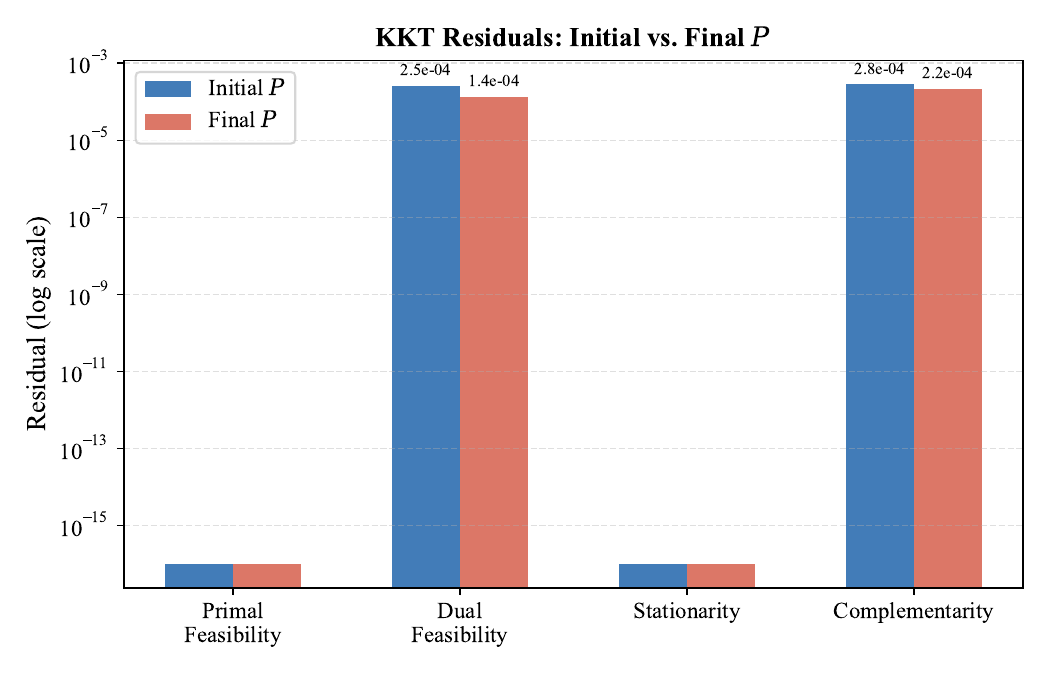}
    \caption{\footnotesize KKT residual.}
    \label{fig:nnls_signal_recovery_kkt_residual}
  \end{subfigure}
  %\hfill
  \begin{subfigure}{0.2\textwidth}
    \centering
    \includegraphics[width=\linewidth]{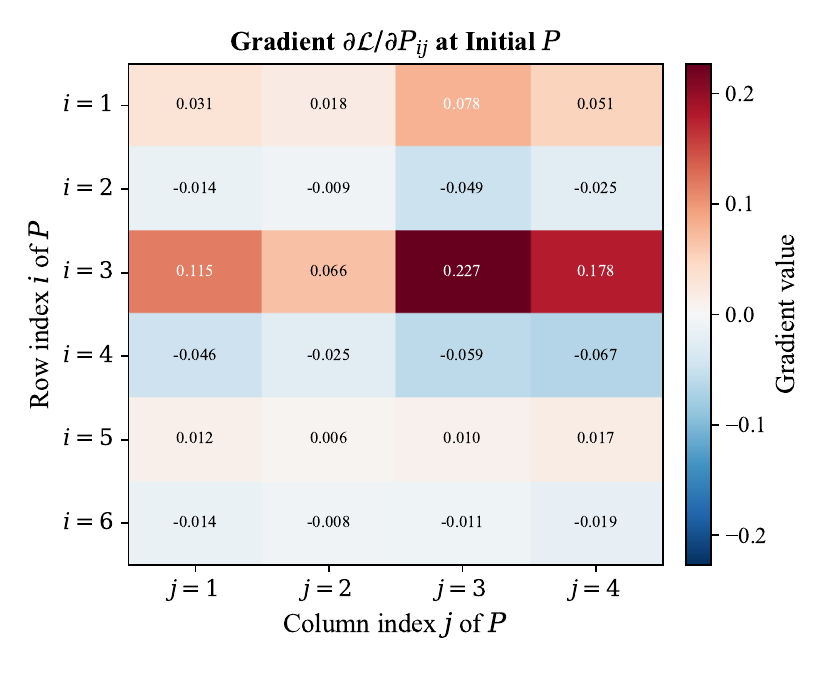}
    \caption{\footnotesize Gradient heatmap.}
\label{fig:nnls_signal_recovery_gradient_heatmap}
  \end{subfigure}
  %\hfill
  \begin{subfigure}{0.26\textwidth}
    \centering
    \includegraphics[width=\linewidth]{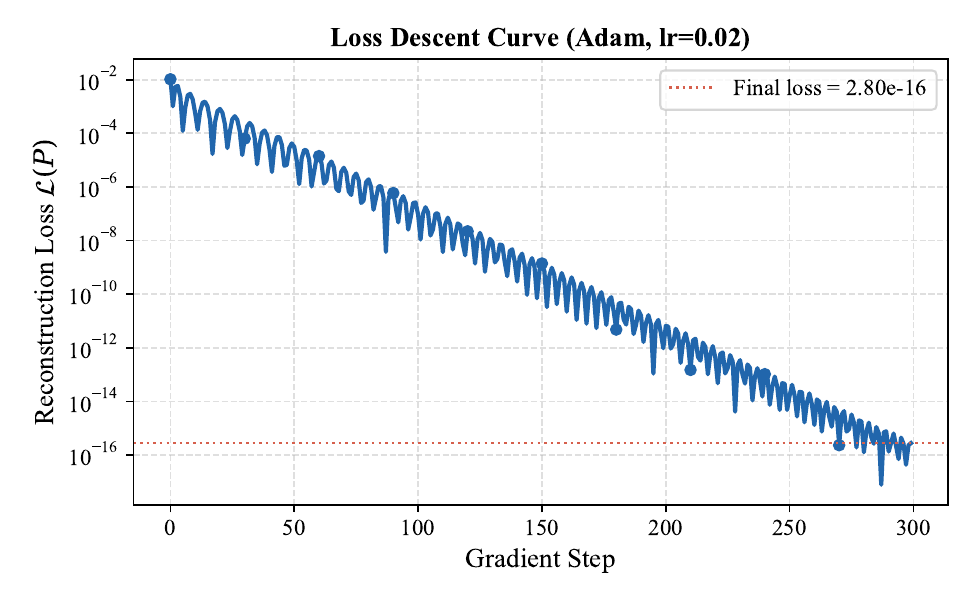}
    \caption{\footnotesize Loss descent curve.}
    \label{fig:nnls_signal_recovery_loss_descent_curve}
  \end{subfigure}
  \caption{Differentiable NNLS signal recovery via CVXPYLayers.
(a) Primal solution $\mathbf{x}^\star$ before and after learning $\mathbf P$, compared with $\mathbf{x}_{\mathrm{true}}$.
(b) KKT residuals of the recovered primal-dual pair under the convention $\boldsymbol{\lambda}^\star=\mathbf b-\mathbf P\mathbf x^\star$.
(c) Gradient heatmap of $\partial\mathcal{L}/\partial P_{ij}$ at the initial iterate.
(d) Reconstruction loss during end-to-end learning of $\mathbf P$ (Adam, 300 steps, lr$=0.02$), decreasing to numerical precision.
}
\label{fig:nnls_signal_recovery_cvxpylayers_numerical_results}
\end{figure*}

\textbf{(1) CVXPYLayers Implementation:}
The NNLS problem \eqref{prob:DP-CVXPYLayers-NNLS-primal} is first expressed as a DPP in CVXPY and then wrapped as a CVXPYLayer that exposes the primal solution $\mathbf{x}^\star$ as a differentiable output  (cf. Figure~\ref{fig:nnls_signal_recovery_cvxpylayers}). The dual solution is subsequently recovered via the KKT relation $\boldsymbol{\lambda}^\star = \mathbf{b} - \mathbf{P}\mathbf{x}^\star$. As illustrated in the code block below, the learnable matrix $\mathbf{P}$ is iteratively updated via gradients backpropagated through the optimization layer, minimizing the reconstruction loss $\mathcal{L}(\mathbf{P})$ in \eqref{eq:nnls-signal-recovery-loss}. The primal solutions, KKT residuals, gradient map, and loss trajectory are illustrated in Fig. ~\ref{fig:nnls_signal_recovery_cvxpylayers_numerical_results}.

\vspace{2mm}
\begin{lstlisting}[language=Python, basicstyle=\ttfamily\scriptsize]
import cvxpy as cp, torch
from cvxpylayers.torch import CvxpyLayer

P_param, b_param = cp.Parameter((m, n)), cp.Parameter(m)
x_var = cp.Variable(n)
problem = cp.Problem(cp.Minimize(0.5*cp.sum_squares(P_param@x_var - b_param)), [x_var >= 0])
layer = CvxpyLayer(problem, parameters=[P_param, b_param], variables=[x_var])
P = torch.nn.Parameter(torch.tensor(P_init, dtype=torch.float64))
x_star, = layer(P, b_torch)
lambda_star = b_torch - P @ x_star
loss = 0.5 * (x_star - x_true_t).pow(2).sum()
loss.backward()
\end{lstlisting}

\section{Learning to Optimize by Differentiable Programming: Case Studies}
\label{sec:case-studies}
\begin{figure*}[t]
    \centering
    \includegraphics[trim=0.75cm 0cm 0.75cm 0cm, clip, width=\linewidth]{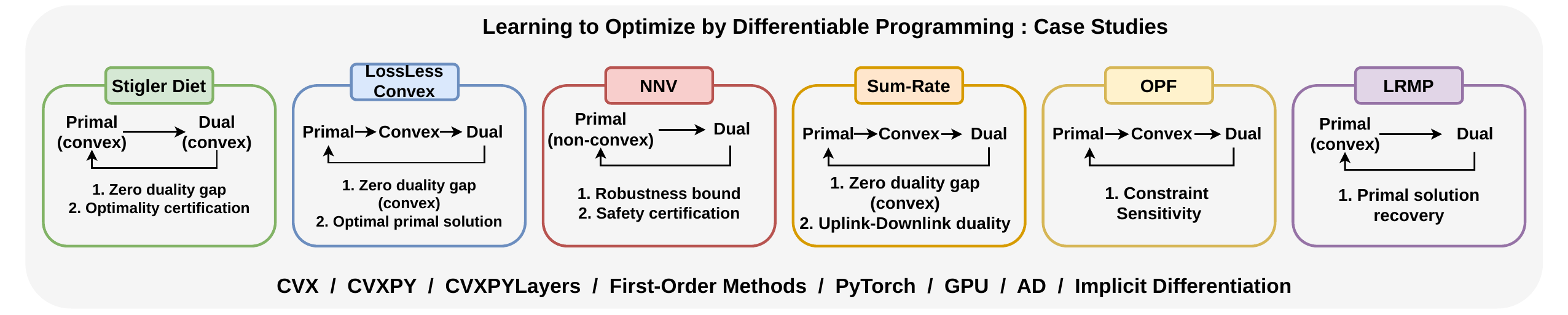}
\caption{Overview of the case studies in this section. Despite arising from diverse applications, all examples share a common primal--dual structure within differentiable programming frameworks, highlighting the broad applicability of differentiable optimization and its ability to unify optimization, duality theory, automatic differentiation, and first-order methods within a common computational framework.
}
\label{fig:diffprog_case_studies}
\end{figure*}

In this section, we present case studies showing how differentiable programming, combined with duality theory and first-order methods, provides a unified framework for solving LP, QP, SOCP, and ECP problems. Differentiable programming generates feasible primal solutions while simultaneously solving the dual problem to assess and refine their quality. The applications span aerospace and optimal control (lossless convexification of nonconvex optimal control problem \cite{accikmecse2013lossless, blackmore2012lossless, accikmecse2011lossless}), wireless networks (sum-rate maximization problem \cite{wei2015wireless, chiang2007power, tan2013fast, zheng2014maximizing}), power systems (optimal power flow~\cite{tan2014resistive, dommel2007optimal, lavaei2011zero}), machine learning verification (neural network verification~\cite{dvijotham2018training}), and graph-based learning (Laplacian-regularized minimization~\cite{tuck2019distributed}), as summarized in Table \ref{tab:method-summary-merged} and Figure~\ref{fig:diffprog_case_studies}, illustrating the broad versatility and impact of differentiable programming.

\subsection{Case Study: Stigler Diet Problem}
Stigler Diet problem ~\cite{garille2001stigler, joannopoulos2015diet} is a classical benchmark in operations research and linear programming. Formulated by George Stigler during World War II, it seeks the lowest-cost diet that satisfies all essential nutritional requirements. Due to the computational limits of the 1940s, Stigler obtained only an approximate solution, and the exact optimum was later computed by Jack Laderman using the then-new simplex method—a process requiring nine staff members and nearly 120 man-days. With modern computational advances, solvers such as Google OR-Tools~\cite{google_ortools_github} now solve this problem in seconds. Below, we review the Stigler Diet Problem and discuss emerging opportunities for applying differentiable programming to this canonical LP task.

\begin{problem}[Stigler Diet Problem]
Let $\mathbf{x} \in \mathbb{R}^n$ denote the quantities of $n$ food items, and let $\mathbf{c} \in \mathbb{R}^n$ be their unit costs. The nutrient matrix $A \in \mathbb{R}^{m \times n}$ contains entries $a_{ij}$ giving the amount of nutrient $j$ in food $i$, and $\mathbf{b} \in \mathbb{R}^m$ specifies the minimum daily requirements. The problem can be written in standard LP form:
\begin{equation}\label{eq:stigler-primal}\tag{Stigler-Diet-Primal}
\begin{aligned}
\min_{\mathbf{x}} \quad & \mathbf{c}^\top \mathbf{x} \\
\text{s.t.} \quad & A\mathbf{x} \ge \mathbf{b}, \quad \mathbf{x} \ge \mathbf{0}.
\end{aligned}
\end{equation}
The constraint $A\mathbf{x} \ge \mathbf{b}$ enforces nutritional adequacy, while $\mathbf{x} \ge 0$ ensures nonnegative food quantities. The objective $\mathbf{c}^\top \mathbf{x}$ is the total daily cost to be minimized.
\end{problem}

To form the Lagrangian, we introduce dual variables $\boldsymbol{\lambda} \ge 0$ for the constraint $\mathbf{b} - A\mathbf{x} \le 0$ and $\boldsymbol{\nu} \ge 0$ for $-\mathbf{x} \le 0$. The Lagrangian becomes:
\begin{equation}
\label{eq:stigler-lagrangian}
\mathcal{L}(\mathbf{x}, \boldsymbol{\lambda}, \boldsymbol{\nu})
= \mathbf{c}^\top \mathbf{x} 
+ \boldsymbol{\lambda}^\top(\mathbf{b} - A\mathbf{x})
- \boldsymbol{\nu}^\top \mathbf{x}.
\end{equation}

The dual function is obtained by taking the infimum of the Lagrangian over all feasible \(\mathbf{x}\). This yields \(g(\boldsymbol{\lambda}, \boldsymbol{\nu}) = \mathbf{b}^\top \boldsymbol{\lambda}\) when \(A^\top \boldsymbol{\lambda} + \boldsymbol{\nu} = \mathbf{c}\), and \(g = -\infty\) otherwise. The dual problem is therefore:
\begin{equation}\label{eq:stigler-dual}\tag{Stigler-Diet-Dual}
\begin{aligned}
\max_{\boldsymbol{\lambda}, \boldsymbol{\nu}} \quad & \mathbf{b}^\top \boldsymbol{\lambda} \\
\text{s.t.} \quad & A^\top \boldsymbol{\lambda} + \boldsymbol{\nu} = \mathbf{c}, \\
& \boldsymbol{\lambda} \ge \mathbf{0}, \quad \boldsymbol{\nu} \ge \mathbf{0}.
\end{aligned}
\end{equation}

Finally, the primal--dual pair can be equivalently expressed as the following saddle-point problem:
\begin{align}
\label{eq:stigler-saddle}
\min_{\mathbf{x} \in \mathbb{R}^n} \ 
\max_{\boldsymbol{\lambda} \succeq \mathbf{0},\, \boldsymbol{\nu} \succeq \mathbf{0}}
\mathcal{L}(\mathbf{x}, \boldsymbol{\lambda}, \boldsymbol{\nu})
\; = \;
\min_{\mathbf{x} \in \mathbb{R}^n}
\max_{\boldsymbol{\lambda} \succeq \mathbf{0},\, \boldsymbol{\nu} \succeq \mathbf{0}}
[
\mathbf{c}^\top \mathbf{x}
+ \boldsymbol{\lambda}^\top (\mathbf{b} - A\mathbf{x})
- \boldsymbol{\nu}^\top \mathbf{x}
].
\end{align}

\begin{figure*}[!t]
  \centering
  \begin{subfigure}{0.24\textwidth}
    \centering
    \includegraphics[width=\linewidth]{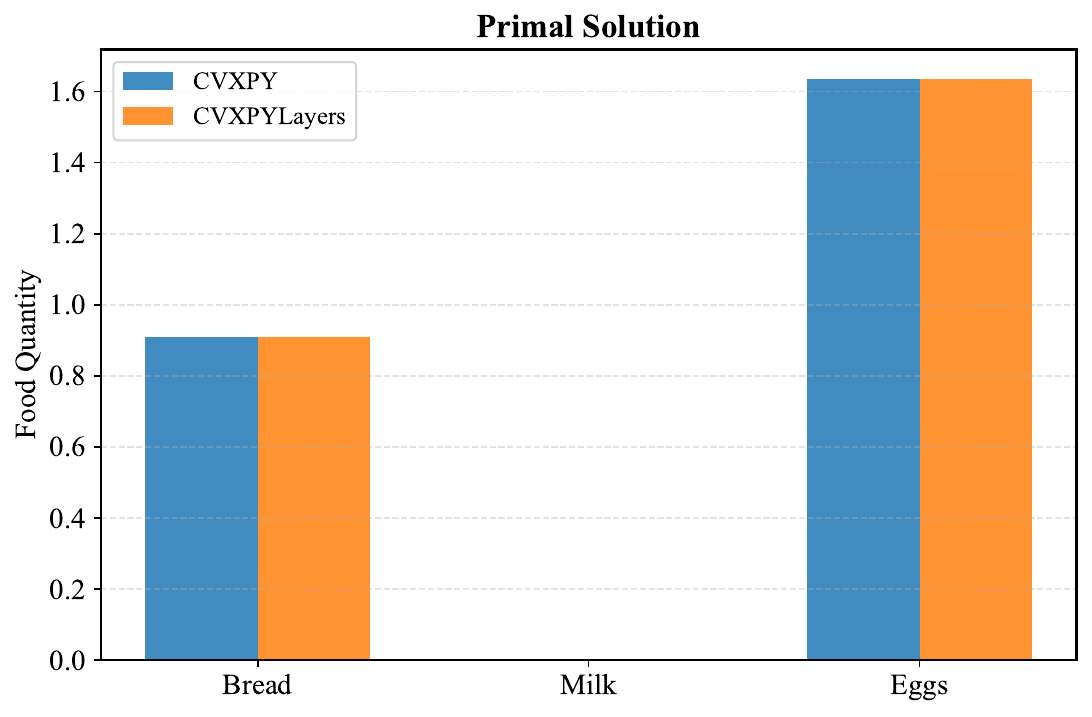}
    \caption{\footnotesize Primal solution.}
    \label{fig:diet_primal_solution}
  \end{subfigure}\hfill
  %\hfill
  \begin{subfigure}{0.24\textwidth}
    \centering
    \includegraphics[width=\linewidth]{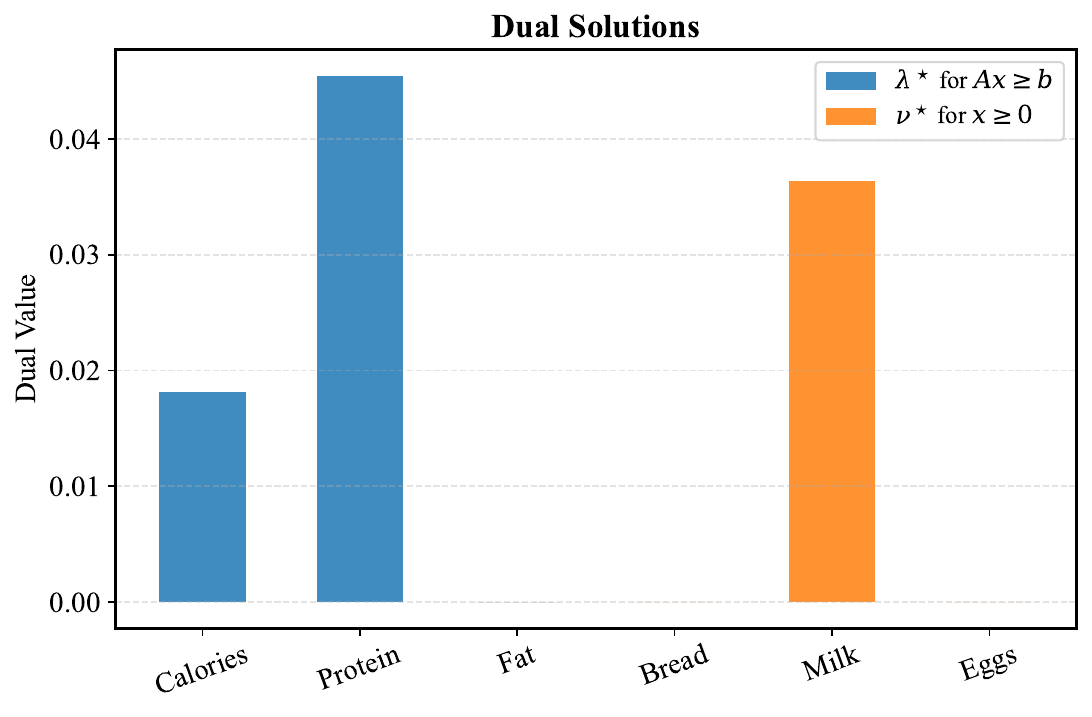}
    \caption{\footnotesize Dual solution.}
    \label{fig:diet_dual_solution}
  \end{subfigure}\hfill
  %\hfill
  \begin{subfigure}{0.24\textwidth}
    \centering
    \includegraphics[width=\linewidth]{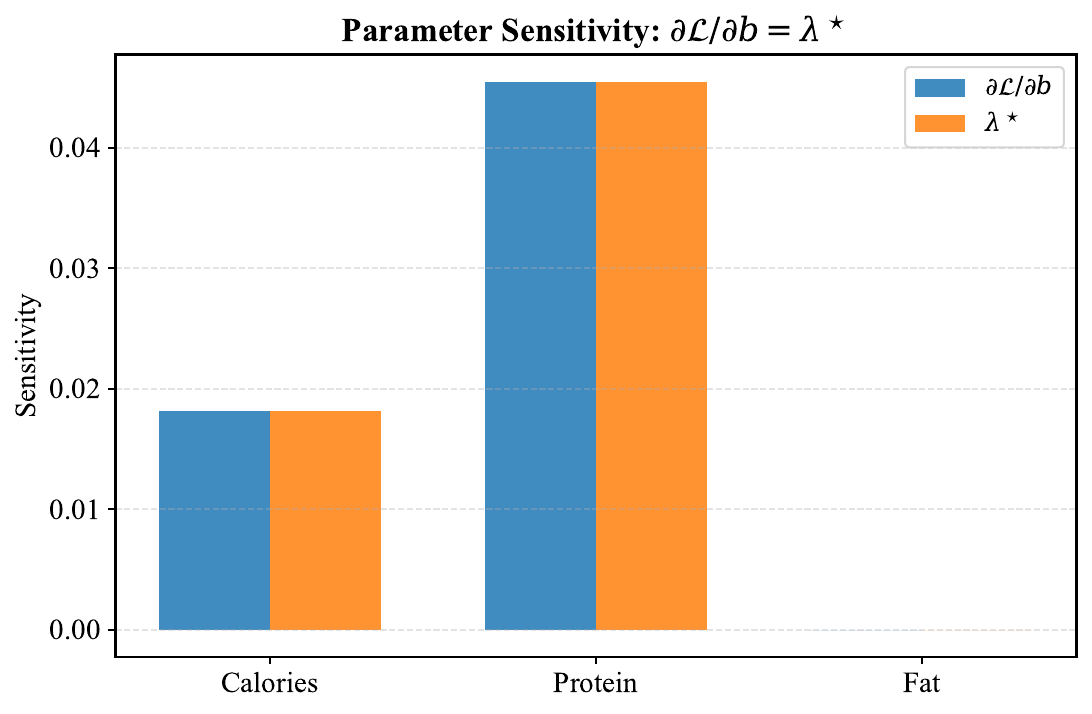}
    \caption{\footnotesize Parameter sensitivity.}
    \label{fig:diet_sensitivity}
  \end{subfigure}\hfill
  %\hfill
  \begin{subfigure}{0.24\textwidth}
    \centering
    \includegraphics[width=\linewidth]{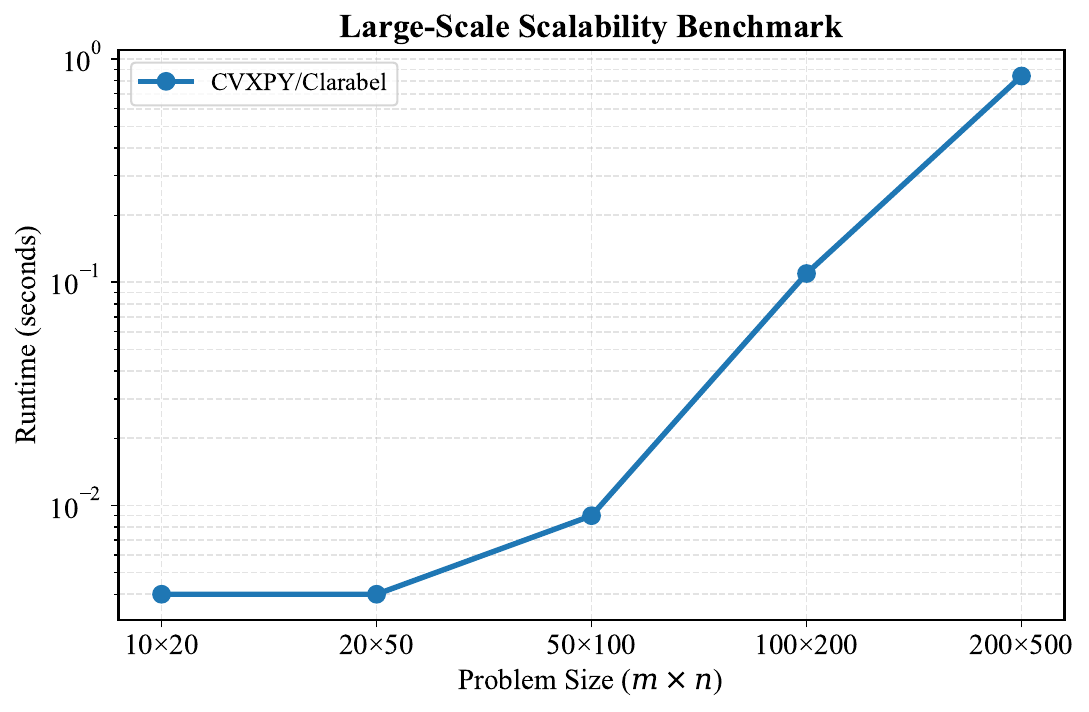}
    \caption{\footnotesize Large scale scalability.}
    \label{fig:diet_scalability}
  \end{subfigure}
  \caption{
Numerical results for the differentiable Stigler Diet problem using CVXPYLayers.
(a) Primal solutions.
(b) Dual solutions.
(c) Gradient verification via $\partial\mathcal{L}/\partial\mathbf{b}$.
(d) Runtime versus problem size.
The close agreement between CVXPY and CVXPYLayers validates the correctness of the optimization layer, while the consistency between gradients and dual variables confirms its differentiability.
}
  \label{fig:case_study_diet_numerical_results}
\end{figure*}

The saddle point characterizes the primal and dual optima and guarantees strong duality under standard regularity conditions. This formulation can be solved numerically using PDHG or ADMM, both of which fit naturally into the PyTorch framework for end-to-end differentiable optimization. These methods iteratively update primal and dual variables and can be implemented as differentiable operators, enabling gradients to flow through parameters such as costs and constraint matrices. This allows the Stigler Diet problem to be solved efficiently while integrating seamlessly into differentiable programming, unifying data-driven models with mathematical optimization. Alternatively, CVXPYLayers enables direct modeling of \eqref{eq:stigler-primal} as a differentiable convex layer. The corresponding numerical results are reported in Fig.~\ref{fig:case_study_diet_numerical_results}.

\subsection{Case Study: Lossless Convexification of Nonconvex Optimal Control}
Lossless convexification is a classical optimization-reformulation technique that transforms a class of Nonconvex Optimal Control Problems (NOCPs) into equivalent convex conic programs while preserving global optimality \cite{accikmecse2013lossless, blackmore2012lossless, accikmecse2011lossless}. A representative application arises in fuel-optimal powered-descent guidance, where thrust lower-bound constraints create a nonconvex feasible region.  Unlike generic convex relaxations, lossless convexification yields an exact SOCP reformulation whose optimal solution is also optimal for the original nonconvex problem. This case study introduces a differentiable implementation of lossless convexification for a nonconvex optimal control problem.

\begin{problem}[Nonconvex Optimal Control Problem]
Consider a discretized fuel-optimal control problem over $N$ time steps. Let $\mathbf{u}_i\in\mathbb{R}^d$ denote the control input at step $i$, and let $\mathbf{z}$ collect all state variables. The original nonconvex optimal control problem is:
\begin{equation}
\begin{aligned}
\min_{\mathbf{z},\{\mathbf{u}_i\}}
\quad &
\sum_{i=1}^{N}
\|\mathbf{u}_i\|_2
\\
\text{s.t.}
\quad &
G\mathbf{z}
=
\mathbf{h},
\\
&
\rho_1
\le
\|\mathbf{u}_i\|_2
\le
\rho_2,
\quad
i=1,\ldots,N,
\\
&
\mathbf{n}^{\top}\mathbf{u}_i
\ge
\|\mathbf{u}_i\|_2
\cos\theta,
\quad
i=1,\ldots,N,
\end{aligned}
\tag{NOCP-Primal}
\label{eq:lcvx_primal}
\end{equation}
where $G\mathbf{z}=\mathbf{h}$ collects the discretized dynamics and boundary conditions, $\rho_1>0$ and $\rho_2$ are the lower and upper thrust bounds, respectively, $\mathbf n$ is a prescribed pointing direction, and $\theta$ denotes the maximum allowable pointing angle.
\end{problem}

For the Problem~\eqref{eq:lcvx_primal}, the lower-bound constraint $\|\mathbf{u}_i\|_2 \ge \rho_1$ removes the interior of a ball and produces a hollow feasible region $\mathcal U =\left\{\mathbf u: \rho_1 \le \|\mathbf u\|_2 \le \rho_2 \right\}$, which is nonconvex. Consequently, Problem~\eqref{eq:lcvx_primal} cannot be solved directly using standard convex optimization techniques. In order to address the above challenge, we use the lossless convexification method in the work \cite{accikmecse2013lossless}. Specifically, we first introduce auxiliary variables $\Gamma_i\in\mathbb R_+$ and replace the nonconvex constraints with $\|\mathbf u_i\|_2 \le \Gamma_i$, $\rho_1\le\Gamma_i\le\rho_2$, and $\mathbf n^\top\mathbf u_i\ge\Gamma_i\cos\theta$ for $i=1,\ldots,N$. Then the relaxed problem of \eqref{eq:lcvx_primal} becomes:
\begin{equation}
\begin{aligned}
\min_{\mathbf z,\{\mathbf u_i,\Gamma_i\}}
\quad &
\sum_{i=1}^{N}
\Gamma_i
\\
\text{s.t.}
\quad &
G\mathbf z
=
\mathbf h,
\quad
\|\mathbf u_i\|_2
\le
\Gamma_i,
\quad
\rho_1
\le
\Gamma_i
\le
\rho_2,
\\
&
\mathbf n^\top\mathbf u_i
\ge
\Gamma_i\cos\theta,
\qquad
i=1,\ldots,N.
\end{aligned}
\tag{NOCP-Relaxed}
\label{eq:lcvx_relaxed}
\end{equation}
Here, $\|\mathbf u_i\|_2 \le \Gamma_i$ is a SOC constraint, where $(\Gamma_i,\mathbf u_i) \in \mathcal Q^{d+1}$ and $\mathcal Q^{d+1} = \left\{(\gamma,\mathbf v):\|\mathbf v\|_2\le\gamma\right\}$. Since the remaining constraints are linear, Problem~\eqref{eq:lcvx_relaxed} is an SOCP.

A fundamental result of lossless convexification states that, under suitable regularity conditions, every optimal solution of the relaxed problem satisfies $\Gamma_i^\star =\|\mathbf u_i^\star\|_2$ for $i=1,\ldots,N$ (See \cite{accikmecse2013lossless, blackmore2012lossless, accikmecse2011lossless} for more details). Consequently, the SOC constraint is active at optimality, and the relaxed SOCP shares the same globally optimal solution as the original nonconvex problem~\eqref{eq:lcvx_primal}. Hence, the convexification is lossless. Collecting all primal variables into a vector $\mathbf x$, Problem~\eqref{eq:lcvx_relaxed} can be written in the standard conic form:
\begin{equation}
\begin{aligned}
\min_{\mathbf x}
\quad &
\mathbf c^\top\mathbf x
\\
\text{s.t.}
\quad &
A\mathbf x
=
\mathbf b,
\quad
\mathbf x
\in
\mathcal K,
\end{aligned}
\tag{NOCP-SOCP}
\label{eq:lcvx_socp}
\end{equation}
where $\mathcal K$ denotes the Cartesian product of second-order cones and linear cones.

Introduce the dual variable $\mathbf y$ associated with the equality constraint $A\mathbf x=\mathbf b$ and the conic dual variable $\mathbf s\in\mathcal K^\star$ associated with $\mathbf x\in\mathcal K$. The Lagrangian is:
\begin{equation}
\mathcal L(\mathbf x,\mathbf y,\mathbf s)
=
\mathbf c^\top\mathbf x
+
\mathbf y^\top
(\mathbf b-A\mathbf x)
-
\mathbf s^\top\mathbf x,
\qquad
\mathbf s\in\mathcal K^\star.
\label{eq:lcvx_lagrangian}
\end{equation}
Therefore, the dual problem becomes:
\begin{equation}
\begin{aligned}
\max_{\mathbf y,\mathbf s}
\quad &
\mathbf b^\top\mathbf y
\\
\text{s.t.}
\quad &
A^\top\mathbf y
+
\mathbf s
=
\mathbf c,
\quad
\mathbf s
\in
\mathcal K^\star.
\end{aligned}
\tag{NOCP-Dual}
\label{eq:lcvx_dual}
\end{equation}
Since second-order cones are self-dual, i.e., $\mathcal K^\star = \mathcal K$, the primal and dual problems share the same conic structure. Under Slater's condition, strong duality holds and the duality gap vanishes.

\begin{figure*}[t]
    \centering
    \subfloat[\footnotesize Lossless SOC tightness.]{
        \includegraphics[width=0.235\textwidth]
        {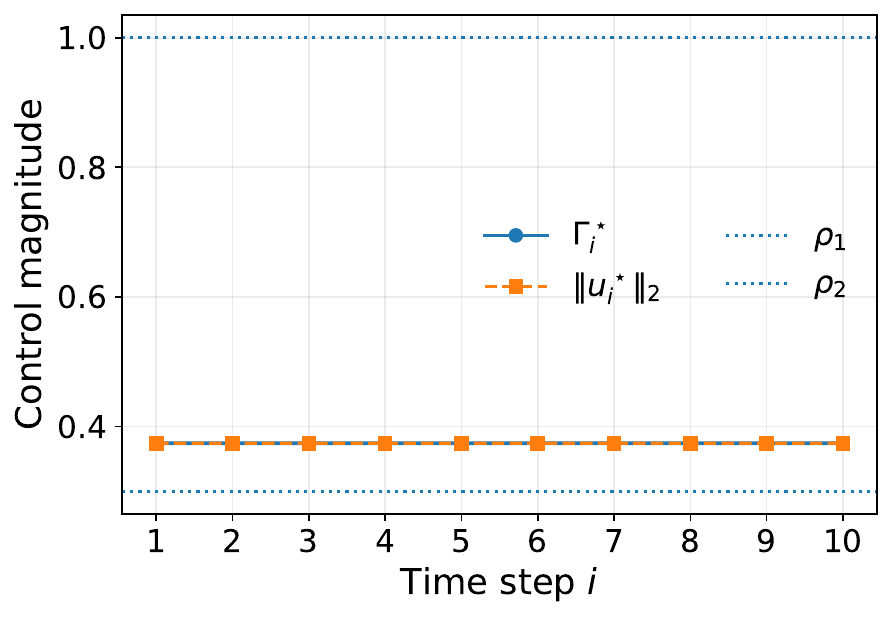}
        \label{fig:lcvx_lossless}
    }%\hfill
    \subfloat[\footnotesize Optimal state evolution.]{
        \includegraphics[width=0.235\textwidth]
    {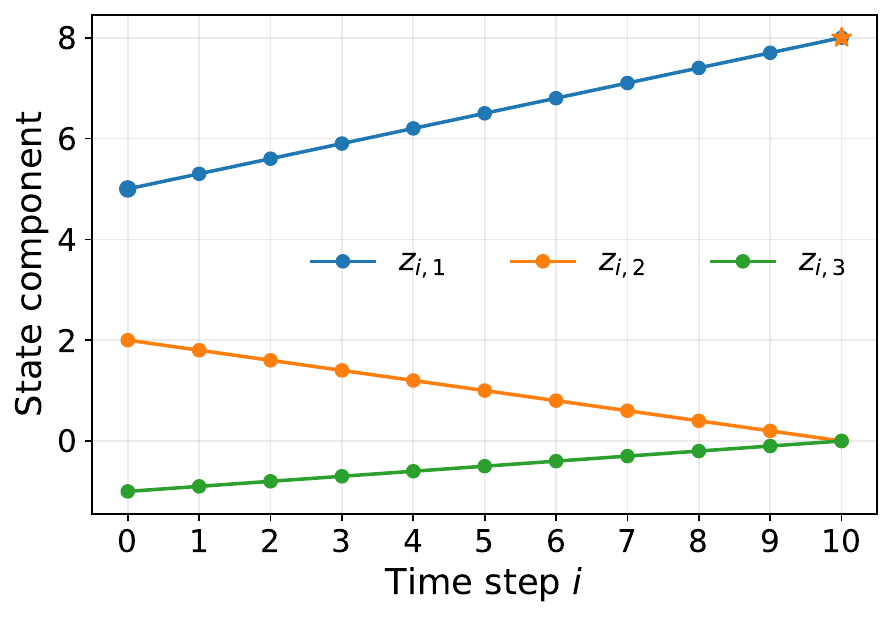}
        \label{fig:lcvx_state}
    }%\hfill
    \subfloat[\footnotesize KKT residuals.]{
        \includegraphics[width=0.245\textwidth]
        {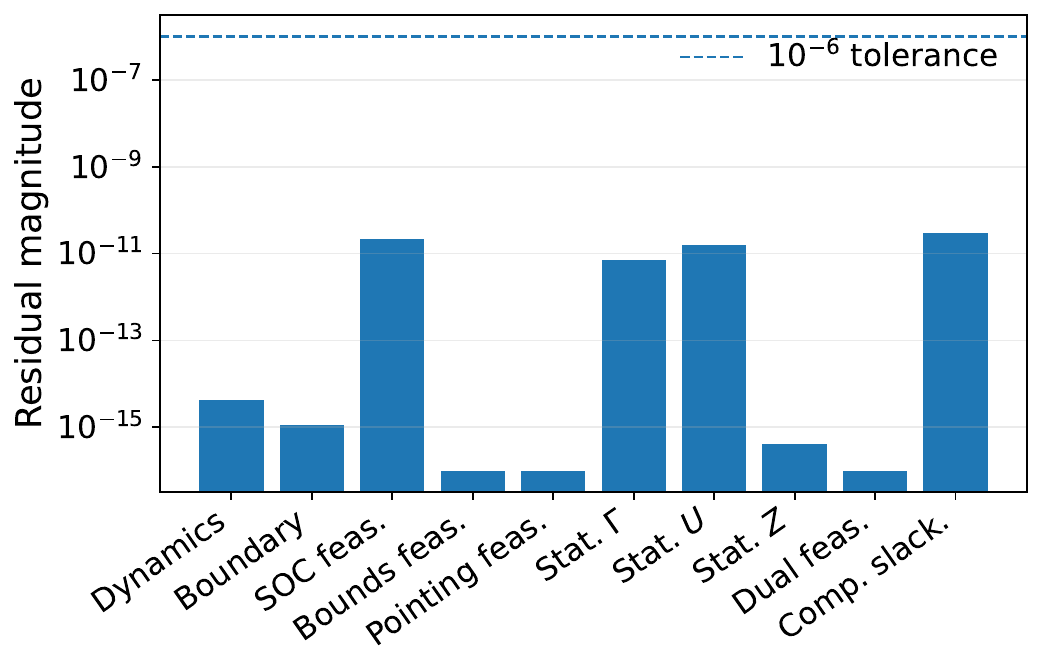}
        \label{fig:lcvx_kkt}
    }%\hfill
    \subfloat[\footnotesize Gradient validation.]{
        \includegraphics[width=0.23\textwidth]
        {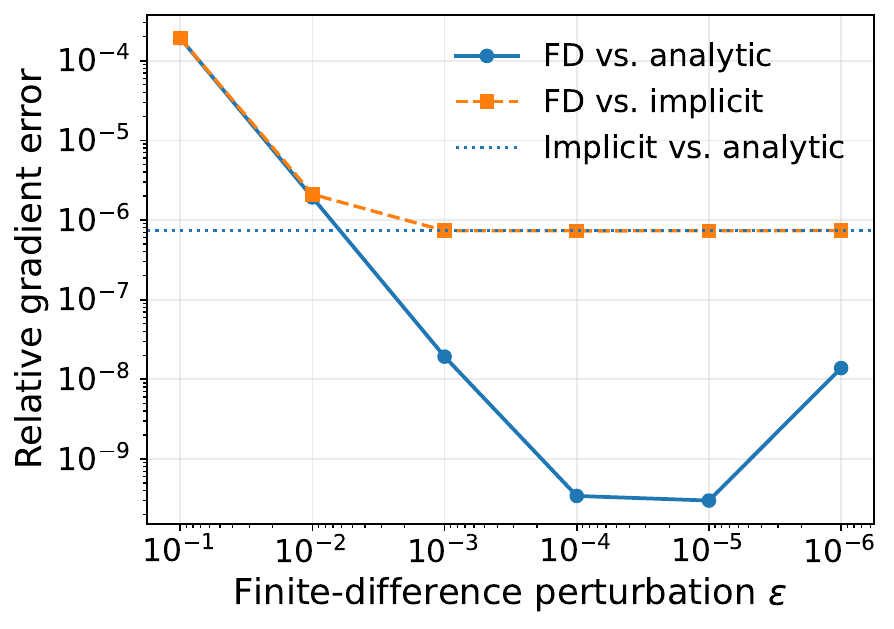}
        \label{fig:lcvx_gradient}
    }
    \caption{
Numerical validation of differentiable lossless convexification.
(a) $\Gamma_i^\star$ coincides with $\|\mathbf u_i^\star\|_2$, confirming tightness of the SOC relaxation.
(b) A representative optimal state trajectory satisfies the prescribed boundary conditions.
(c) The selected KKT residuals remain well below the numerical tolerance.
(d) Finite-difference and implicit gradients closely agree with the analytic reference.
}
    \label{fig:case_study_lossless_results}
\end{figure*}

Lossless convexification demonstrates how carefully designed reformulations can convert nonconvex optimal control problems into equivalent SOCPs that are both computationally tractable and differentiable. Fig.~\ref{fig:case_study_lossless_results} summarizes the numerical validation. Specifically, the results confirm the lossless property and KKT optimality conditions, while the agreement between CVXPY and CVXPYLayers solutions and the implicit-gradient validation demonstrate compatibility with end-to-end learning frameworks. Beyond optimal control, such formulations have broad relevance to modern aerospace applications, including spacecraft trajectory planning, satellite formation flying, autonomous rendezvous, and resource allocation in large-scale LEO satellite constellations such as Starlink. These results further illustrate the potential of integrating conic optimization, differentiable programming, and learning-based decision-making into a unified computational framework.

\subsection{Case Study: Neural Network Verification}
Deep neural networks have achieved remarkable success in areas such as image recognition and natural language processing but remain vulnerable to adversarial attacks, which are small and imperceptible perturbations that cause misclassifications \cite{papernot2016distillation}. This vulnerability underscores the importance of Neural Network Verification (NNV) problem \cite{albarghouthi2021introduction, zhang2018efficient, wang2021beta}, which aims to formally prove that models satisfy specific mathematical properties ensuring reliability, safety, and robustness. However, existing verification methods face scalability and computational challenges with limited formal guarantees. Differentiable programming offers a promising solution by framing verification as an optimization problem that unites learning with formal optimization principles~\cite{dvijotham2018training}. Through the use of optimization duality, automatic differentiation, and predictor–verifier training, differentiable programming enables the development of neural networks that are provably robust, computationally efficient, and equipped with formal certificates of correctness.

\begin{problem}[Neural Network Verification Problem]
Feed-forward neural networks trained with losses such as cross-entropy or squared error operate on input--output pairs \((x_{\text{nom}}, y_{\text{true}})\). The network is modeled as a sequence of layer transformations \(h_k\), producing activations \(x_k = h_k(x_{k-1})\) for \(k = 1,\dots,K\), mapping the input \(x_0\) to the final output \(x_K\). Verification seeks to ensure that, for all inputs within a prescribed neighborhood \(S_{\text{in}}(x_{\text{nom}})\), the output satisfies a linear property $c^\top x_K + d \le 0$, where \(c\) and \(d\) depend on \(x_{\text{nom}}\) and the true label \(y_{\text{true}}\). This formulation captures tasks such as adversarial robustness and monotonicity certification. To search for counterexamples, the verification task is cast as the optimization problem:
\begin{equation}\label{prob:NNV-primal}\tag{NNV-Primal}
\begin{aligned}
\max_{x} \quad & c^\top x_K + d \\
\text{s.t.}\quad & x_0 \in S_{\text{in}}(x_{\text{nom}}), \\
& x_{k+1} = h_k(x_k), \quad k = 0,\dots,K-1.
\end{aligned}
\end{equation}
If the optimal value of \eqref{prob:NNV-primal} is strictly negative, the network is guaranteed to satisfy the specification under all admissible input perturbations.
\end{problem}

Introducing dual variables $\lambda_k$ associated with the equality constraints $x_{k+1} = h_k(x_k)$, the Lagrangian is defined as:
\begin{align}
L(x, \lambda) &= c^\top x_K + d + \sum_{k=0}^{K-1} \lambda_k^\top \big(x_{k+1} - h_k(x_k)\big) \notag \\
&= d
+ \big( c^\top x_K + \lambda_{K-1}^\top x_K \big)
+ \sum_{k=1}^{K-1} \big( \lambda_{k-1}^\top x_k - \lambda_k^\top h_k(x_k) \big)
- \lambda_0^\top h_0(x_0).
\end{align}

The dual function is $g(\lambda) = \sup_x L(x,\lambda)$, where the feasible region reflects the admissible input set and propagated activation bounds. The resulting dual problem is:
\begin{equation}\label{prob:NNV-dual}\tag{NNV-Dual}
\begin{aligned}
\min_{\lambda_0,\dots,\lambda_{K-1}} \quad & g(\lambda)=
d + \psi_0(\lambda_0) + \sum_{k=1}^{K-1} \phi_k(\lambda_{k-1}, \lambda_k) \\
\text{s.t.} \quad & \lambda_{K-1} = -c,
\end{aligned}
\end{equation}
where $\phi_k(\lambda_{k-1}, \lambda_k) = \sup_{x_k} [\lambda_{k-1}^\top x_k - \lambda_k^\top h_k(x_k) ]$ and $\psi_0(\lambda_0) = \sup_{x_0 \in S_{\text{in}}} [ -\lambda_0^\top h_0(x_0) ]$. By solving \eqref{prob:NNV-dual}, if $g(\lambda) < 0$, then $c^\top \phi(x_0) + d \le 0$ holds for all $x_0 \in S_{\text{in}}(x_{\text{nom}})$, thereby certifying that the network satisfies the specification.

\vspace{2mm}
\textbf{SOS-Based Polynomial and SDP Relaxations for NNV Problem.}
Beyond first-order primal–dual approaches, neural network verification can also be framed algebraically by encoding all network constraints within a global optimization model. This viewpoint enables convex relaxation techniques that approximate the network’s nonlinear and nonconvex behavior with tractable surrogates. A prominent example is the family of Sum-of-Squares (SOS) relaxations~\cite{lasserre2009moments, zheng2021chordal}, which reformulate verification as a polynomial program and relax it into a hierarchy of semidefinite programs. Each relaxation level provides a certified lower bound that converges under standard assumptions: positive bounds confirm the property, while nonpositive ones may expose violations. Although theoretically powerful, SOS methods grow rapidly in complexity and are typically applied with care to preserve scalability.

A more scalable alternative is the semidefinite relaxation introduced by Raghunathan et al.~\cite{raghunathan2018semidefinite}. This method employs a rank-one lifting of network variables, expressing quadratic interactions in a higher-dimensional space and relaxing only the nonconvex rank constraint. The resulting structured SDP preserves linear relations dictated by the architecture and input set, producing relaxations that are significantly tighter than linear bounds yet far more tractable than higher-order SOS hierarchies~\cite{miller2022decomposed, chiu2025sdp, brown2022unified, luo2010semidefinite}. A positive SDP certificate verifies the specification, offering a practical balance between accuracy and efficiency. Together, SOS hierarchies and rank-one lifting SDPs complement first-order optimization methods, expanding the toolkit for rigorous and scalable neural network verification.

\begin{table*}[t]
\centering
\caption{Representative Case Studies Integrating Differentiable Programming and First-Order Methods.}
\label{tab:method-summary-merged}
\renewcommand{\arraystretch}{1.6}
\scriptsize
\newcolumntype{Y}{>{\RaggedRight\arraybackslash}X}
\newcolumntype{L}[1]{>{\RaggedRight\arraybackslash}p{#1}}

\begin{tabularx}{\textwidth}{L{0.12\textwidth} L{0.04\textwidth} L{0.10\textwidth} Y L{0.16\textwidth}}
\toprule
\textbf{Case Study} 
& \textbf{Type} 
& \textbf{Method} 
& \textbf{Mechanism \& Benefit} 
& \textbf{Implementation} \\
\midrule

Stigler Diet \cite{garille2001stigler, joannopoulos2015diet}
& LP 
& ADMM, \newline PDHG
& \textbf{Mechanism:} Proximal map splitting. \newline
  \textbf{Benefit:} Differentiable for end-to-end learning.
& D-LADMM \cite{xie2019differentiable} \newline
  CVXPYLayers \cite{agrawal2019differentiable, agrawal2020differentiating}\\
\cmidrule(l){1-5}

Lossless Convexification \cite{accikmecse2013lossless, blackmore2012lossless, accikmecse2011lossless}
& SOCP
& PDHG, \newline ADMM
& \textbf{Mechanism:} Tight relaxation of SOCP reformulation. \newline
  \textbf{Benefit:} Global optimal solution to nonconvex problem.
& CVXPYLayers \cite{agrawal2019differentiable, agrawal2020differentiating}\\
\cmidrule(l){1-5}

NNV \cite{dvijotham2018training}
& LP
& PDG, \newline PGD 
& \textbf{Mechanism:} Convex relaxation of ReLUs. \newline
  \textbf{Benefit:} Enhance neural network training stability and robustness.
& Auto\_LiRPA \cite{xu2020fast}; \newline
  $\beta$-CROWN \cite{wang2021beta} \\
\cmidrule(l){1-5}

Sum-Rate \cite{zheng2014maximizing, tan2013fast}
& ECP
& PDG, \newline PDHG
& \textbf{Mechanism:} Perron--Frobenius spectral-radius convexification
  via exponential-cone reformulation. \newline
  \textbf{Benefit:} Certification of global optimality.
& CVXPYLayers \cite{agrawal2019differentiable, agrawal2020differentiating} \\
\cmidrule(l){1-5}

OPF \cite{tan2014resistive, lavaei2011zero}
& QP
& PDG 
& \textbf{Mechanism:} SOCP reformulation and relaxation. \newline
  \textbf{Benefit:} Global optimal solution to nonconvex problem.
& DeepOPF-U \cite{liang2023deepopf} \newline
  CVXPYLayers \cite{agrawal2019differentiable, agrawal2020differentiating} \\
\cmidrule(l){1-5}

LRMP \cite{wong2016quantifying}
& QP 
& ADMM 
& \textbf{Mechanism:} Preconditioner via duality and regularization. \newline
  \textbf{Benefit:} Differentiable for adaptive regularization.
& CVXPYLayers \cite{agrawal2019differentiable, agrawal2020differentiating}; \newline
  AA\_mDLAM \cite{ebrahimi2025aa} \\
\bottomrule

\end{tabularx}
\end{table*}

\subsection{Case Study: Sum-Rate Maximization in Wireless Networks}
Wireless network optimization \cite{wei2015wireless} seeks to allocate transmit powers efficiently in the presence of mutual interference among coexisting communication links. Among various formulations, weighted sum-rate maximization is one of the most fundamental problems, as it captures the tradeoff between spectral efficiency and interference management \cite{chiang2007power, tan2013fast, zheng2014maximizing, tan2016optimal}. Due to the coupled interference structure in the Signal-to-Interference-plus-Noise Ratio (SINR) expressions, the resulting optimization problem is generally nonconvex and NP-hard \cite{luo2008dynamic, tan2011nonnegative}. In this case study, we consider a spectral-radius convexification of the weighted sum-rate maximization problem based on nonnegative matrix theory and the Perron--Frobenius theorem \cite{krause2001concave, zheng2016wireless, tan2011spectrum}. This formulation provides a useful example of how classical wireless optimization can be transformed into a convex problem that is suitable for differentiable conic optimization.

\begin{problem}[Sum-Rate Maximization Problem]
Consider a wireless network consisting of $L$ transmitter--receiver pairs sharing a common spectrum. Let $\mathbf p=(p_1,\ldots,p_L)^T\in\mathbb R_+^L$ denote the transmit power vector. The SINR of link $l$ is given by \cite{zheng2018max}:
\begin{equation}
\mathrm{SINR}_l(\mathbf p)
=
\frac{G_{ll}p_l}
{\sum_{j\neq l}G_{lj}p_j+n_l},
\end{equation}
where $G_{lj}$ denotes the channel gain from transmitter $j$ to receiver $l$, and $n_l>0$ is the receiver noise power.
Define the normalized interference matrix $\mathbf F$ and normalized noise vector $\mathbf v$ as:
\begin{equation}
F_{lj}
=
\begin{cases}
0, & l=j,\\
G_{lj}/G_{ll}, & l\neq j,
\end{cases}
\qquad
v_l
=
\frac{n_l}{G_{ll}}.
\end{equation}

Then $\mathrm{SINR}_l(\mathbf p) = \frac{p_l} {(\mathbf F\mathbf p+\mathbf v)_l}$. The weighted sum-rate maximization problem under weighted power constraints is:
\begin{equation}\tag{Sum-Rate-Primal}
\label{prob:pf_sumrate}
\begin{aligned}
\max_{\mathbf p\ge 0}
\quad
&
\sum_{l=1}^{L}
w_l
\log\!\left(
1+
\frac{p_l}
{(\mathbf F\mathbf p+\mathbf v)_l}
\right)
\\
\textnormal{s.t.}
\quad
&
\mathbf a_k^T\mathbf p
\le
\bar p_k,
\qquad
k=1,\ldots,K,
\end{aligned}
\end{equation}
where $\mathbf w\in\mathbb R_+^L$ is the link-priority weight vector, $\mathbf a_k\in\mathbb R_+^L$ is the power-weight vector, and $\bar p_k>0$ is the corresponding power budget.
\end{problem}

Introduce the achievable-rate vector $\mathbf r=(r_1,\ldots,r_L)^T$ and the interference-temperature vector $\mathbf q = \mathbf F\mathbf p+\mathbf v$. Since $r_l = \log(1+\mathrm{SINR}_l)$, it follows that $e^{r_l}-1 = \frac{p_l}{q_l}$ and $p_l = (e^{r_l}-1)q_l$. In vector form, we get $\mathbf p = \operatorname{diag}(e^{\mathbf r}-\mathbf 1)\mathbf q$ and $\mathbf p+\mathbf q = \operatorname{diag}(e^{\mathbf r})\mathbf q$. For each weighted power constraint, we define $\mathbf B_k = \mathbf F + \frac{1}{\bar p_k}\mathbf v\mathbf a_k^T$ for $k=1,\ldots,K$. Using $\mathbf q=\mathbf F\mathbf p+\mathbf v$ together with $\mathbf a_k^T\mathbf p\le \bar p_k$, one obtains $\mathbf B_k\mathbf p \le \mathbf q$. Substituting $\mathbf p = \operatorname{diag}(e^{\mathbf r})\mathbf q-\mathbf q$ gives $\mathbf B_k \operatorname{diag}(e^{\mathbf r}) \mathbf q \le (\mathbf I+\mathbf B_k)\mathbf q$ (see \cite{zheng2014maximizing} for the detailed derivation). Therefore, the original problem \eqref{prob:pf_sumrate} can be equivalently reformulated as:
\begin{equation}
\label{prob:rate_domain_nonconvex}
\begin{aligned}
\max_{\mathbf r,\mathbf q}
\quad
&
\mathbf w^T\mathbf r
\\
\textnormal{s.t.}
\quad
&
\mathbf B_k
\operatorname{diag}(e^{\mathbf r})
\mathbf q
\le
(\mathbf I+\mathbf B_k)\mathbf q,
\qquad
k=1,\ldots,K.
\end{aligned}
\end{equation}

Although the objective becomes linear in $\mathbf r$, the feasible set remains nonconvex because the constraints contain the coupled terms $e^{r_l}q_l$. To remove this coupling, define $\widetilde{\mathbf B}_k = (\mathbf I+\mathbf B_k)^{-1}\mathbf B_k$ for $k=1,\ldots,K$. Assume that each $\widetilde{\mathbf B}_k$ is an irreducible nonnegative matrix, i.e., $\widetilde{\mathbf B}_k \ge \mathbf 0$ for $k=1,\ldots,K$. Then the constraint in \eqref{prob:rate_domain_nonconvex} is equivalent to $\widetilde{\mathbf B}_k \operatorname{diag}(e^{\mathbf r})\mathbf q\le\mathbf q$. By the Perron--Frobenius Subinvariance Theorem \cite{seneta2006non, wei2015wireless}, the inequality $\widetilde{\mathbf B}_k \operatorname{diag}(e^{\mathbf r})\mathbf q\le\mathbf q$ holds if and only if $\rho\;(\widetilde{\mathbf B}_k\operatorname{diag}(e^{\mathbf r}))\le 1$, where $\rho(\cdot)$ denotes the spectral radius, i.e., the eigenvalue with the largest absolute value. Consequently, the original nonconvex weighted sum-rate maximization problem admits the convex reformulation:
\begin{equation}
\label{prob:spectral_radius_convex}
\begin{aligned}
\max_{\mathbf r}
\quad
&
\mathbf w^T\mathbf r
\\
\textnormal{s.t.}
\quad
&
\log
\rho
\!\bigl(
\widetilde{\mathbf B}_k
\operatorname{diag}(e^{\mathbf r})
\bigr)
\le
0,
\qquad
k=1,\ldots,K.
\end{aligned}
\end{equation}

The convexity of \eqref{prob:spectral_radius_convex} follows from the log-convexity property of the Perron--Frobenius eigenvalue \cite{boyd2004convex} for irreducible nonnegative matrices \cite{berman1994nonnegative}. Therefore, under the quasi-inverse nonnegativity condition, the original NP-hard weighted sum-rate maximization problem can be solved through a convex optimization problem in the achievable-rate domain.

The Lagrange dual structure of \eqref{prob:spectral_radius_convex} also provides a useful interface for primal--dual verification in differentiable optimization solvers. Define $\phi_k(\mathbf r) = \log \rho (\widetilde{\mathbf B}_k \operatorname{diag}(e^{\mathbf r}))$. Then the spectral-radius constraints in \eqref{prob:spectral_radius_convex} can be written as $\phi_k(\mathbf r)\le 0$ for $k=1,\ldots,K$. Introducing Lagrange multipliers $\lambda_k\ge 0$ for the spectral-radius constraints, the Lagrangian is:
\begin{equation}
\mathcal L(\mathbf r,\boldsymbol\lambda)
=
\mathbf w^T\mathbf r
-
\sum_{k=1}^{K}
\lambda_k
\phi_k(\mathbf r).
\end{equation}
The corresponding Lagrange dual problem can be written in the abstract form:
\begin{equation}\tag{Sum-Rate-Dual}
\label{prob:pf_spectral_dual}
\begin{aligned}
\min_{\boldsymbol\lambda\ge 0}
\quad
&
g(\boldsymbol\lambda) = \sup_{\mathbf r}
\mathcal L(\mathbf r,\boldsymbol\lambda).
\end{aligned}
\end{equation}
The optimal dual variables $\lambda_k^\star$ quantify the sensitivity of the weighted sum-rate to the active Perron--Frobenius spectral constraints. Moreover, the resulting primal--dual formulation is amenable to first-order methods such as PDG and PDHG, and is closely related to the classical uplink--downlink duality, where the dual variables admit a virtual power-allocation interpretation and the zero duality gap guarantees an equivalent achievable SINR region in both domains.

\begin{figure*}[!t]
  \centering
  \begin{subfigure}{0.25\textwidth}
    \centering
    \includegraphics[width=\linewidth]{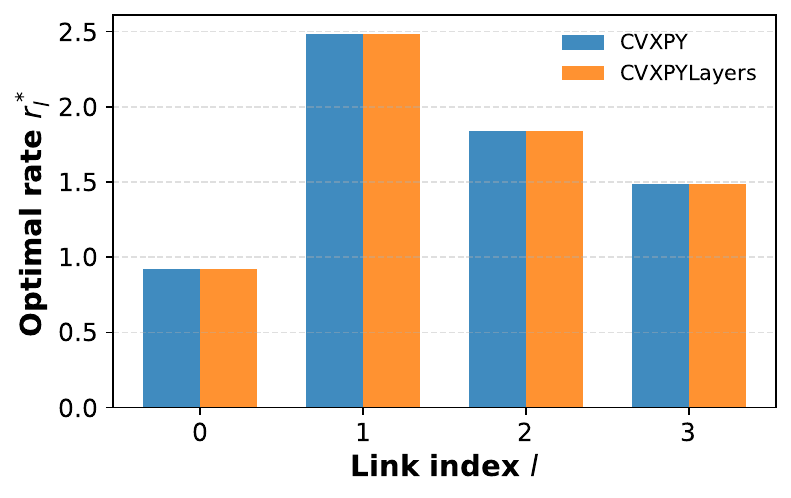}
    \caption{\footnotesize Optimal rate vector.}
    \label{fig:sumrate_rate_vector}
  \end{subfigure}\hfill
  %\hfill
  \begin{subfigure}{0.25\textwidth}
    \centering
    \includegraphics[width=\linewidth]{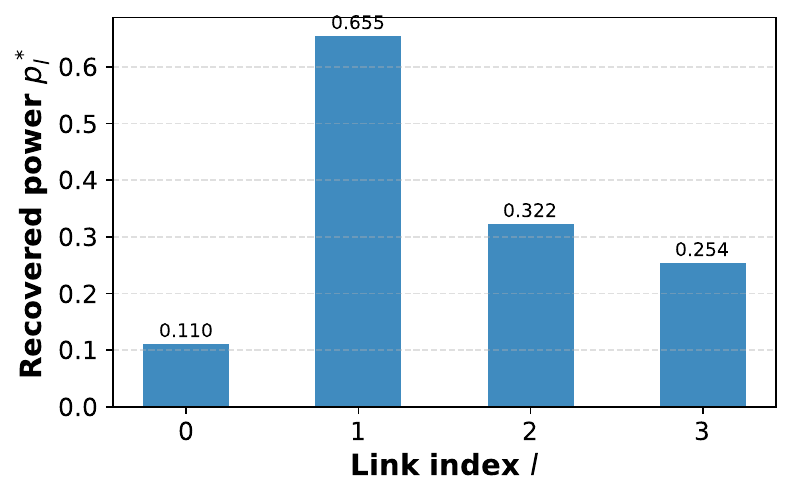}
    \caption{\footnotesize Recovered power vector.}
    \label{fig:sumrate_power_vector}
  \end{subfigure}\hfill
  %\hfill
  \begin{subfigure}{0.22\textwidth}
    \centering
    \includegraphics[width=\linewidth]{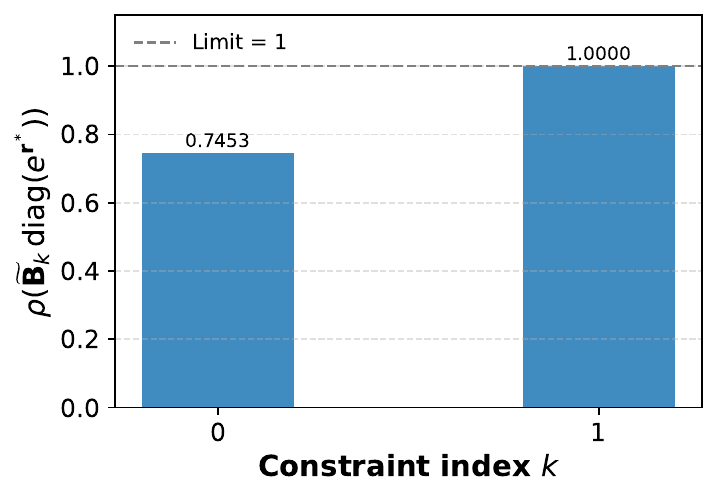}
    \caption{\footnotesize Spectral radius.}
    \label{fig:sumrate_spectral_radius}
  \end{subfigure}\hfill
  %\hfill
  \begin{subfigure}{0.25\textwidth}
    \centering
    \includegraphics[width=\linewidth]{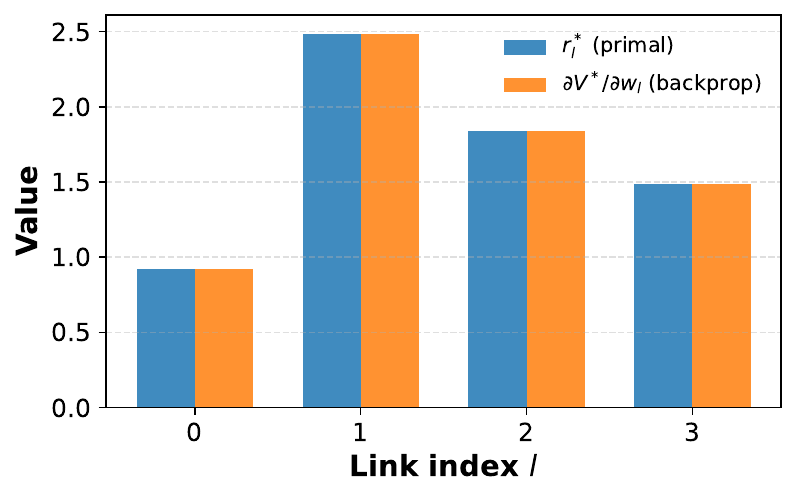}
    \caption{\footnotesize Gradient verification.}
    \label{fig:sumrate_gradient_backpropa}
  \end{subfigure}
\caption{Numerical results for the sum-rate maximization case study via CVXPYLayers.
(a) Optimal rate vector $\mathbf{r}^*$.
(b) Recovered transmit-power vector $\mathbf{p}^*$.
(c) Spectral-radius constraints $\rho(\widetilde{\mathbf{B}}_k\operatorname{diag}(e^{\mathbf{r}^*})) \le 1$: $k{=}1$ is active and $k{=}0$ is slack, consistent with $\boldsymbol{\mu}^*_{k=0}=\mathbf{0}$ and $\boldsymbol{\mu}^*_{k=1}\ne\mathbf{0}$.
(d) The backpropagated gradient $\partial V^*/\partial w_l$ matches $r^*_l$, confirming correct differentiation through the CVXPYLayers optimization layer.
}
  \label{fig:case_study_sumrate_numerical_results}
\end{figure*}

In particular, for differentiable programming implementations, the spectral-radius constraints can be further transformed using the Collatz--Wielandt theorem from nonnegative matrix theory \cite{berman1994nonnegative, friedland1975some, wei2015wireless}. Specifically, the Perron--Frobenius constraint $\rho(\widetilde{\mathbf B}_k \operatorname{diag}(e^{\mathbf r})) \le 1$ is equivalent to the existence of a positive vector $\mathbf u_k>0$ satisfying $\widetilde{\mathbf B}_k \operatorname{diag}(e^{\mathbf r})\mathbf u_k \le \mathbf u_k$. Introducing the logarithmic change of variables $\mathbf u_k=e^{\mathbf s_k}$, the $i$-th row can be written as:
\begin{align}
\sum_{j=1}^{L} (\widetilde B_k)_{ij}
\exp\!\left(r_j+s_{k,j}-s_{k,i}\right)\le 1,
\end{align}
which admit equivalent exponential-cone representations and can therefore be modeled directly in CVXPY and embedded as a differentiable optimization layer using CVXPYLayers. After obtaining the optimal achievable-rate vector $\mathbf r^\star$, the corresponding transmit-power vector can be recovered by solving $\mathbf q =(\mathbf I+\mathbf F - \mathbf F\operatorname{diag}(e^{\mathbf r^\star}))^{-1}\mathbf v$ and $\mathbf p^\star = \operatorname{diag}(e^{\mathbf r^\star}-\mathbf 1)\mathbf q$, which follows directly from the rate-domain transformation introduced earlier. The numerical results in Fig.~\ref{fig:case_study_sumrate_numerical_results} present the optimal rate vector, the recovered transmit-power vector, the Perron--Frobenius spectral-radius verification, and the gradient consistency obtained through CVXPYLayers. The close agreement between the CVXPY solution and the CVXPYLayers output confirms the correctness of the Perron--Frobenius convexification. Furthermore, the consistency between the backpropagated gradients $\partial V^*/\partial w_l$ and the optimal rate vector $\mathbf{r}^\star$ validates the differentiability of the optimization layer.

\subsection{Case Study: Optimal Power Flow Problem}
Optimal Power Flow (OPF) problem \cite{tan2014resistive, dommel2007optimal, lavaei2011zero} is a fundamental optimization problem in power systems and is widely regarded as one of the most important operational tasks. The goal of OPF is to determine how to adjust the controllable variables in a power network in order to minimize a chosen objective such as generation cost or transmission losses while strictly enforcing the governing physical laws given by the power flow equations and respecting all engineering limits including voltage magnitudes, generator capacities, and line flow constraints. The OPF formulation is intrinsically nonlinear and nonconvex, which makes it a challenging problem to solve. Since Carpentier introduced the original formulation in 1962, a large body of research has focused on efficient solution strategies. Numerous optimization methods have been explored, including nonlinear programming, quadratic programming, and Newton based algorithms, each contributing to the development of modern OPF solution techniques. In this section, we examine how differentiable programming can be applied to solve the OPF problem.

\begin{problem}[Optimal Power Flow Problem]
Consider a power network with bus set $\mathcal{N}$ and transmission line set $\mathcal{E}$. Let $v \in \mathbb{R}^N$ denote the vector of bus voltages, constrained by lower and upper bounds $\underline{V}$ and $\overline{V}$. The network is modeled by a weighted Laplacian conductance matrix $G \in \mathbb{R}^{N\times N}$, where each line $(i,j)\in\mathcal{E}$ has a positive conductance $g_{ij}>0$. For each bus $i$, we define the nodal power surrogate $p_i(v) = v_i (Gv)_i$. Let $d_i$ denote the demand at load bus $i \in \mathcal{L}$ and $\overline{p}_i$ the generation capacity limit at generator bus $i \in \mathcal{G}$. In addition, each line $(i,j) \in \mathcal{E}$ is subject to a thermal current limit $I^{\max}_{ij}$. The optimal power flow problem is then formulated as:
\begin{align}
\min_{v \in \mathbb{R}^N} \quad & 
    f(v) = v^\top G v \;=\; \sum_{(i,j)\in\mathcal{E}} g_{ij}\,(v_i - v_j)^2
    \label{eq:opf_obj} \tag{OPF-Primal}\\
\text{s.t.} \quad 
    & \underline{V} \;\le\; v \;\le\; \overline{V}, \qquad  \forall i \in \mathcal{N}, \label{eq:opf_voltage}\\
    & p_i(v) \;\ge\; d_i, \qquad  \forall i \in \mathcal{L}, \label{eq:opf_load}\\
    & p_i(v) \;\le\; \overline{p}_i, \qquad  \forall i \in \mathcal{G}, \label{eq:opf_gen}\\
    & \big|\,g_{ij}(v_i - v_j)\,\big| \;\le\; I^{\max}_{ij}, \qquad  \forall (i,j) \in \mathcal{E}. \label{eq:opf_line}
\end{align}
The objective \eqref{eq:opf_obj} minimizes the total resistive losses of the network. Constraint \eqref{eq:opf_voltage} enforces operational voltage bounds at all buses. Constraints \eqref{eq:opf_load}--\eqref{eq:opf_gen} ensure that the nodal power surrogate meets the load demands while respecting generator capacity limits. Finally, constraint \eqref{eq:opf_line} imposes thermal current limits on each transmission line, thereby ensuring safe operation of the network.
\end{problem}

Here, we temporarily disregard Constraint~\eqref{eq:opf_voltage} on voltage bounds in order to focus on the load balance, generation, and line-flow constraints. For each line $(i,j)\in\mathcal{E}$, define the line current surrogate $i_{ij}(v) = g_{ij}(v_i - v_j)$, which allows the thermal limit constraint to be written as $|i_{ij}(v)| \le I^{\max}_{ij}$. Introducing dual variables for these constraints yields the Lagrangian:
\begin{equation}
\begin{aligned}
L(v,\lambda,\gamma,\mu) 
&= v^\top G v 
+ \sum_{i\in\mathcal{L}} \lambda_i \big(d_i - p_i(v)\big)  
+ \sum_{i\in\mathcal{G}} \gamma_i \big(p_i(v) - \overline{p}_i\big) 
+ \sum_{(i,j)\in\mathcal{E}} \mu_{ij} \big(\,|i_{ij}(v)| - I^{\max}_{ij}\,\big),
\end{aligned}
\end{equation}
where $\lambda_i$, $\gamma_i$, and $\mu_{ij}$ denote the dual variables associated with the load, generation, and line-flow constraints, respectively. 

Accordingly, the associated saddle-point formulation is given by:
\begin{equation}
\label{prob:OPF-Saddle}
\min_{v \in \mathbb{R}^N} \; 
\max_{\substack{\lambda \geq 0, \; \gamma \geq 0, \; \mu \geq 0}} 
L(v,\lambda,\gamma,\mu).
\end{equation}

For analytical insight, it is convenient to first examine the dual problem obtained when line-flow limit \eqref{eq:opf_line} is ignored, that is, when $\mu \equiv 0$. In this simplified setting, the Lagrangian reduces to $L(v,\lambda,\gamma,0)$, and the corresponding dual problem takes the form:
\begin{equation} \label{prob:OPF-Dual}\tag{OPF-Dual}
\begin{aligned}
\max_{\lambda,\gamma} \quad & 
\sum_{i\in\mathcal{L}} \lambda_i d_i
- \sum_{i\in\mathcal{G}} \gamma_i \overline{p}_i \\
\text{s.t.} \quad &
G - \tfrac{1}{2}\mathrm{diag}(\lambda - \gamma) G
- \tfrac{1}{2} G\,\mathrm{diag}(\lambda - \gamma) \succeq 0, \\
& \lambda_i \ge 0, \quad i \in \mathcal{L}, \\
& \gamma_i \ge 0, \quad i \in \mathcal{G}.
\end{aligned}
\end{equation}
This semidefinite dual highlights the role of the nodal power constraints; in the full model with line-flow limits, we instead work directly with the saddle-point formulation~\eqref{prob:OPF-Saddle}.

We next employ a projected primal--dual gradient algorithm to solve the saddle-point formulation \eqref{prob:OPF-Saddle}. At each iteration, the primal variable $v$ is updated by taking a gradient descent step on the Lagrangian, where the gradient is automatically computed via PyTorch’s autograd. This step is followed by a projection onto the feasible voltage box to enforce operational limits:
\begin{equation}
v^{k+1} = \Pi_{[\underline V,\,\overline V]}\!(v^{k} - \tau \,\nabla_v L(v^{k},\lambda^{k},\gamma^{k},\mu^{k})),
\label{eq:primal_step}
\end{equation}
where $\tau>0$ is the primal step size and $\Pi_{[\underline V,\overline V]}(\cdot)$ denotes the projection operator implemented via project\_voltage. The dual variables are updated by projected gradient ascent:
\begin{align}
\lambda_i^{k+1} &= [\lambda_i^k + \eta \,\big(d_i - p_i(v^{k+1})\big)]_+,  \qquad i \in \mathcal{L}, \label{eq:dual_lambda}\\
\gamma_i^{k+1}  &= [\gamma_i^k + \eta \,\big(p_i(v^{k+1}) - \overline{p}_i\big)]_+, \qquad i \in \mathcal{G}, \label{eq:dual_gamma}\\
\mu_{ij}^{k+1}  &= [\mu_{ij}^k + \eta \,\big(|g_{ij}(v_i^{k+1}-v_j^{k+1})| - I^{\max}_{ij}\big)]_+, \qquad (i,j)\in\mathcal{E}, \label{eq:dual_mu}
\end{align}
where $\eta>0$ is the dual step size, $[\cdot]_+$ denotes elementwise projection onto the nonnegative orthant, and $p_i(v) = v_i(Gv)_i$ represents the nodal power surrogate.

\begin{figure*}[t]
    \centering
    \subfloat[\footnotesize Objective evolution.]{
        \includegraphics[width=0.32\textwidth]
        {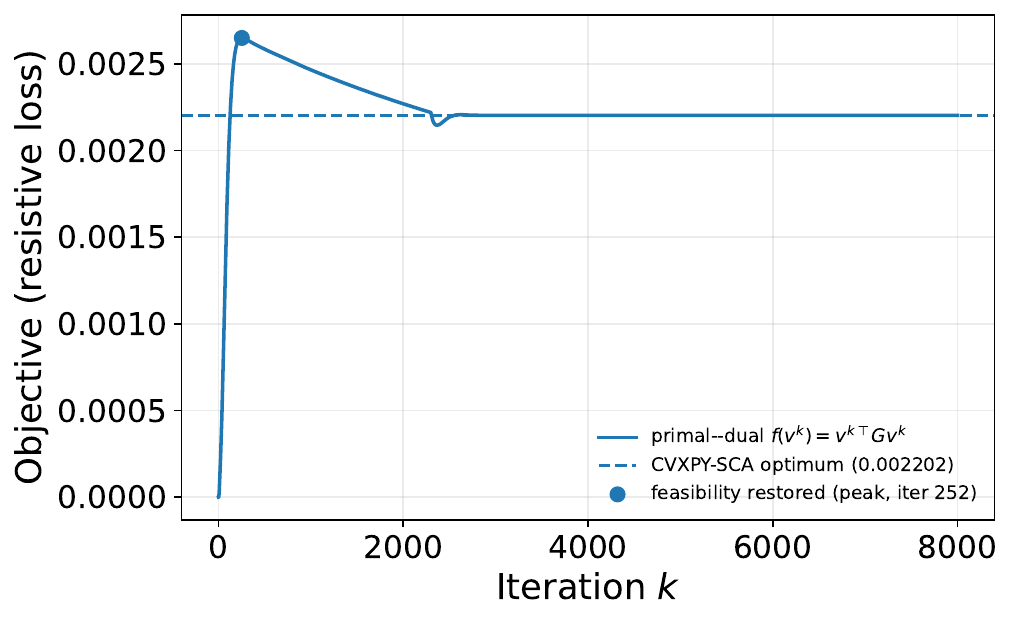}
        \label{fig:opf_objective}
    }%\hfill
    \subfloat[\footnotesize Constraint-violation convergence.]{
        \includegraphics[width=0.32\textwidth]
        {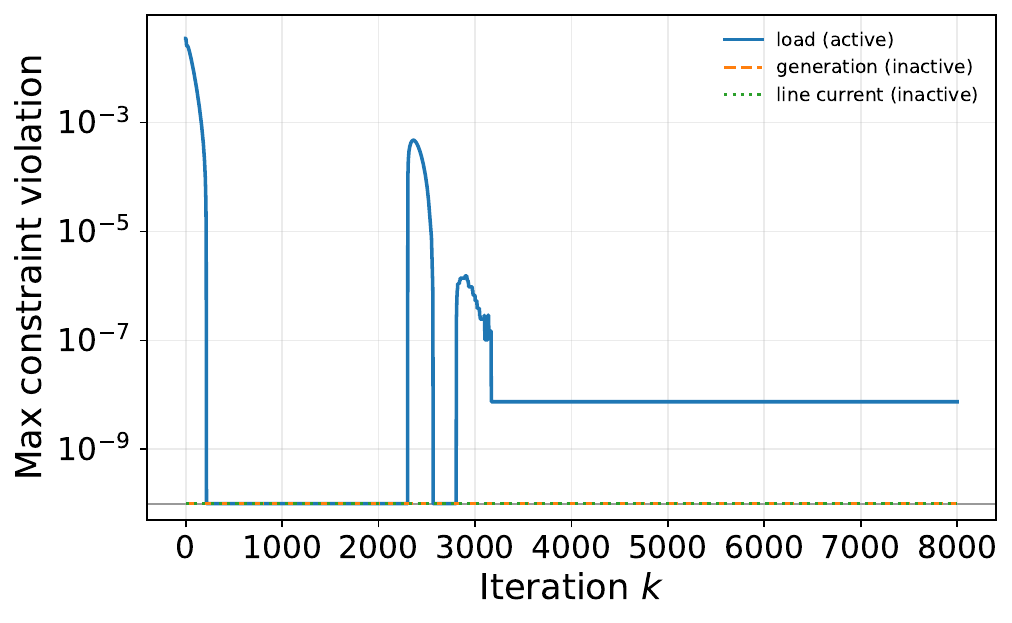}
        \label{fig:opf_violation}
    }%\hfill
    \subfloat[\footnotesize Optimal voltage profiles.]{
        \includegraphics[width=0.32\textwidth]
        {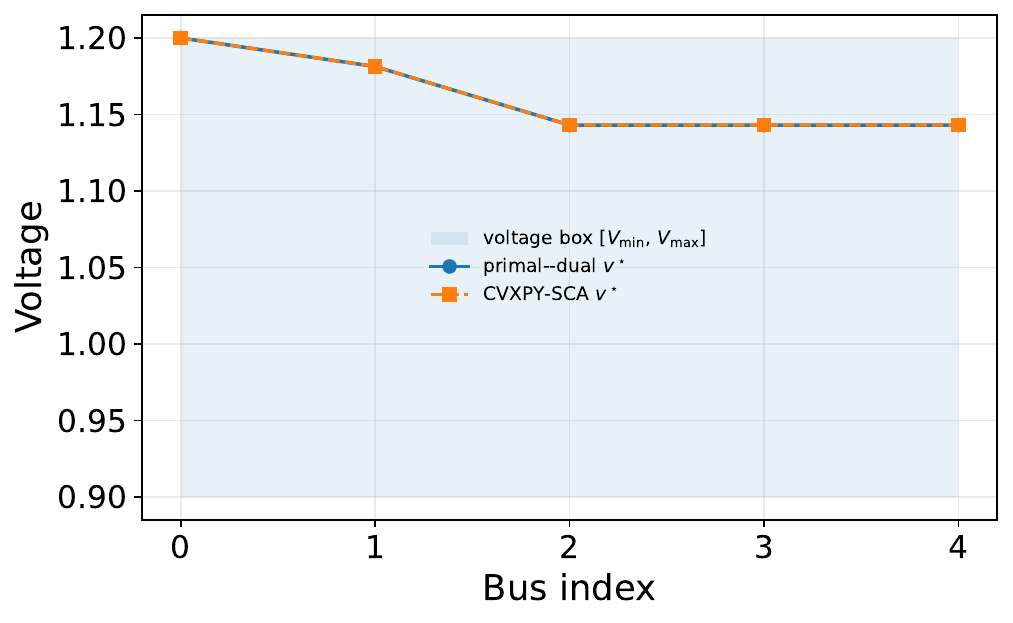}
        \label{fig:opf_voltage}
    }
    \caption{
Numerical results for the OPF case study.
(a) The projected primal--dual method restores feasibility and converges toward the CVXPY-SCA reference.
(b) The load-constraint violation converges to numerical tolerance, while the generation and line-flow constraints remain inactive.
(c) The primal--dual and CVXPY-SCA methods yield nearly identical feasible voltage profiles.
}
    \label{fig:case_study_opf_results}
\end{figure*}

In practice, the gradient $\nabla_vL$ is computed directly using PyTorch's autograd, which automatically provides appropriate subgradients for nonsmooth components such as absolute-value terms. This enables a streamlined, fully differentiable implementation of the primal--dual dynamics within a differentiable programming framework. Alternatively, the nonconvex power constraints $p_i(v)\ge d_i$ and $p_i(v)\le\overline{p}_i$ can be linearized around the current iterate via Successive Convex Approximation (SCA)~\cite{chiang2007power}, reducing each subproblem to a convex quadratic program. These convex subproblems can then be solved using CVXPY for conventional optimization or embedded as differentiable optimization layers using CVXPYLayers, producing primal and dual solutions while enabling end-to-end gradient backpropagation. Fig.~\ref{fig:case_study_opf_results} illustrates the numerical behavior of the projected primal--dual method and its comparison with CVXPY-SCA. The primal--dual iterations recover feasibility and converge to a near-KKT solution, while the final objective value and voltage profile closely agree with those obtained by CVXPY-SCA.

\subsection{Case Study: Laplacian Regularized Minimization Problem}
This section examines the application of differentiable programming to the Laplacian Regularized Minimization Problem (LRMP)~\cite{tuck2019distributed, chen2024manifold, wong2016quantifying, nk2012lx}, which minimizes a convex objective composed of a general convex function and a Laplacian-based smoothness regularizer. As a subclass of convex optimization, LRMP admits various solution strategies depending on its structural properties~\cite{boyd2004convex}.

\begin{problem}[Laplacian Regularized Minimization Problem]
Given a weighted graph with vertices $1,\dots,n$ and edge weights $w_{ij}\geq 0$, the regularizer takes the form $\mathcal{L}(x) = \sum_{(i,j)\in \mathcal{E}} w_{ij}(x_i - x_j)^2$, which enforces smoothness of the solution across neighboring nodes. Equivalently, this can be expressed using the weighted graph Laplacian $L \in \mathbb{R}^{n\times n}$, satisfying $L=L^\top$, $L_{ij}\leq 0$ for $i\neq j$, and $L\mathbf{1}=0$ for the all-ones vector $\mathbf{1}$. The LRMP is then formulated as:
\begin{equation}
\min_{x \in \mathbb{R}^n} \; F(x) = f(x) + \tfrac{1}{2} x^\top L x,
\label{prob:lrmp}
\end{equation}
where $f : \mathbb{R}^n \to \mathbb{R}\cup\{+\infty\}$ denotes a proper, closed, convex function.
\end{problem}

For~\eqref{prob:lrmp}, a point $x$ is optimal if and only if there exist $g\in\mathbb{R}^n$ and $g \in \partial f(x)$ such that:
\begin{align} 
g + \nabla \mathcal{L}(x) = g + L x = 0,
\quad \Rightarrow \quad
Lx = -g,
\label{eq:lrmp-kkt}
\end{align}
where $\partial f(x)$ denotes the subdifferential of $f$ at $x$~\cite{rockafellar1997convex, borwein2006convex, clarke1990optimization, tuck2019distributed}.

In particular, by standard results from linear algebra theory~\cite{boyd2004convex, strang2022introduction}, for any matrix $A \in \mathbb{R}^{m \times n}$ and linear equation $Qx=q$, the range space and null space of $Q$ are given by $\mathcal{R}(Q) = \{Qx \mid x \in \mathbb{R}^n\}$ and $\mathcal{N}(Q) = \{x \mid Qx = 0\}$, respectively. Whenever a solution exists, the general solution can be written as $x = x_p + x_n$, where $x_p$ is a particular solution satisfying $Qx_p = q$ and $x_n \in \mathcal{N}(Q)$ is an arbitrary null-space component. Applying this result to the Laplacian system $Lx = -g$, the solution decomposes into a component in $\mathcal{R}(L)$ and a component in $\mathcal{N}(L)$. Since $L\mathbf{1} = 0$, for some scalar $c \in \mathbb{R}$, the solution admits the form:
\begin{equation}
\label{eq:lrmp-x-solution}
x^\star = -L^{\dagger} g + c\,\mathbf{1}.
\end{equation}

We next consider a quadratic optimization problem with graph Laplacian regularization under nonnegativity constraints, which naturally falls into the class of LRMPs and arises in a variety of applications, including hyperspectral data unmixing with graph Laplacian regularization and nonnegativity constraints~\cite{ammanouil2015graph}, graph-regularized nonnegative matrix factorization~\cite{cai2010graph}, and semi-supervised graph Laplacian PageRank on large graphs~\cite{gao2011semi}. Specifically, let $A \in \mathbb{R}^{m \times n}$ and $b \in \mathbb{R}^{m}$ denote the data matrix and observation vector, respectively, and let $x \in \mathbb{R}^{n}$ be the decision variable subject to $x \succeq 0$. The resulting Laplacian-Regularized Nonnegative Least Squares (LR-NNLS) problem is formulated as:
\begin{equation}\label{prob:Learning_LRMP_primal} \tag{LR-NNLS-primal}
\min_{x \succeq 0} \;
\frac{1}{2}\|A x - b\|_2^2 + \frac{1}{2}\,x^\top L x,
\end{equation}
where $L \in \mathbb{R}^{n \times n}$ is the graph Laplacian encoding dependencies among the components of $x$. When the graph structure is fixed and $L$ is regarded as constant, the problem can be efficiently solved using CVXPYLayers. In contrast, if the graph structure changes and $L$ is no longer fixed, the problem cannot be directly formulated and solved in CVXPYLayers. 

To address this issue, we introduce the auxiliary variable $y = Ax - b$ and the dual variables $\mu \succeq 0$ and $\lambda \in \mathbb{R}^m$. The corresponding Lagrangian is given by:
\begin{equation}
\label{eq:Lrnnls_lagrangian}
\mathcal{L}(x, y; \lambda, \mu)
= \tfrac{1}{2}\|y\|_2^2 + \tfrac{1}{2}x^\top L x 
+ \lambda^\top (A x - b - y) - \mu^\top x .
\end{equation}

Based on the KKT conditions and~\eqref{eq:lrmp-x-solution}, the optimal primal solutions admit the representation:
\begin{align}
y^\star &= \lambda^\star, \label{eq:lrnnls-y-solution}\\
x^\star &= L^{\dagger}(\mu^\star - A^\top \lambda^\star) + c\,\mathbf{1}, \label{eq:lrnnls-x-solution}
\end{align}
where $\mu^\star$ and $\lambda^\star$ denote the optimal dual solutions, and the scalar $c$ is uniquely determined by primal feasibility and the complementary slackness conditions. To numerically validate~\eqref{eq:lrnnls-x-solution}, we solve an LR-NNLS instance with $(m,n)=(50,30)$ using CVXPY and reconstruct $x^\star$ from the dual solutions via a least-squares estimate of $c$, namely $c = \frac{1}{n}\mathbf{1}^\top(x^\star - z)$. The resulting reconstruction error is on the order of $10^{-6}$, confirming the theoretical consistency.

The above procedure can be implemented using modern differentiable programming frameworks such as PyTorch, providing a practical and scalable approach for solving both unconstrained and constrained LRMPs. As shown in~\eqref{eq:lrmp-x-solution} and~\eqref{eq:lrnnls-x-solution}, the primal solutions typically involve the Moore--Penrose pseudoinverse $L^{\dagger}$, whose explicit computation is costly and numerically delicate due to the singularity of graph Laplacians. PyTorch mitigates these challenges via optimized tensor operations, automatic differentiation, GPU acceleration, and support for iterative linear algebra routines, enabling efficient evaluation of dual objectives and gradients without explicit matrix factorizations. The resulting dual iterates can then be mapped back to primal variables through the corresponding optimality conditions. This perspective yields a flexible and scalable computational framework for large-scale LRMPs that fully exploits their dual structure.

\section{Conclusion}
\label{sec:conclu}
Differentiable programming provides a powerful foundation for large-scale optimization by representing algorithms as differentiable computation graphs and enabling end-to-end optimization. When combined with duality theory and first-order methods, it introduces principled modeling structure, verifiable guarantees, and new pathways for improving convergence. This paper presented the theoretical foundations of differentiable programming, analyzed its integration with cone programming, and demonstrated how first-order methods can be embedded within modern automatic-differentiation frameworks to achieve adaptive and accelerated performance on complex, high-dimensional problems. Through case studies in linear programming, aerospace and optimal control, wireless network optimization, optimal power flow, Laplacian regularization, and neural network verification, we illustrated how primal–dual updates and iterative schemes such as ADMM and PDHG can be efficiently implemented in PyTorch. Taken as a whole, differentiable programming, duality theory, and scalable first-order methods form a coherent toolkit for building adaptive, efficient, and certifiable optimization pipelines, pointing toward a new paradigm for solving the massive-scale problems that characterize modern applications.

\bibliographystyle{ACM-Reference-Format}
\bibliography{diffprog}

\end{document}